\documentclass[sigconf,balance=false,authorversion]{acmart}
\usepackage{pdfpages}
\usepackage{acmart-taps}
\usepackage{svg}

\usepackage[utf8]{inputenc}
\usepackage[T1]{fontenc} 

\usepackage{mathtools}
\usepackage{amsmath}
\usepackage{amsfonts}
\usepackage{url}
\usepackage[super]{nth}
\usepackage{wasysym}

\usepackage{booktabs}
\usepackage{multirow}
\usepackage{colortbl}

\usepackage{graphicx}
\usepackage{color}

\usepackage{balance}

\usepackage{ifthen}

\usepackage{xspace}

\usepackage{eurosym}

\usepackage{soul}

\definecolor{fixme}{rgb}    {0.8,0.3,0.3} 
\definecolor{fixed}{rgb}    {0.6,0.2,0.2} 
\definecolor{removed}{rgb}  {0.9,0.9,0.9} 
\definecolor{draftonly}{rgb}{0.5,0.5,0.5}
\definecolor{finalonly}{rgb}{0.5,0.5,0.5}

\newenvironment{SUBENVfixme}{\color{fixme}}{\color{black}}
\newenvironment{SUBENVfixed}{\color{fixed}}{\color{black}}
\newenvironment{SUBENVremoved}{\color{removed}}{\color{black}}
\newenvironment{SUBENVcomment}[2]{\color{#1}[#2:~}{]\color{black}}
\newenvironment{SUBENVdraftonly}{\color{draftonly}}{\color{black}}

\newcommand{\ignore}[1]{}

\newcommand{\fixme}[1]{\begin{SUBENVfixme}#1\end{SUBENVfixme}}
\newcommand{\fixed}[1]{\begin{SUBENVfixed}#1\end{SUBENVfixed}}
\newcommand{\removed}[1]{\begin{SUBENVremoved}#1\end{SUBENVremoved}}
\newcommand{\mycomment}[3]{\begin{SUBENVcomment}{#1}{#2}~#3\end{SUBENVcomment}}
\newcommand{\draftonly}[1]{\begin{SUBENVdraftonly}#1\end{SUBENVdraftonly}}
\newcommand{\finalonly}[1]{}
\newcommand{\draftorsubmissiononly}[1]{#1}
\newcommand{\lorem}[0]{\color{draftonly}Lorem ipsum dolor sit amet, consectetuer adipiscing elit. Aenean ante dolor, suscipit porta, posuere sit amet, dapibus non, turpis.\color{black}}
\newcommand{\loremipsum}[0]{\color{draftonly}Lorem ipsum dolor sit amet, consectetuer adipiscing elit. Aenean ante dolor, suscipit porta, posuere sit amet, dapibus non, turpis. Vivamus elit felis, pharetra a, tempor sit amet, aliquam aliquet, sapien. Cum sociis natoque penatibus et magnis dis parturient montes, nascetur ridiculus mus. Suspendisse potenti. Curabitur augue diam, gravida eget, malesuada et, sollicitudin a, felis. Morbi vitae dui. Vivamus eros risus, tempor id, rhoncus vitae, fringilla ac, sem. Praesent vehicula gravida quam. Aliquam aliquam. Nam volutpat blandit urna. Nam quis nunc non diam mattis vestibulum. Mauris non orci et urna molestie congue. Aliquam erat volutpat.\color{black}}

\newcommand{\finalAdd}[1]{\textcolor{blue}{#1}}
\newcommand{\finalDel}[1]{\textcolor{red}{\protect\sout{#1}}}
\newcommand{\finalChanged}[1]{\textcolor{author1}{#1}}
\newcommand{\finalReplace}[2]{\finalDel{#1} \finalAdd{#2}}
\newcommand{\finalHL}[1]{\textcolor{blue}{#1}}
\newcommand{\finalDelImage}[3]{
  \begin{figure}[H]
    \colorbox{red}{\makebox[\linewidth][c]{\includegraphics[width=\linewidth]{#1}}}
    \caption{\textcolor{red}{REMOVED: \sout{#2}}}
    \label{#3}
  \end{figure}
}

\newcommand{\finalDelEnumerate}[1]{%
  \begingroup
  \color{red} 
  \begin{tabular}[t]{@{}p{\linewidth}@{}} 
    \begin{enumerate}
      \setlength\itemsep{0pt} 
      \setlength\parskip{0pt} 

      #1
    \end{enumerate}
  \end{tabular}
  \endgroup
}

\newcommand{\finalDelItemize}[1]{
  \begingroup
  \color{red} 
  \renewcommand{\labelitemi}{\textcolor{red}{\sout{$\bullet$}}} 
  \renewcommand{\labelitemii}{\textcolor{red}{\sout{$\circ$}}} 
  \renewcommand{\labelitemiii}{\textcolor{red}{\sout{$\diamond$}}} 
  \begin{itemize}
    \color{red} 
    \sout{#1} 
  \end{itemize}
  \endgroup
}

\newcommand{\finalDelTable}[1]{
  \begingroup
  \captionsetup{font={color=red}} 
  \begin{table*}[!htbp]
    \centering
    \setlength{\fboxsep}{6pt} 
    \fcolorbox{red}{white}{ 
      \begin{minipage}{\dimexpr\textwidth-12pt} 
        #1 
      \end{minipage}
    }
  \end{table*}
  \endgroup
}

\newcommand{\setmode}[1]{
  \ifthenelse {\equal{#1}{final}} {
    \pagestyle{empty}
    \renewcommand{\fixme}[1]{##1}
    \renewcommand{\fixed}[1]{##1}
    \renewcommand{\removed}[1]{}
    \renewcommand{\mycomment}[3]{}
    \renewcommand{\draftonly}[1]{}
    \renewcommand{\loremipsum}[0]{}
    \renewcommand{\lorem}[0]{}
    \renewcommand{\finalonly}[1]{##1}
    \renewcommand{\draftorsubmissiononly}[1]{}
    \renewcommand{\finalAdd}[1]{##1}
    \renewcommand{\finalDel}[1]{}
    \renewcommand{\finalDelImage}[3]{}
    \renewcommand{\finalDelTable}[1]{}
    \renewcommand{\finalDelItemize}[1]{}
    \renewcommand{\finalDelEnumerate}[1]{}
    \renewcommand{\finalChanged}[1]{##1}
    \renewcommand{\finalHL}[1]{#1}

  } {
    \ifthenelse {\equal{#1}{submission}} {
      \renewcommand{\fixme}[1]{##1}
      \renewcommand{\fixed}[1]{##1}
      \renewcommand{\removed}[1]{}
      \renewcommand{\mycomment}[3]{}
      \renewcommand{\draftonly}[1]{}
      \renewcommand{\loremipsum}[0]{}
      \renewcommand{\lorem}[0]{}
    } {
      \ifthenelse {\equal{#1}{changes}} {
        \renewcommand{\fixme}[1]{##1}
        \renewcommand{\fixed}[1]{##1}
        \renewcommand{\removed}[1]{}
        \renewcommand{\mycomment}[3]{}
        \renewcommand{\draftonly}[1]{}
        \renewcommand{\loremipsum}[0]{}
        \renewcommand{\lorem}[0]{}
      } {
        \ifthenelse {\equal{#1}{draftbubblecomments}} {
      } { }
    }
  }
 }
}

\makeatletter
\def\url@tnystyle{%
  \@ifundefined{selectfont}{\def\UrlFont{\sf}}{\def\UrlFont{\footnotesize}}}
\makeatother
\newcommand{\footurl}[1]{\footnote{\url{#1}}}

\usepackage{colortbl}
\usepackage{makecell}
\usepackage{acmart-taps}

\setmode{final}

\usepackage{subcaption}
\usepackage{cleveref}
\usepackage[justification=centering]{caption}

\usepackage{booktabs}
\usepackage{comment}
\usepackage{siunitx}
\usepackage{listings}

\definecolor{author1}{rgb}     {0.9,0.5,0.0}
\definecolor{author2}{rgb}     {0.6,0.0,0.8}
\definecolor{author3}{rgb}     {0.0,0.5,0.0}
\definecolor{author4}{rgb}     {0.8,0.3,0.5}
\definecolor{author5}{rgb}     {0.5,0.0,0.0}
\definecolor{author6}{rgb}     {0.0,0.6,0.8}
\definecolor{author7}{rgb}     {0.1,0.5,1.0}

\definecolor{percentage2}{rgb}{0.9490, 0.9804, 0.9961} %
\definecolor{percentage3}{rgb}{0.8824, 0.9451, 0.9804} %
\definecolor{percentage4}{rgb}{0.7961, 0.8980, 0.9608} %
\definecolor{percentage5}{rgb}{0.8235, 0.8667, 0.9098} %
\definecolor{percentage6}{rgb}{0.6039, 0.7922, 0.8941} %
\definecolor{percentage7}{rgb}{0.5686, 0.7608, 0.8667} %
\definecolor{percentage8}{rgb}{0.3804, 0.7137, 0.8471} %
\definecolor{percentage9}{rgb}{0.3804, 0.6314, 0.7922}

\definecolor{skillRate1}{rgb}{0.949, 0.983, 0.934}
\definecolor{skillRate2}{rgb}{0.829, 0.914, 0.802}
\definecolor{skillRate3}{rgb}{0.655, 0.859, 0.663}
\definecolor{skillRate4}{rgb}{0.392, 0.739, 0.429}
\definecolor{skillRate5}{rgb}{0.2, 0.627, 0.373}

\newcommand{\ie}{\emph{i.e.~}}
\newcommand{\eg}{\emph{e.g.~}}

\newcommand{\ea}{{\xspace}\emph{et~al.}\xspace}
\newcommand{\os}{{\xspace}OpenSCAD\xspace}
\newcommand{\pb}{{\xspace}programming-based\xspace}
\newcommand{\Pb}{{\xspace}Programming-based\xspace}

\newcommand{\dd}{drag-and-drop\xspace}
\newcommand{\csg}{CSG\xspace}
\newcommand{\cadapps}{CAD applications\xspace}
\newcommand{\cadapp}{CAD application\xspace}
\newcommand{\cad}{CAD\xspace}
\newcommand{\cadsh}{\cad}

\newcommand{\code}[1]{\texttt{#1}\xspace}

\newcommand{\uquote}[1]{“\textit{#1}”}

\newcounter{goal}
\renewcommand{\thegoal}{G\arabic{goal}}
\newcounter{feature}
\renewcommand{\thefeature}{F\arabic{feature}}
\usepackage{tikz}

\newenvironment{itemize*}%
{\begin{compactitem}%
\setlength{\plparsep}{1em}}%
{\end{compactitem}}

\newenvironment{enumerate*}[1][]%
{\begin{compactenum}[#1]%
\setlength{\plitemsep}{2pt}}%
{\end{compactenum}}

\usepackage{wrapfig}

\newcolumntype{R}[2]{%
    >{\adjustbox{angle=#1,lap=\width-(#2)}\bgroup}%
    l%
    <{\egroup}%
}

\newcommand{\greyhline}{\arrayrulecolor[rgb]{0.753,0.753,0.753}\hline}

\definecolor{darkred}{RGB}{139,0,0}
\definecolor{darkgreen}{RGB}{0,100,0}
\definecolor{darkblue}{RGB}{0,0,139}

\lstdefinelanguage{OpenSCAD}{
  keywords={
    module, function, include, use,
    for, if, else, echo, projection, intersection, union, difference,
    circle, square, polygon, polyhedron, sphere, cube, cylinder, cone, torus,
    linear_extrude, rotate, translate, scale, resize, mirror, color, offset, hull, minkowski, assign
  },
  keywordstyle=\color{darkblue}\bfseries,
  ndkeywords={parameter},
  ndkeywordstyle=\color{teal}\bfseries,
  identifierstyle=\color{darkgreen},
  sensitive=false,
  comment=[l]{//},
  morecomment=[s]{/*}{*/},
  commentstyle=\color{teal}\itshape,
  stringstyle=\color{darkred},
  morestring=[b]',
  morestring=[b]",
  numberstyle=\color{black},
  numbers=left,
  keywords=[2]{true, false},
  keywordstyle=[2]\color{darkred},
  literate={0}{{\textcolor{darkred}{0}}}{1}%
    {1}{{\textcolor{darkred}{1}}}{1}%
    {2}{{\textcolor{darkred}{2}}}{1}%
    {3}{{\textcolor{darkred}{3}}}{1}%
    {4}{{\textcolor{darkred}{4}}}{1}%
    {5}{{\textcolor{darkred}{5}}}{1}%
    {6}{{\textcolor{darkred}{6}}}{1}%
    {7}{{\textcolor{darkred}{7}}}{1}%
    {8}{{\textcolor{darkred}{8}}}{1}%
    {9}{{\textcolor{darkred}{9}}}{1}%
    {.0}{{\textcolor{darkred}{.0}}}{2}%
    {.1}{{\textcolor{darkred}{.1}}}{2}%
    {.2}{{\textcolor{darkred}{.2}}}{2}%
    {.3}{{\textcolor{darkred}{.3}}}{2}%
    {.4}{{\textcolor{darkred}{.4}}}{2}%
    {.5}{{\textcolor{darkred}{.5}}}{2}%
    {.6}{{\textcolor{darkred}{.6}}}{2}%
    {.7}{{\textcolor{darkred}{.7}}}{2}%
    {.8}{{\textcolor{darkred}{.8}}}{2}%
    {.9}{{\textcolor{darkred}{.9}}}{2}%
    ,
}

\copyrightyear{2024}
\acmYear{2024}
\setcopyright{acmlicensed}\acmConference[UIST '24]{The 37th Annual ACM
Symposium on User Interface Software and Technology}{October 13--16,
2024}{Pittsburgh, PA, USA}
\acmBooktitle{The 37th Annual ACM Symposium on User Interface Software
and Technology (UIST '24), October 13--16, 2024, Pittsburgh, PA, USA}
\acmDOI{10.1145/3654777.3676417}
\acmISBN{979-8-4007-0628-8/24/10}

\begin{document}

\title{Facilitating the Parametric Definition of Geometric Properties in Programming-Based CAD}

\author{J. Felipe Gonzalez}
\orcid{0000-0002-0716-1689}
\affiliation{%
  \institution{Carleton University}
  \institution{Univ. Lille, CNRS, Inria, Centrale Lille, UMR 9189 CRIStAL}
  \postcode{F-59000}
  \city{Lille}
  \country{France}
}
\email{johannavila@cmail.carleton.ca}

\author{Thomas Pietrzak}
\orcid{0000-0002-2013-7253}
\affiliation{%
  \institution{Univ. Lille, CNRS, Inria, Centrale Lille, UMR 9189 CRIStAL}
  \postcode{F-59000}
  \city{Lille}
  \country{France}
}
\email{thomas.pietrzak@univ-lille.fr}

\author{Audrey Girouard}
\orcid{0000-0003-3223-105X}
\affiliation{%
  \institution{Carleton University}
  \city{Ottawa}
  \state{ON}
  \country{Canada}
}
\email{audrey.girouard@carleton.ca}

\author{G\'ery Casiez}
\orcid{0000-0003-1905-815X}
\affiliation{%
  \institution{Univ. Lille, CNRS, Inria, Centrale Lille, UMR 9189 CRIStAL}
  \postcode{F-59000}
  \city{Lille}
  \country{France}}
\additionalaffiliation{%
  \institution{Institut Universitaire de France}
  \city{Paris}
  \country{France}}
\email{gery.casiez@univ-lille.fr}

\renewcommand{\shortauthors}{J. Felipe Gonzalez, \ea}

\begin{abstract}

\finalReplace{Parametric Computer-aided design (CAD) allows users to create reusable models by integrating variables into the geometric properties of objects.
These variables can be adjusted to generate new versions of the models, offering significant advantages in fields like digital personal fabrication, where users can customize models without the need for a complete redesign.
However, creating parametric designs poses significant challenges, particularly within program\-ming based CAD applications.
In these applications, users define models using a programming language within a code editor while the application generates a visual representation in a viewport.
Users must employ complex programming and arithmetic expressions to describe geometric properties, linking various object properties to create parametric designs.
Unfortunately, these applications lack assistance in this process, making it unnecessarily demanding.
To address this issue, we propose a solution that enables users to retrieve parametric expressions from the visual representation for reuse in the code, thereby streamlining the parametric design process.}{
Parametric Computer-aided design (CAD) enables the creation of reusable models by integrating variables into geometric properties, facilitating customization without a complete redesign.
However, creating parametric designs in \pb CAD presents significant challenges.
Users define models in a code editor using a programming language, with the application generating a visual representation in a viewport.
This process involves complex programming and arithmetic expressions to describe geometric properties, linking various object properties to create parametric designs.
Unfortunately, these applications lack assistance, making the process unnecessarily demanding.
We propose a solution that allows users to retrieve parametric expressions from the visual representation for reuse in the code, streamlining the design process.}
We demonstrated this concept through a proof-of-concept implemented in the \pb CAD application, \os, and conducted an experiment with 11 users.
Our findings suggest that this solution could significantly reduce design errors, improve interactivity and engagement in the design process, and lower the entry barrier for newcomers by reducing the mathematical skills typically required in \pb CAD applications.

\end{abstract}



\maketitle

\section{Introduction}

Parametric \textit{Computer-Aided Design} (CAD) uses parameters to define object geometry, allowing quick modifications and reusability \cite{camba_parametric_2016}.
This flexibility supports practices like digital personal fabrication \cite{berman_anyone_2020}, enabling users to create and share customizable models \cite{stemasov_mixmatch_2020}.
For instance, web applications such as Customizer \cite{makerbot_industries_customizer_2024} from Thingiverse \cite{makerbot_industries_thingiverse_2024} and MakeWithTech \cite{shapiro_makewithtech_2024} enable users to upload models with exposed parameters, allowing others to generate various versions of the base models. Most CAD applications use direct manipulation, allowing users to edit models through visual elements like \dd, menus, and buttons \cite{shneiderman_direct_1983}.
They incorporate parametric features via \textit{constraints}, which are rules applied to control dimensions, positions, or relationships within the model using modifiable parameters \cite{coons_outline_1963, zou_review_2022}.
For example, FreeCAD \cite{tahiriyo_freecad_2024} lets users set line lengths as constraints linked to spreadsheet cells, updating the sketch when these values are changed.

Applications like \os \cite{kintel_openscad_2024} and JSCAD \cite{mueller_jscad_2024} follow a \textit{\Pb} approach \cite{gonzalez_understanding_2024}.
In \Pb \cad, models are defined textually using programming languages, with the application rendering a visual representation in a viewport after compilation.
Parametric designs are created by defining geometric properties through variables and arithmetic expressions, keeping relationships consistent regardless of parameter values.
For example, a variable \texttt{height} can set both the height of one cube and the position of another on top of it, ensuring both adjust correctly when the variable's value changes.
Creating parametric designs in \pb \cadapps is challenging due to the complexity of deriving expressions for geometric properties, requiring math skills and spatial reasoning \cite{gonzalez_understanding_2024}.
Even experts face difficulties aligning parameters with spatial axes, formulating correct expressions, and navigating nested transformations \cite{gonzalez_understanding_2024}.

\begin{figure}[h]
\centering
\includegraphics[width=\linewidth]{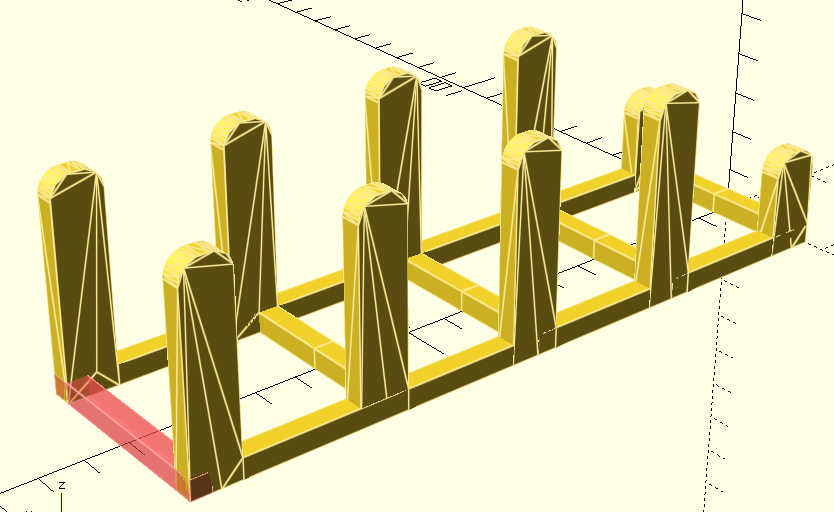}
\caption{Pot lid holder model from Thingiverse. ID 6402905}
\Description{FreeCAD, an example of direct manipulation CAD with parametric features. The figure shows the interface of the direct manipulation CAD application FreeCAD. It shows the view of a model and a spreadsheet panel with some parameters and values on the left. The model is parametric and controlled by the values in the spreadsheet.}
\label{fig:motivationConstraints}
\end{figure}

The geometric properties of different model parts are interrelated.
Consider the \os model 6402905 of a parametric \textit{"Pot lid holder"} from Thingiverse\footnote{\url{https://www.thingiverse.com/thing:6402905} accessed on 01/09/2024}, shown in Figure \ref{fig:motivationConstraints}.
The position of the highlighted part must be defined relative to the two \textit{rounded spikes}.
The user must determine the spikes' positions through spatial transformation definitions to align them with the center of the highlighted part, as shown in Listing \ref{lst:exampleGeomOS}.
This expression ensures that changes in the parameters controlling the spikes' positions automatically adjust the highlighted part's placement.
In the scenario where the highlighted part is about to be placed, the application already has the parametric definitions for the spikes' sizes and positions.
Allowing users to select components directly within the view could simplify the process, as visual identification is easier than code navigation \cite{gonzalez_introducing_2023, victor_bret_2013}.
For example, clicking on a spike in Figure \ref{fig:motivationConstraints} could reveal its parametric definition, enabling users to adjust it to place the highlighted part accurately.
This approach reduces the need for manual code analysis, speeding up the process.

\begin{lstlisting}[basicstyle=\small, firstnumber=65, caption={Example of parametric definition of a translate in OpenSCAD},label=lst:exampleGeomOS]
translate([-width/2,num*(length+spike_thickness)+spike_thickness-thickness,0])
cube([width,thickness,thickness]);
\end{lstlisting}

We aim to improve the parametric design capabilities within \pb \cadapps by introducing \textit{bidirectional programming interactions}.
Bidirectional programming allows users to directly leverage the information from the view in the code in \pb applications.
By analyzing 30 \os models from Thingiverse, we found that geometric properties are mainly linear combinations of variables.
Building on the work of Gonzalez \ea \cite{gonzalez_introducing_2023}, which enabled selecting elements directly in the view, we extended \os to allow users to extract parametric definitions from the view, simplifying model creation.
An evaluation with 11 \os users demonstrated the effectiveness of the solution, showing that it reduces design errors, improves interactivity, and lowers entry barriers by reducing the mathematical skill needed in \pb \cad.
Our contributions involve a formative study to understand better how geometric properties are defined in \pb \cadapps, a design goal to facilitate the process of parametric design in these applications, a proof of concept in \os, and a validation of the proposed solution.

Our implementation is available at \href{http://ns.inria.fr/loki/bp}{http://ns.inria.fr/loki/bp}.

\section{Related work}

First, we define the interaction paradigms of this work to frame our findings.
Then, we explain different approaches for parametric design in CAD applications.

\subsection{Interaction paradigms}

\textit{Programming-based CAD} refers to applications where users describe models entirely through coded instructions, with the system rendering the result in a viewport \cite{gonzalez_understanding_2024}.
In this paradigm, the code serves as the complete model description, and any model edits are reflected in the code.
Text-based applications such as JSCad~\cite{mueller_jscad_2024}, BRL-CAD~\cite{sekhar_nayak_brl-cad_2024}, and \os~\cite{kintel_openscad_2024} fall into this category. In addition, CAD applications that use visual programming, such as BlockSCAD~\cite{blockscad_inc_blockscad_2023} or Grasshopper \cite{davidson_grasshopper_2023}, are also considered \pb CAD.

\finalDel{
Some applications allow users to modify models using code within a direct manipulation interface, a concept known as \textit{Content} \textit{Oriented} \textit{Programming} \cite{mcguffin_categories_2020}.
FreeCAD~\cite{freecad_python_2023} and Blender~\cite{foundation_blenderorg_2023}, for instance, enable users to create models through direct manipulation interactions and provide text editors for executing scripts to modify the model.
These approaches differ fundamentally from programming-based workflow; code here is used to perform specific actions rather than comprehensively describing the visually represented 3D model.
In programming-based paradigms, modifying the model entails editing the code coherently and re-executing all scripts to regenerate the model, whereas, in other paradigms, the code executes actions to modify the model's current state.}

McGuffin and Fuhrman \cite{mcguffin_categories_2020} introduce the concept of \textit{Bidirectional Programming} applications, where users can modify the output using both direct manipulation and instructions.
In such applications, changes made to the output through direct manipulation trigger updates in the code to maintain coherence, as seen in scalable vector graphics (SVG) environments like Sketch-N-Sketch \cite{hempel_sketch-n-sketch_2019} or Twoville \cite{johnson_computational_2023}.
Bidirectional programming CAD applications \cite{cascaval_differentiable_2022,keeter_antimony_2024,keeter_libfive_2024,gonzalez_introducing_2023} adhere to the programming-based CAD paradigm by keeping the code as the full model description, while also extending the interaction capabilities to include direct manipulation in the viewport,\finalReplace{Antimony \cite{keeter_antimony_2023} is a programming-based \cadapp that allows users to utilize visual programming to edit models directly in the viewport, with the system updating the instructions coherently}{as seen in applications such as Antimony \cite{keeter_antimony_2024}.}

CadQuery \cite{urbanczyk_cadquery_2024} is a \pb framework for modeling design, using Boundary Representation (BREP) \cite{hoffmann_geometric_1989} to specify geometric information of objects' faces, vertices, or edges.
Users can modify a part by selecting it through a query and applying editing commands.
However, deriving queries in complex models can be challenging and require users to connect the code with the view mentally.
\finalDel{
Furthermore, selecting a part is notably easier to do directly in the view. }
Mathur \ea \cite{mathur_interactive_2020} developed features allowing users to extract queries through mouse clicks directly from the view where selecting parts is notably easier \cite{hutchins_direct_1985}.
We draw on this concept to facilitate the retrieval of information from the view.

Gonzalez \ea \cite{gonzalez_introducing_2023} present a modified version of \os with editing and navigation features and direct manipulation interactions.
The application allows users to edit the model directly in the view, applying spatial transformation through drag-and-drop interactions.
Although features facilitate the editing of the model, modifications only support spatial transformation with raw numbers without using arithmetic expressions and do not facilitate the design of parametric models.
Navigation features allow users to connect the code and the view with visual cues.

We leverage the concept of bidirectional programming, specifically in Mathur \ea \cite{mathur_interactive_2020} and Gonzalez \ea's work \cite{gonzalez_introducing_2023}, as the foundation for developing our solution.

\vspace*{-3pt}

\subsection{Parametric design in direct manipulation \cadapps}

Parametric \cadapps implementing a direct manipulation approach fix geometric properties through \textit{constraints} \cite{hillyard_analysis_1978, coons_outline_1963}.
A constraint is a rule applied to geometric elements within a model to control dimensions, positions, or relationships between components.
There are two types of constraints: Geometric and Dimensional \cite{lin_variational_1981}.
Geometric constraints define the relationship between two or more elements in the scene.
For example, to force two lines always to keep the same length.
Dimensional constraints fix the values of geometric properties of elements such as positions, sizes, or angles.
Table \ref{tab:typeConstraints} describes common constraints in \cadapps such as FreeCAD \cite{tahiriyo_freecad_2024,riegel_sketcher_2024}, Fusion360 \cite{autodesk_inc_fusion_2024,autodesk_inc_fusion_2024-1}, or AutoCAD~\cite{autodesk_inc_autocad_2024,autodesk_inc_autocad_2024-1}.

Constraints help define and enforce specific geometric relationships between different design parts.
In parametric design, these constraints are often expressed using variables, which users can adjust to create various versions of the design \cite{joan-arinyo_combining_1999}.
For instance, consider a box design where a variable defines the width.
By exposing this variable as a model parameter, users can easily modify its value as needed.
Subsequently, the application will regenerate the box with the updated width value.
\finalDel{Figure \ref{fig:FreeCADParametric} shows a typical case in FreeCAD of a parametric design.
Geometric properties are linked to the value of cells in a spreadsheet.
Users can modify cells' values to re-generate a new version of the model.}
\finalDel{In \cadapps, constraints are typically expressed as algebraic equations \cite{lin_variational_1981}.
These equations form a system that the application solves using a \textit{geometric} \textit{constraint} \textit{system} \cite{zou_review_2022}. The outcome of this system depends on the consistency of the defined equations.
When all geometric properties are uniquely and coherently defined with constraints, the system is called \textit{"fully} \textit{constrained,"} indicating a well-defined equation system with a unique solution.
In cases where not all geometric properties are constrained but those that are defined coherently, the system is considered consistent and under-constrained.
In such instances, multiple solutions exist, and the application must decide which solution to apply.
For example, a user applies a constraint to make them parallel two lines initially pointing in different directions.
The application then has several options to resolve this constraint.
It can maintain one line in its original position while adjusting the other or vice versa.
Alternatively, the application may choose to relocate both lines to make them parallel.
Geometric properties can also be defined with conflicting constraints, making the system inconsistent.
An illustrative case is applying a parallel constraint followed by an orthogonal constraint to two lines \cite{zou_review_2022}.}
Even for experts, creating constraints can be challenging.
Solutions like CODA \cite{veuskens_coda_2021} assist by suggesting applicable constraints based on elements in the view.
We drew inspiration from this concept, recognizing that leveraging existing application information can enhance our solution's definition of new geometric properties.

\begin{table}[!htbp]
\centering
\caption{Common Constraints in Direct Manipulation \cad}
\Description{A table describing the common constraints that direct manipulation CAD applications allowing parametric modeling provides. The table is split into two sections, the first describing Geometric constraints and the second describing dimensional constraints.}
\begin{tabular}{@{}p{8.5cm}@{}}
\toprule
\textbf{Geometric }\\
\midrule
\textbf{Coincident:} Forces two points or objects to share the same location. \\
\textbf{Collinear:} Requires two elements to lie on the same straight line. \\
\textbf{Concentric:} Enforces a common center point for two circles. \\
\textbf{Parallel:} Aligns two lines or edges to be parallel. \\
\textbf{Perpendicular:} Forces two lines or edges to meet at a right angle.
\finalDel{\textbf{Tangent:} Ensures that a curve or circle is tangent to another curve or circle at a specified point. \\
\textbf{Symmetry:} Requires elements to be symmetric with respect to a specified axis or plane. \\
\textbf{Perpendicular Bisector:} Defines a line that is perpendicular to and passes through the midpoint of another line. \\ \textbf{Midpoint:} Forces two points to share the same midpoint. \\} \\
\hline
\textbf{Dimensional } \\
\hline
\textbf{Distance:} Specifies the distance between two points or objects.\\
\textbf{Angle:} Defines the angle between two lines or edges.\\
\textbf{Radius/diameter:} Sets the radius/diameter of a circle or arc.\\
\textbf{Length:} Determines the length of a line or the size of an object.\\
\textbf{Width/height:} Specifies the width or height of an object.
\finalDel{\textbf{Depth:} Sets the depth or thickness of an object.\\
\textbf{Chamfer:} Creates a beveled edge or corner.\\} \\
\bottomrule
\end{tabular}\label{tab:typeConstraints}
\end{table}

It is noteworthy that in direct manipulation \cadapps users express design intents through tools that are described in a more explicit language compared to \pb \cad.
As seen in Table \ref{tab:typeConstraints}, constraints include high-level definitions such as making two lines collinear.
This forces the application to interpret them and propose a solution.
In other words, the user expresses \textit{\textbf{WHAT}} they want, and the system seeks a solution to provide it.
This differs from \pb \cadapps where users need to describe \textit{\textbf{HOW}} the models are built.

\finalDelImage{gfx/FreeCADPar.png}{In FreeCAD, users can define geometric properties referencing variable cells of a spreadsheet (on the right) and re-execute the features in the history tree (on the left).}{fig:FreeCADParametric}

\subsection{Parametric design in \pb \cadapps}

When designing in \pb \cadapps, users define all the geometric properties of the model, except when the application provides default values to information not provided in the code.
For example, creating a cube without specifying its size parameter results in a default cube of size 1$\times$1$\times$1 being generated.
\finalDel{Under the definition of the constrained system, we could arguably say that \pb models are always consistent and fully constrained, although constraints definitions are different than in direct manipulation applications.
The equivalents of the dimension constraints (Table \ref{tab:typeConstraints}) would define the code's sizes, positions, and orientations.
For instance, the "distance" constraint could be expressed as applying a \code{translate} transformation.
On the other hand, the user must describe geometric constraints, such as keeping two lines parallel within the code.} Users describe \textit{\textbf{HOW}} geometric properties are formed through programming and mathematical expressions.
Understanding how users in \pb environments define geometric properties in parametric designs is crucial to facilitating the design process.
However, beyond code comments, there is often no clear indication of the users' intentions behind these definitions.
For example, in the provided example in Listing \ref{lst:exampleGeomOS}, the rationale behind specific choices for location and size definitions is not immediately apparent.
Furthermore, there is a lack of research investigating design patterns in defining geometric properties in \pb \cad.
Chytas and Tsilingiris \cite{chytas_learning_2018} study how 13 to 17-year-old students create \pb models.
Later, Chytas \ea \cite{chytas_exploring_2019} studied several \os models from websites to identify programming patterns and design preferences. Their research provides statistics on various code statements (\eg frequency of loops, conditionals, or spatial transformation usage) but does not delve into how geometric properties are interrelated with other objects.
Previous research \cite{gonzalez_understanding_2024} with \os users has shown that users often struggle to formulate mathematical expressions for parametric models, a process scarcely supported by existing tools.
Furthermore, users indicated that the definition of objects' positions and sizes is frequently relative to the positions and sizes of other objects, highlighting a complex interdependence in design decisions.

We draw on these works to address the identified challenges.

\section{Method}

We aim to facilitate the parametric design in \pb \cadapps.
First, we conducted a formative study analyzing 30 \os models from Thingiverse to identify how the geometric properties are defined.
Then, based on our findings, we define the design goals for a bidirectional application that facilitates the definition of geometric properties in parametric models.
Later, we reused and modified the source code of Gonzalez \ea \cite{gonzalez_introducing_2023} to allow users to retrieve parametric definitions of objects directly from the view to be reused in the code.
Finally, we tested our modified version with \os users and analyzed their user experience.

\subsection{Formative study}

\Pb \cadapps allow users to define geometric properties with programming expressions.
For example, the size of a cube can be defined with a raw number, a variable, an arithmetic expression, or more complex programming structures such as conditionals.
Based on Gonzalez \ea's survey about the challenges of \os users~\cite{gonzalez_understanding_2024}, we hypothesize that users usually define the positions and sizes of elements as a linear combination of the positions and sizes of other elements.
For example, a common operation is placing a box on top of another box. In such a case, the position of the second box is defined in terms of the size and position of the first cube, as depicted in Listing \ref{ex2Connstraints}.

\begin{lstlisting}[firstnumber =4, basicstyle=\small, caption={Parametric model of a cube on top of another cube.}, label=ex2Connstraints]
// First cube
cube(size = size_cube_a, center = true);
// Second cube
translate([0,0,size_cube_a/2 + size_cube_b/2])
cube(size = size_cube_b, center = true);
\end{lstlisting}

With more complex models, the definition of geometric properties considers the position and orientation of multiple parts.
As a result, we hypothesize that, often, they are described as linear combinations defined as
\texttt{translate([tx, ty, tz])} where \(t_i\) = \(\sum \alpha_j x_j + c\), with \(\alpha_j, c\) constants, and \(x_j\) a variable.

To assess this assumption, we have analyzed 30 models from Thingiverse.
\finalDel{Accessing the website on 02/11/2023, w} We filtered the customizable models with the option \uquote{Popular Last 7 Days} and downloaded the first ten models.
We repeated the same process with the filters \uquote{Popular Last 30 Days} and \uquote{Popular This Year}.
Duplicated models were discarded and replaced with the next ones on the list.
Figure \ref{fig:figmodels} in the appendix \ref{sec::appendixModels} contains all the models used in the formative study.

We modified \os to analyze the definition of geometric properties.
The application parses the code into an \textit{Abstract Syntax tree} (AST) \cite{aho_compilers_2006}.
Then, it scans the AST to identify code statements responsible for generating primitive geometries (\eg spheres or cylinders) and executing spatial transformations (\ie translations, rotation, scale).
The application identifies the nature of parameter definitions, either \textit{C1} for non-default raw numerical values, \finalDel{(\eg \texttt{5})} \textit{C2} for a single variable, \finalDel{(\eg \texttt{var1})} \textit{C3} for a linear combination of variables, \finalDel{(\eg \texttt{3 + 2*var1 - var2})} \textit{C4} for a non-linear combination, \finalDel{(\eg \texttt{3 + 2*var1*var2})} or \textit{C5} for a structure involving more complex programming constructs, \finalDel{(\eg \texttt{(var1>3)?1:2})} as delineated in Table \ref{tab:classExpressions}.
For example, a cube defined as \texttt{cube(size = [5, size\_y, size\_z+3])} would be classified under C1, C2, and C3, whereas a spatial transformation like \texttt{translate([0,0, size\_x*i])} would be allocated solely to the C4 category.
C1 and C2 are included in the C3 definition, but we keep the difference seeking a detailed analysis.

\vspace*{-6pt}

\begin{table}[h]
\centering
\caption{Categories of expressions in \os models.}
\Description{Categories used in the formative study. The table explains the categories used in the formative study to classify expressions in OpenSCAD models. It has three columns: the first is an identifier for the category starting in C1 and finishing in C5, the second is the category number, and the third column gives a short description of the category with an example.}
\begin{tabular}{c p{1.8cm} p{4.7cm}}
\toprule
\textbf{ID} & \textbf{Classification} & \textbf{Description} \\
\midrule
C1 & Raw number &  Non-default numeric. e.g., \texttt{4.0} \\
C2 & One variable & A variable call. e.g., \texttt{var1}\\
C3 & Linear combination & Linear combination of variables $\sum \alpha_i \cdot x_i + c$. \eg,~\texttt{3 + 2*var1 - var2}\\
C4 & Polynomial expression & Non-linear polynomial expressions $\sum \alpha_i \cdot x_i \cdot y_i$.  \eg,~\texttt{3 + 2*var1*var2}\\
C5 & Other & Other programming structures such as conditionals. \eg,~\texttt{(var1>3)?1:2}\\
\bottomrule
\end{tabular}
\label{tab:classExpressions}
\end{table}

The results of the analysis are detailed in Table \ref{tab:resultsFormative}.
It is important to note that the \texttt{scale} statement is barely used in the models.
Its participation in the total of expressions analyzed is only 1\%.
Furthermore, most expressions within the \texttt{rotate} statements are raw numbers, with 140 out of 286 \texttt{rotate} statements.
Indeed, when validating the results, we confirmed that in most cases, rotations are performed at standard angles such as 45 or 90 degrees.
Finally, we confirmed our hypothesis, verifying that most of the positions (through \texttt{translate} statements) and sizes (through primitive definitions) are defined as raw numbers (C1), one variable call (C2), or a linear combination of existing variables (C3).
These categories represent 71\% of the total parameters analyzed in primitives definition (44\%) and spatial transformations (27\%).

\begin{table}[!ht]
\centering
\small
\caption{Formative study results. Total and percentual occurrences per category used to define parameters in primitives, translation, rotation, and scale statements}
\Description{Formative study results. The table presents findings from a formative study featuring 6 columns and 7 rows. The top row delineates code statements, encompassing primitives and various transformations analyzed. The first column categorizes different aspects. Cells in rows 2 to 6 and columns 2 to 5 display occurrences of each category (row) within the corresponding code statement (column), accompanied by percentage values relative to the total expressions analyzed. These cells are shaded in blue, with intensifying shades denoting higher percentages. The final column tabulates total occurrences per category alongside percentages. Similarly, the last row showcases total occurrences per code statement, complete with percentages. The last column and row are highlighted in green, with darker hues representing higher percentages. The cell at the intersection of the last column and row displays the total number of analyzed expressions.}
\begin{tabular}{lrrrr|r}
\toprule
& \textbf{Primitive} & \textbf{Translate} & \textbf{Rotate} & \textbf{Scale} & \textbf{Total}\\
\midrule
\textbf{C1} & \cellcolor{percentage5}196 (11\%) & \cellcolor{percentage4}130 (7\%) & \cellcolor{percentage5}140 (8\%) & 8 (0.4\%) & \cellcolor{skillRate2}474 (25\%) \\
\textbf{C2} & \cellcolor{percentage9}294 (16\%) & \cellcolor{percentage4}126 (7\%) & \cellcolor{percentage2}35 (2\%) & 2 (0.1\%) & \cellcolor{skillRate2}457 (25\%)\\
\textbf{C3} & \cellcolor{percentage9}312 (17\%) & \cellcolor{percentage7}234 (13\%) & \cellcolor{percentage2}29 (2\%) & 5 (0.3\%) & \cellcolor{skillRate3} 580 (31\%)\\
\textbf{C4} & 0 (0\%) & \cellcolor{percentage2}48 (3\%) & \cellcolor{percentage2}31 (2\%) & 4 (0.2\%) & 83 (4\%)\\
\textbf{C5} & 26 (1\%) & \cellcolor{percentage6}191 (10\%) & \cellcolor{percentage2}51 (3\%) & 0 (0\%) &\cellcolor{skillRate1} 268(14\%)\\
\hline
\textbf{Total} & \cellcolor{skillRate4}828 (45\%) & \cellcolor{skillRate3}729 (39\%) & \cellcolor{skillRate1}286 (15\%) & 19 (1\%) & \cellcolor{skillRate5} 1862 (100\%) \\
\bottomrule
\end{tabular}
\label{tab:resultsFormative}
\end{table}

\subsection{Design goals}

Users define geometric properties by using the relationships between objects, as outlined in previous research \cite{gonzalez_understanding_2024} and corroborated by the formative study.
Concerning sizes and positions, statements frequently define these properties as linear combinations of variables, corresponding to linear relationships between elements.
The positioning and sizing of a new model element often depend on the location and dimensions of another object.
Moreover, identifying a position is straightforward in the visual representation \cite{mathur_interactive_2020, gonzalez_introducing_2023}.
A valuable tool would enable the extraction of model information to define the geometric properties of other elements within the code.
This information must be readily accessible, allowing users to understand its spatial implications directly in the visual representation, where identification is the simplest.
In essence, the design goal is to \emph{facilitate the extraction of geometric information from objects' parametric definitions in the view for code reuse}.

\section{Bidirectionnal programming to define geometric properties}

We have implemented a proof-of-concept that introduces features that enhance the creation of parametric models using bidirectional programming.
We have re-used the modified version of \os from Gonzalez \ea \cite{gonzalez_introducing_2023}.
Specifically, we reused the implemented feature to select an element in the view by clicking on it.
We modified this version based on Mathur \ea \cite{mathur_interactive_2020} work to fulfill the design goal by allowing users to retrieve information from the view.

\os parses code into an AST and later into the Abstract \csg Tree, evaluating all programming structures and variables and replacing them with the raw values, retaining only numeric information after evaluation.
No information about the parametric definition of objects is stored at this stage.
We modified the source code of \os to ensure that \csg tree nodes store the parametric definition of the geometric properties used to define them.
Primitives store the definition of the size, whereas spatial transformation stores the parametric definition of the transformation.

When users position an object relative to another, they are often concerned with specific locations around it.
For instance, when placing cube \textit{A} on top of cube \textit{B}, the \texttt{translate} statement needs to consider the top of cube \textit{B} and the bottom of cube \textit{A}.
Unfortunately, most \pb \cadapps work with a \csg representation where definitions are abstract, and there is no information on vertices, faces, or edges \cite{requicha_representations_1980, hoffmann_geometric_1989}.
We redefined \csg node definitions in \os to include \textit{handles} that the user can use to retrieve the parametric definition of the position of an object.
Handles were added, creating a grid of 3$\times$3$\times$3 points distributed symmetrically around the object's center in 3D primitives.
For 2D primitives, the application created a 3$\times$3 grid.
Details about the distribution of handles are shown in Table \ref{tab:control_points} in  Appendix \ref{sec::appendixModels}.

\begin{figure}[h]
\centering
\includegraphics[width=0.85\linewidth]{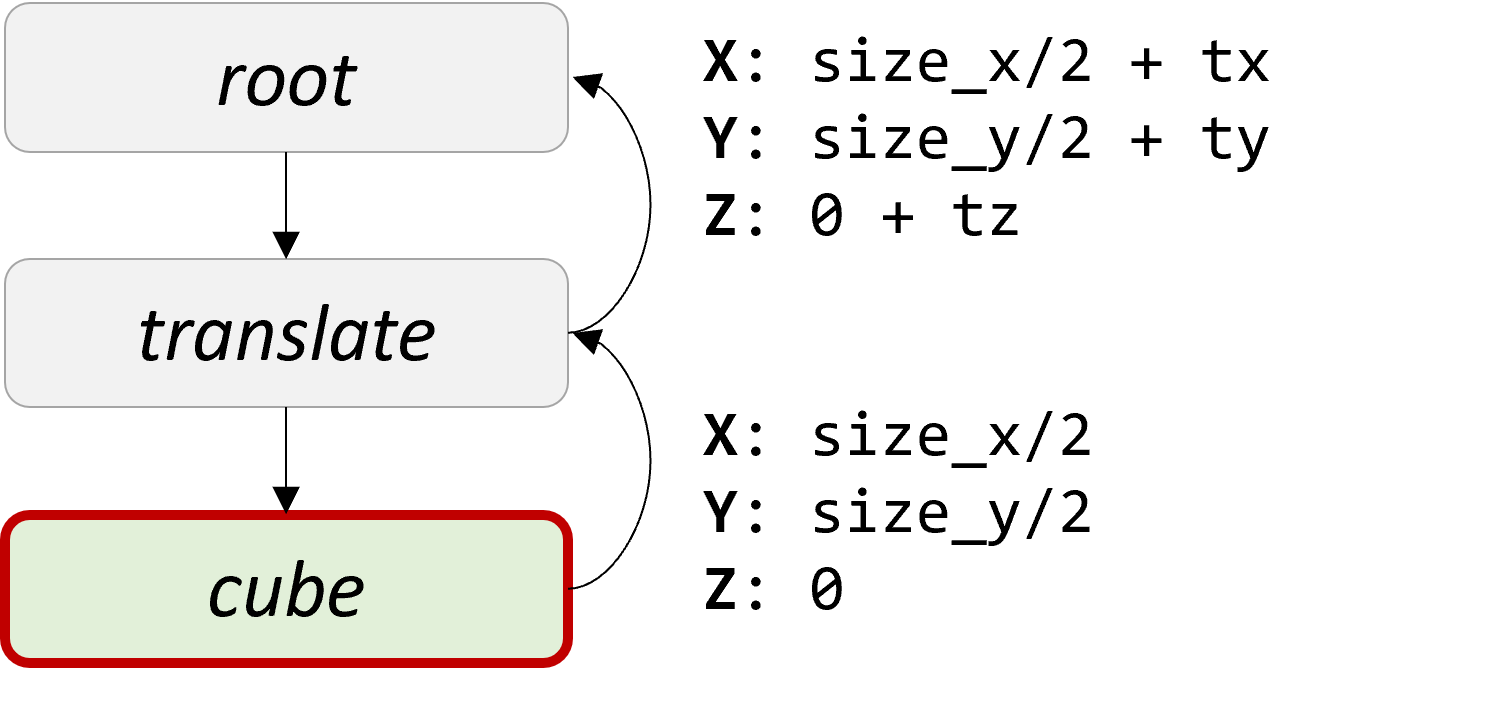}
\caption{Representation of how \os computes the position of the handle placed at the center of the bottom face of a cube.}
\Description{This figure illustrates how OpenSCAD features extract information from the view. The diagram depicts the CSD (Constructive Solid Geometry) nodes of the code example provided in Listing 3. With three nodes represented, each row illustrates the process of computing the value of a handle selected within a cube. The computation progresses backward through "translate" statements, culminating in the root. Each step in the calculation of expressions is visually delineated, showcasing the iterative process employed by the system.}
\label{subfig:controlPointExample}
\end{figure}

Non-primitive nodes include a single handle at the node's position, covering boolean operations, spatial transformations, or programming structures.
Given a determined handle of a selected node, the application can define the position of the handle in terms of the variables used in the code.
The application iterates on the \csg tree to locate the selected node.
Then, the selected node provides the definition of the position of the handle relative to the node's center.
Later, the application iterates on \texttt{translate} nodes in the branch of the selected node, adding their definitions to the position of the handle.
Figure \ref{subfig:controlPointExample} describes how \os derives the parametric position of the handle placed in the middle of the bottom face of the cube (axis \textit{z}) created by the code in Listing \ref{lst:exampleControlPoints}.

\begin{lstlisting}[basicstyle=\small, firstnumber =7, caption={Handles example},label=lst:exampleControlPoints]
translate([tx,ty,tz])
cube([size_x,size_y,size_z]);
\end{lstlisting}

\finalDel{\os defines the position of the handle by aggregating all \texttt{translate} definitions from the root to the selected object and incorporating the handle's local position relative to the selected object's center.}
\os gathers information from the CSG nodes without filtering out trivial values, such as translating 0 units in one direction, leading to less readable expressions and requiring an expression simplification.
To streamline the development, we implemented a Python server that exposes a service to simplify arithmetic expressions using the \textit{simpy} library \cite{team_simpy_overview_2024}. \os sends the raw expression to the server, which answers with a simplified expression.
If there is a communication error with the server, \os uses the non-simplified expression instead.
We developed two features to facilitate information retrieval from the model's view.
These features, built on the modified version of Gonzalez \ea \cite{gonzalez_introducing_2023}, enable users to \textit{select} objects within the model and use the capabilities of \emph{position} and \textit{delta vector}.

\subsection{Position}

The position feature allows users to determine the location of a handle in a selected object relative to the origin (\ie, \csg root position \texttt{[0,0,0]}) of the view.
Users activate this feature by selecting the\finalReplace{\emph{Absolute location}}{\emph{Position}} button in the menu bar.
Then, users can select an object \cite{gonzalez_introducing_2023}
and the application displays the handles, marking the object's center with an always-visible purple handle.
The rest of the handles behave like any other geometry and can be hidden behind other geometries.
This feature aims to provide information to the user without automatically editing the code, ensuring user control over the definition of the model \cite{mathur_interactive_2020}.
Users can right-click on any handle to turn it green, indicating that the application has copied the parametric position to the clipboard so the user can use it in the code to define a new element property.

\begin{figure}[!thbp]
\centering
\subfloat[Preview of a cup in progress.]{\begin{minipage}{.49\linewidth}\label{subfig:absLocex1}\centering\includegraphics[trim = 16cm 2cm 1.5cm 8cm, clip, height=2.7cm]{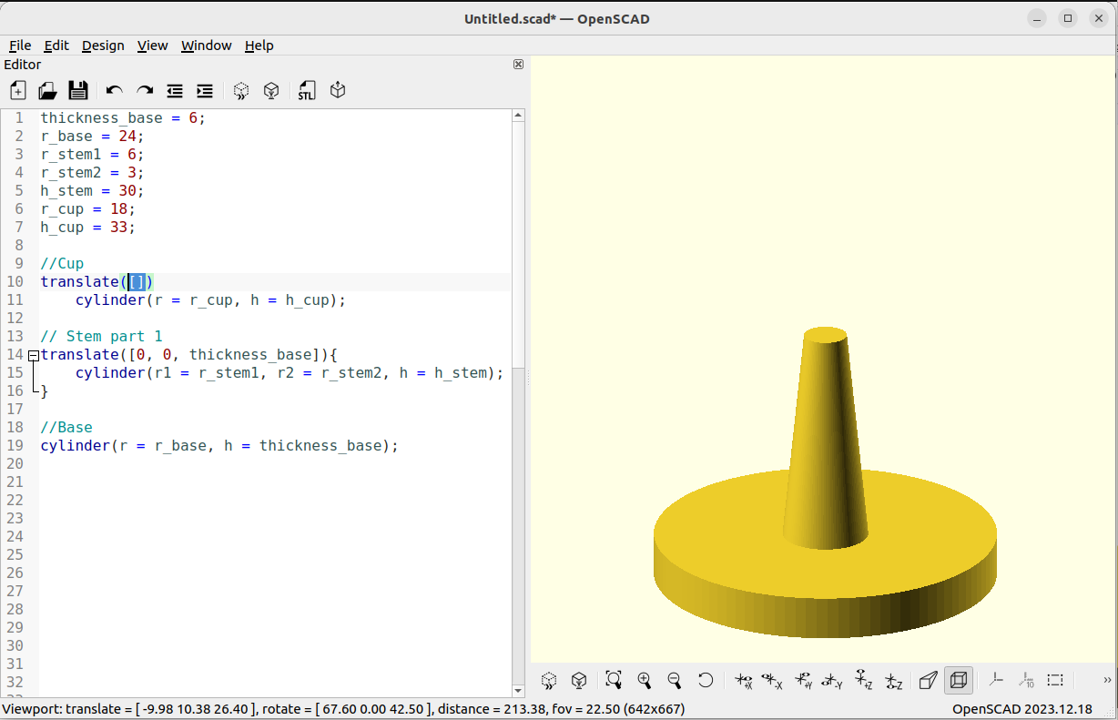}\end{minipage}}
\subfloat[The user select an object]{\begin{minipage}{.49\linewidth}\label{subfig:absLocex2}\centering\includegraphics[trim = 16cm 2cm 1.5cm 7.5cm, clip, height=2.7cm]{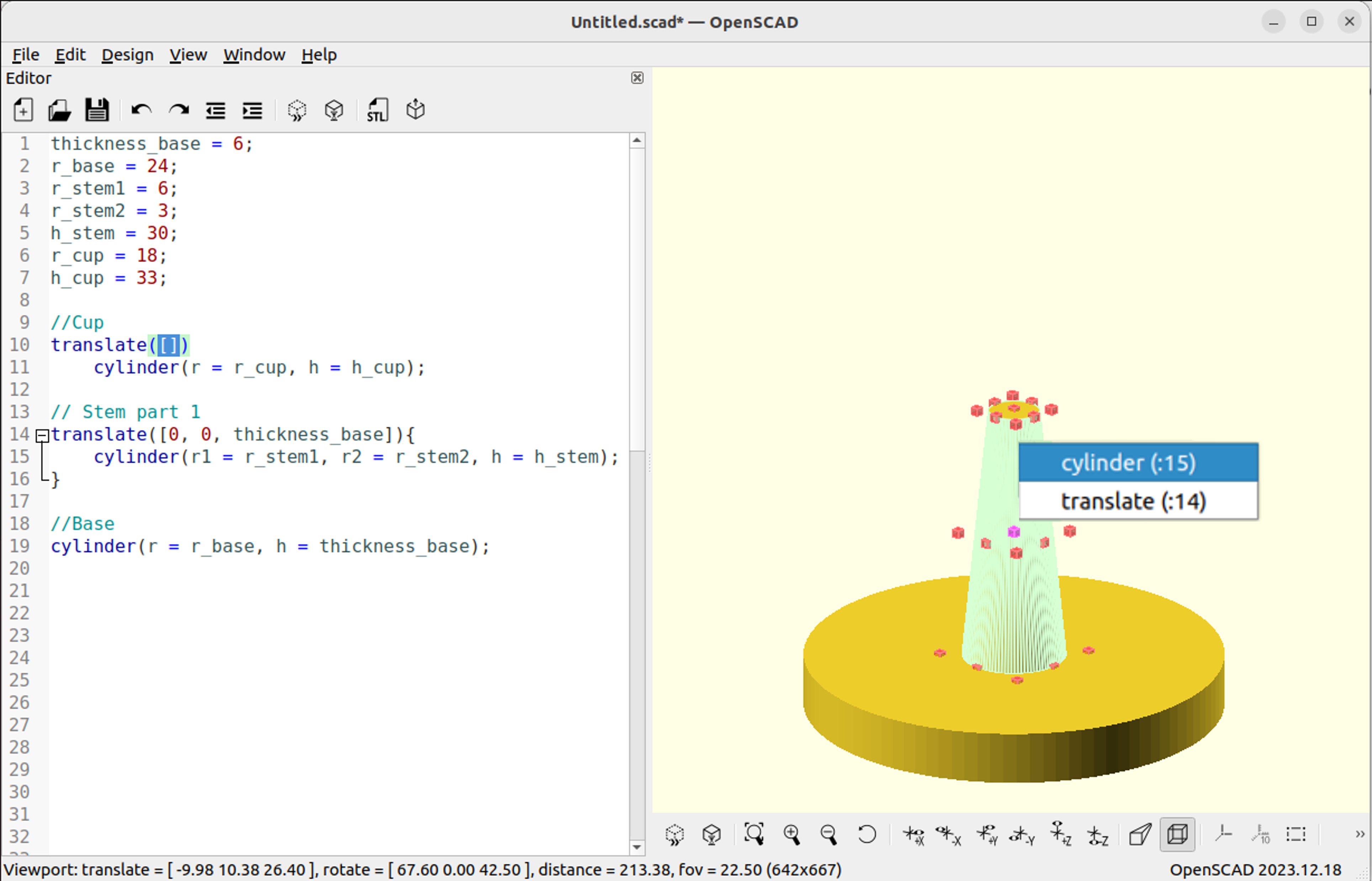}\end{minipage}} \\
\subfloat[A handle turns green after right-clicking it. The definition is stored in the clipboard.]{\begin{minipage}{.49\linewidth}\label{subfig:absLocex3}\centering\includegraphics[trim = 19cm 6cm 4.5cm 8.2cm, clip, height=2.7cm]{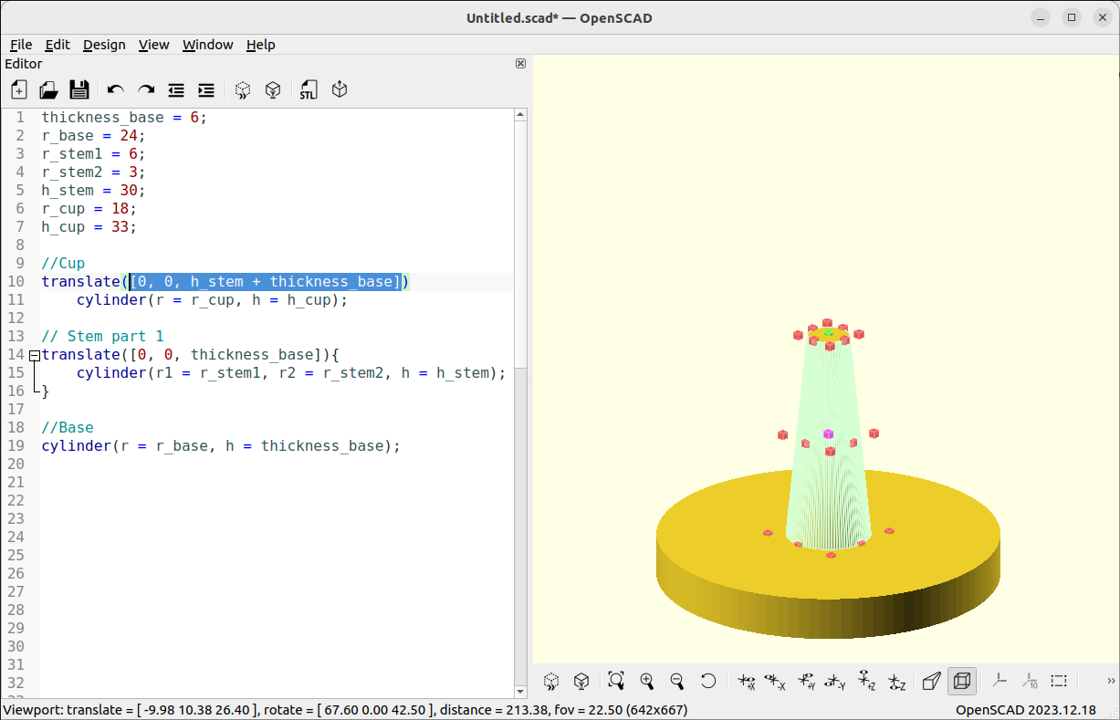}\end{minipage}}
\subfloat[The user place the definition stored in the clipboard in a \texttt{translate} to place another object]{\begin{minipage}{.49\linewidth}\label{subfig:absLocex4}\centering\includegraphics[trim = 15cm 2cm 0.5cm 45
cm, clip, height=2.7cm]{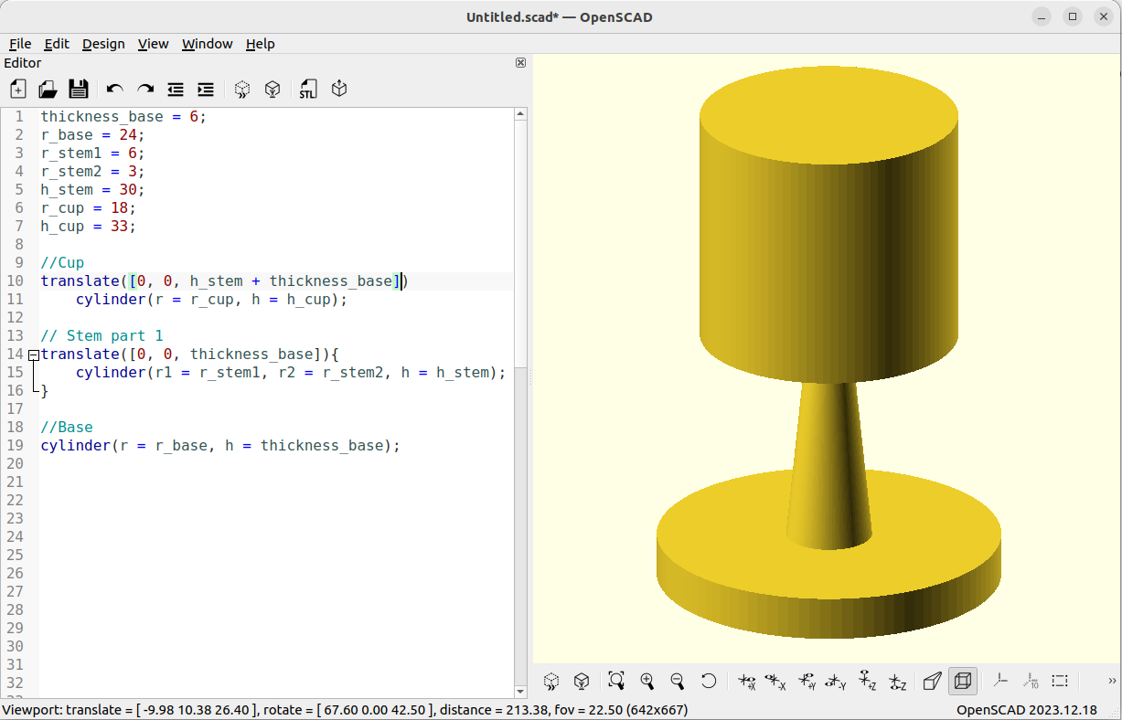}\end{minipage}}
\caption{Position feature allows users to retrieve the location of an object's handle relative to the origin.}
\Description{Position feature example. This figure presents an example illustrating the use of the position feature. It comprises three subfigures, each demonstrating the progression of utilizing the position features. In the first subfigure, the stem is selected, and the system highlights the conic cylinder in green while displaying all associated handles. Moving to the second subfigure, a close-up of the top of the stem reveals the middle top handle highlighted in green, indicating the user's selection. Subsequently, the system places the location definition in the clipboard. Finally, the third subfigure displays the culmination of these actions, showcasing the final result wherein the user has utilized the expression to position a cylinder, thereby completing the cup.}
\label{fig:absLocExample}
\end{figure}

Consider the case where a user is designing a cup.
The user has placed a cylindrical base and a cylindrical stem on top, as depicted in Listing \ref{lst:exampleAbsoluteLoc1}.
Using the \finalReplace{absolute location}{position} feature, the user could place a cylinder for the cup on top of the stem (Figure \ref{subfig:absLocex1}).

\begin{lstlisting}[basicstyle=\small, numbers=left, caption={Example before using \finalReplace{absolute location}{position} feature}, label=lst:exampleAbsoluteLoc1]
thickness = 6;
r_base = 24;
r_stem1 = 6;
r_stem2 = 3;
h_stem = 30;
r_top = 18;
h_top = 33;
// Cup


// Stem
translate([0,0,thickness]){
cylinder(r1=r_stem1,r2=r_stem2,h=h_stem);
}
//Base
cylinder(r=r_base,h=thickness);
\end{lstlisting}

The user could first select the stem's cylinder by right-clicking on it.
The user could select the cylinder in the menu displaying the \csg nodes involved in that part, as depicted in Figure \ref{subfig:absLocex2}.
\os would display the handles, and the user could select the one in the middle of the top where the cup cylinder will be placed.
The handle would turn green so that the user would have in the clipboard the definition of the position of that point in terms of the variables used in the model (Figure \ref{subfig:absLocex3}).
The user can then create a translation, paste the definition stored in the clipboard as shown in Listing \ref{lst:exampleAbsoluteLoc}, and add a cylinder to create the cup (Figure \ref{subfig:absLocex4}).

\begin{lstlisting}[basicstyle=\small, firstnumber= 7, numbers=left, caption={Example after using \finalReplace{Absolute location}{position} feature}, label=lst:exampleAbsoluteLoc]
h_top = 33;
// Cup
translate((*@\colorbox{green}{\texttt{[0,0,h\_stem+thickness]}}@*))
cylinder(r = r_top, h = h_top);
\end{lstlisting}

\subsection{Delta Vector}

The \finalReplace{relative location}{delta vector} feature calculates the \finalAdd{vector between two handles. This} arithmetic expression allows aligning a handle of one object with a handle of another object, establishing a \emph{coincidence} or a \emph{snapping} effect.
The process is similar to the \finalReplace{absolute location}{position} feature but in a two-step process.
Users enable this feature by pressing the\emph{\finalReplace{relative location}{delta vector}} button in the menu bar and then selecting the object they want to move.
The application marks the center of the object with always visible purple handles.
The rest of the handles behaves like any other geometry and can be hidden behind other geometries.
The user selects a destination object, and the application also shows its handles, with centers in purple and others in blue. In this setup, red points indicate origin points, and blue points mark destination points.
After right-clicking on a red handle, which turns it white, the user selects the destination point by right-clicking on it. This action turns both handles green to inform the user that the system has placed the parametric definition in the clipboard. The\finalReplace{relative location}{delta vector} feature calculates the difference between the destination and origin handles, allowing users to determine the necessary transformation to align the origin and the destination points.

\begin{figure}[!thbp]
\centering
\Description{Delta vector feature example. This figure illustrates an example showcasing the use of the relative position feature. It consists of two subfigures, each depicting the progression of employing the delta vector features. In the first subfigure, returning to the example of the cup, a cup, and a cylinder are displayed, with both objects showcasing their respective handles. Using illustrative markers, the figure indicates that the user has selected a handle of the sphere and one of the cylinder's handles, signifying the user's intention to move the sphere to the cylinder. In the second subfigure, the sphere is accurately positioned following the application's derivation of the expression, which is subsequently used in a translation statement for the sphere.}
\subfloat[The user right-clicks the origin(1) and destination handles(2) creating the delta vector in the clipboard.]{\label{subfig:relaLocex1}\includegraphics[width=.44\linewidth]{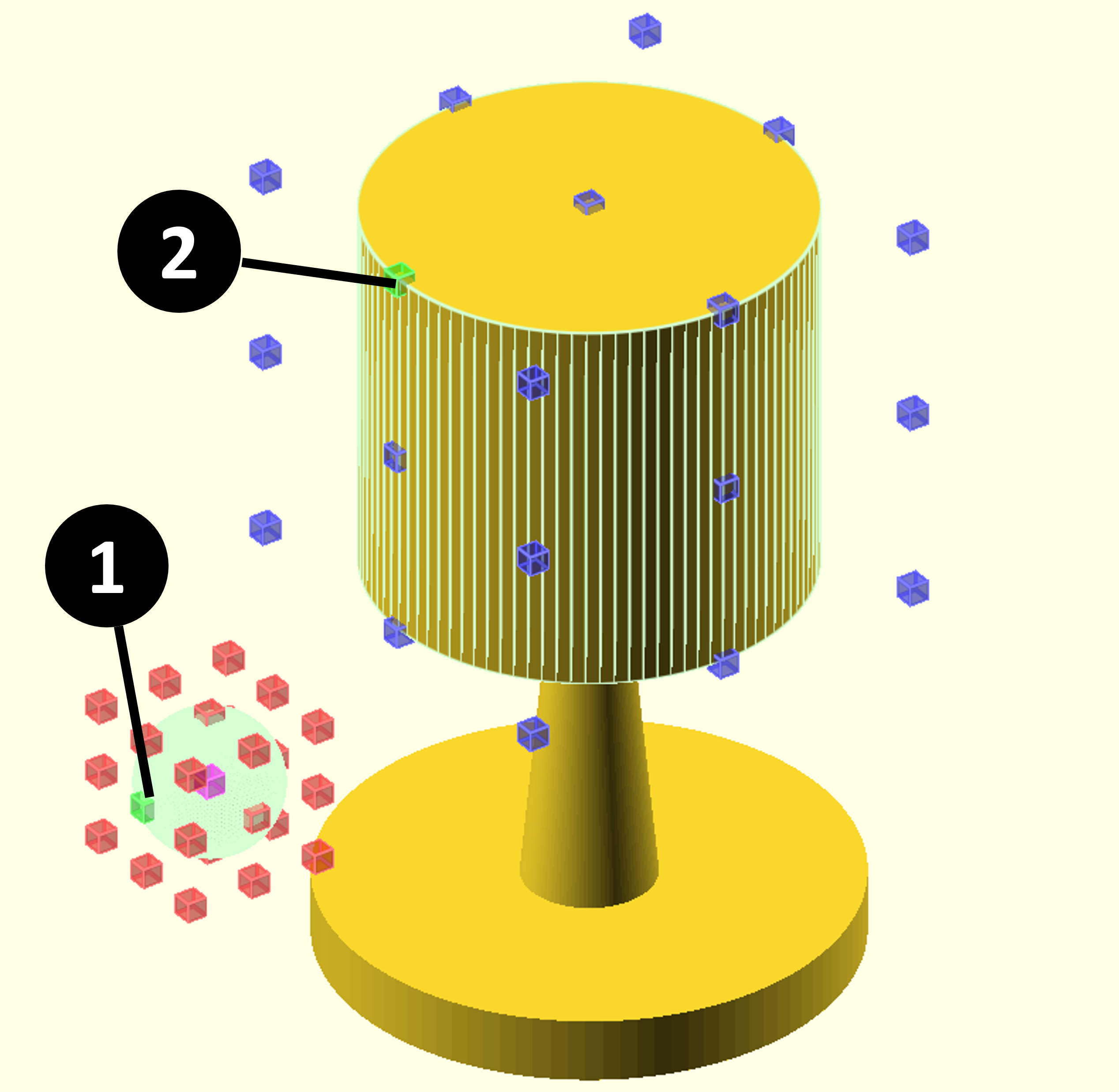}} \quad
\subfloat[The user can place the \texttt{translate} into the sphere definition to locate it parametrically.]{\label{subfig:relaLocex2}\includegraphics[width=.44\linewidth]{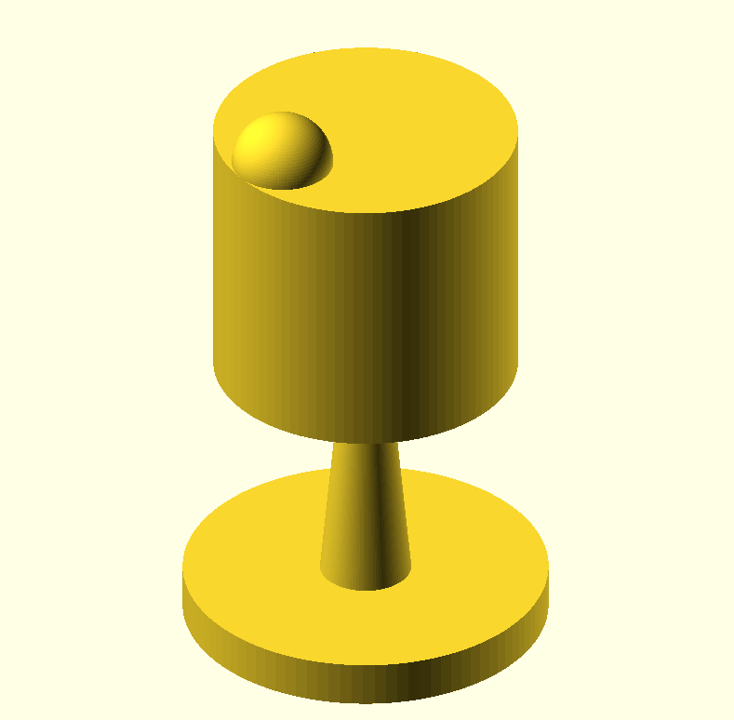}}
\caption{Delta vector allows users to place one object's handle relative to another object's handle.}\label{fig:relLocExample}
\end{figure}

Revisiting the example of a cup, consider the addition of decorative spheres around its upper part.
Using the \finalReplace{relative location}{delta vector} feature allows for precise placement.
The user selects the sphere and the cylinder at the top of the piece, prompting the handles to appear on both.
By right-clicking on corresponding handles intended to align, as shown in Figure \ref{subfig:relaLocex1}, the application generates the exact transformation \texttt{[r\_top - r\_sphere,0, thickness + h\_stem + h\_top]} and stores it in the clipboard.
Inserting it into a \texttt{translate} statement accurately positions the sphere, as seen in Figure \ref{subfig:relaLocex2}. A loop can be later added to place additional decorations symmetrically.

\section{User study}

We conducted an experiment with eleven \os users to \finalDel{analyze authoring strategies from users and to} evaluate the effectiveness of bidirectional programming in simplifying the parametric design process in \pb \cad.

The experiment consisted of three parts.
Firstly, we collected demographic information from participants and asked about their experience with other \cadapps \finalReplace{Participants self-rated their skill levels on a scale from one (novice) to five (expert) for each application.
We also collected information about their experience with general-purpose programming languages and asked them to rate their \os skills on the same scale.}{and general programming languages.}
\finalDel{Additionally, we discussed their understanding of parametric design and interest in creating parametric models.}
In the second part, participants performed a task to create a parametric design using the original \os version.
\finalDel{They verbalized their thought process throughout the task.}
Upon completion, we discussed the challenges encountered and their overall experience.
We then introduced and demonstrated the features implemented in \os.
Participants practiced briefly with the enhanced \os version before creating a second parametric model, utilizing the new features where applicable.
Given the participants' expertise in \os, we deemed any learning effect negligible and thus did not counterbalance the use of the two \os versions.
For consistency, participants utilized the original version first as a control step, followed by the modified version.
The third part involved participants sharing their experiences using the new features and discussing the potential impact of such solutions in \pb \cadapps.

Each experiment session lasted approximately 90 minutes.
We took detailed notes on participants' responses and their design thinking processes.
Additionally, we recorded the screen during the design tasks to assess performance.

\subsection{Recruitment and Participants}

We recruited participants from social media \os channels on Reddit (\verb+r/openscad+) and Facebook (\verb+OpenSCADAcademy+) to conduct the experiment using video conferencing.
The only requirement for participation was proficiency in creating parametric designs with OpenSCAD.
Before the sessions, participants were instructed to install the AnyDesk remote desktop application \cite{anydesk_software_gmbh_anydesk_2024}. They accessed a Linux machine we prepared using AnyDesk to perform parametric design tasks during the experiments.
\finalDel{Participants then performed parametric design tasks on the remote machine. One task involved designing a model using the original version of OpenSCAD, while the other task involved using our modified version of OpenSCAD.}

\finalReplace{We report the demographics of the participants and their previous experience with \cadapps in Table \ref{tab:demographicsExper}.
All participants self identified as male and varied in age: one was between 20 and 29, three were between 30 and 39, four were between 40 and 49, one was between 50 and 59, and two were between 60 and 69}{All participants self-identified as male and varied in age between 20 and 69 years old} (average: 44.5, standard deviation: 14.2).
All participants, except P3, had four or more years of 3D modeling experience (average: 8.9y, standard deviation: 5.8).
Except for P3, all participants self-rated with four or more in at least one programming language.
\finalDel{All participants except P4 and P7 had experience with other \cadapp, but only P3, P6, P8, P9, and P10 self-rated with 3 or more at least one of them.}
Finally, participants self-rated their skill level with \os as follows: Two participants with 2, four participants with 3, four participants with 4, and one participant with 5.

\finalDelTable{
\small
\Description{Demographic information of the participants. The participant's id is placed in the first column, starting with P1 and ending with P11. The second column stores the age range of participants split into 20 to 29, 30 to 39, 40 to 49, 50 to 59, and 60 to 69. The third column stores the 3D printing experience of the participants in years. The fourth column stores the self-rated skill level in OpenSCAD. The fifth to the twelfth columns store the participants' self-rated skill level in the other CAD applications FreeCAD, Rhino, TinkerCAD, Fusion360, SketchUp, AutoCAD, Solidworks and Others. The thirteenth to the nineteenth columns store participants' self-rated skill level in the programming languages of general-purpose Python, C, C++, C#, JavaScript, Java, and Others. Cells are shaded in green with darker hues representing higher scores.}
\caption[Parametric Definition of Geometric Properties - Participants demographic information]{REMOVED: \sout{ Demographics and self-rated skill level in \cadapps and programming languages. \\ Participants self-rated their skill level on the scale: 1 (Novice), 2 (Advanced Beginner), 3 (Competent), 4 (Proficient), 5 (Expert). The level reported in the category \emph{Others} is the highest rank expressed by the participant among the options.
\\ Others*: Onshape, MeshMixer, Inventor, CATIA
\\ Others**: Matlab, PHP, Bash, IBM Assembly, ARM Assembly}}
\label{tab:demographicsExper}
\centering
\arrayrulecolor[rgb]{0.753,0.753,0.753}
\begin{tabular}{l!{\color{black}\vrule}clc!{\color{black}\vrule}llllllll!{\color{black}\vrule}lllllll!{\color{black}\vrule}}
\arrayrulecolor{black}\cline{5-19}
\multicolumn{1}{l}{} &              &                        & \multicolumn{1}{l!{\color{black}\vrule}}{}              & \multicolumn{8}{c!{\color{black}\vrule}}{Other \cadapps}                       & \multicolumn{7}{c!{\color{black}\vrule}}{Programming languages}    \\
\arrayrulecolor{black}\hline
\multicolumn{1}{|l|}{\rotatebox[origin=c]{90}{Participant}} & \rotatebox[origin=c]{90}{Age Range}& \rotatebox[origin=c]{90}{  3D modeling  } \rotatebox[origin=c]{90}{ experience (y)  }& \rotatebox[origin=c]{90}{OpenSCAD} & \rotatebox[origin=c]{90}{FreeCAD}& \rotatebox[origin=c]{90}{Rhino}& \rotatebox[origin=c]{90}{TinkerCAD}& \rotatebox[origin=c]{90}{Fusion360}& \rotatebox[origin=c]{90}{SketchUP}& \rotatebox[origin=c]{90}{AutoCAD}& \rotatebox[origin=c]{90}{Solidworks}& \rotatebox[origin=c]{90}{Others*}&  \rotatebox[origin=c]{90}{Python}& \rotatebox[origin=c]{90}{C}& \rotatebox[origin=c]{90}{C++}& \rotatebox[origin=c]{90}{C\#}&  \rotatebox[origin=c]{90}{Javascript          }& \rotatebox[origin=c]{90}{Java}& \rotatebox[origin=c]{90}{Others**}\\
\arrayrulecolor{black}\hline
\arrayrulecolor{black}\multicolumn{1}{|l|}{P1}&	40-49&	8&	\cellcolor{green!50}3&	\cellcolor{green!10}1&	&	&	&	&	&	&	&	\cellcolor{green!90}5&	&	&	&	&	&	\cellcolor{green!90}5 \\  \greyhline
\arrayrulecolor{black}\multicolumn{1}{|l|}{P2}&	60-69&	9&	\cellcolor{green!90}5&	&	\cellcolor{green!30}2&	&	&	&	&	&	\cellcolor{green!70}2&	\cellcolor{green!60}4&	&	\cellcolor{green!90}5&	&	&	&	\cellcolor{green!90}5 \\  \greyhline
\arrayrulecolor{black}\multicolumn{1}{|l|}{P3}&	60-69&	2&	\cellcolor{green!50}3&	\cellcolor{green!30}2&	&	&	\cellcolor{green!50}3&	&	&	&	&	\cellcolor{green!10}1&	&	&	&	&	&	 \\  \greyhline
\arrayrulecolor{black}\multicolumn{1}{|l|}{P4}&	50-59&	10&	\cellcolor{green!50}3&	&	&	&	&	&	&	&	&	\cellcolor{green!90}5&	\cellcolor{green!90}5&	&	&	\cellcolor{green!90}5&	&	 \\  \greyhline
\arrayrulecolor{black}\multicolumn{1}{|l|}{P5}&	30-39&	18&	\cellcolor{green!70}4&	&	&	&	\cellcolor{green!10}1&	&	&	&	\cellcolor{green!10}1&	\cellcolor{green!50}3&	&	&	&	\cellcolor{green!70}4&	\cellcolor{green!30}2&	\cellcolor{green!30}2 \\  \greyhline
\arrayrulecolor{black}\multicolumn{1}{|l|}{P6}&	30-39&	20&	\cellcolor{green!30}2&	\cellcolor{green!50}3&	\cellcolor{green!70}4&	&	&	&	&	\cellcolor{green!10}1&	\cellcolor{green!30}2&	\cellcolor{green!30}2&	&	\cellcolor{green!50}3&	&	&	&	\cellcolor{green!30}2 \\  \greyhline
\arrayrulecolor{black}\multicolumn{1}{|l|}{P7}&	40-49&	12&	\cellcolor{green!70}4&	&	&	&	&	&	&	&	&	\cellcolor{green!90}5&	\cellcolor{green!90}5&	&	\cellcolor{green!90}5&	&	\cellcolor{green!90}5&	 \\  \greyhline
\arrayrulecolor{black}\multicolumn{1}{|l|}{P8}&	30-39&	5&	\cellcolor{green!70}4&	\cellcolor{green!10}1&	&	&	\cellcolor{green!70}4&	&	&	&	&	\cellcolor{green!70}4&	&	\cellcolor{green!70}4&	&	\cellcolor{green!70}4&	&	 \\  \greyhline
\arrayrulecolor{black}\multicolumn{1}{|l|}{P9}&	20-29&	4&	\cellcolor{green!50}3&	\cellcolor{green!70}4&	&	&	\cellcolor{green!50}3&	&	&	\cellcolor{green!30}2&	\cellcolor{green!50}3&	\cellcolor{green!50}3&	&	\cellcolor{green!50}3&	&	&	\cellcolor{green!30}2&	\cellcolor{green!70}4 \\  \greyhline
\arrayrulecolor{black}\multicolumn{1}{|l|}{P10}&	40-49&	5&	\cellcolor{green!70}4&	&	&	\cellcolor{green!70}4&	&	\cellcolor{green!30}2&	\cellcolor{green!10}1&	&	&	&	&	&	&	\cellcolor{green!90}5&	&	\cellcolor{green!50}3 \\ \greyhline

\arrayrulecolor{black}\multicolumn{1}{|l|}{P11}&	40-49&	5&	\cellcolor{green!30}2&	&	&	&	&	&	\cellcolor{green!50}3&	&	&	\cellcolor{green!70}4&	&	\cellcolor{green!90}5&	&	&	&	 \\

\arrayrulecolor{black}\hline
\end{tabular}
}
\subsection{Design tasks}

\begin{figure}[!htbp]
\centering
\Description{Models used in the experiment. The figure depicts the two models used in the experiment, a chalice and a box.}
\subfloat[Model A. ]{\label{subfig:modelAExperiment}\includegraphics[width=.46\linewidth]{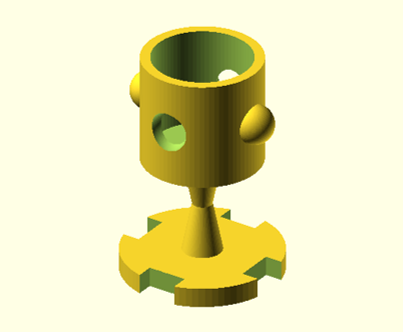}} \quad
\subfloat[Model B.
]{\label{subfig:modelBExperiment}
\includegraphics[width=.46\linewidth]{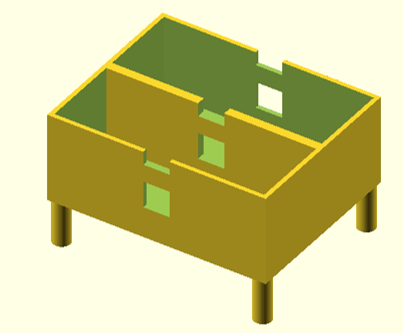}}
\caption{Models used in the experiment. Participants replicated the models, exposing parameters as required}\label{fig:modelsExperiment}
\end{figure}

Participants performed two parametric design tasks, first using the original and later our modified version of \os.
We proposed two models: model A, a chalice-like model (Figure \ref{subfig:modelAExperiment}), and model B, a box (Figure \ref{subfig:modelBExperiment}), aiming to make them comparably challenging in terms of the number of required primitives, spatial transformations, and boolean operations.
However, we designed them with distinct structures to avoid redundancy in the design experience.
For model A, participants were required to \finalReplace{make a parametric design exposing}{expose} parameters for the size of the cutouts in the base, the sizes of the holes in the cup, the height and radius of the cup, and the length of the stem.
In Model B, participants were required to \finalReplace{create a parametric design that exposed}{expose} parameters for the length of the legs, the size of the windows, and the box's height, width, and depth.
We converted the models into STL files and uploaded them to the STL online viewer, 3DViewer~\cite{kovacs_online_2024}, generating shareable links that allowed participants to view the models in 3D on their computers.
\finalDel{The participants used the original version of \os to replicate one model, then our modified version to replicate another model.}
To mitigate order bias, we counterbalanced the sequence: half of the participants worked on Model A first, followed by Model B, while the others started with Model B, and then proceeded to Model A.

In the first task, participants used the original \os version. This exercise had \finalReplace{three}{two} goals: first, to establish a baseline for comparing the performance of the design process and user experience with the modified \os version; second, to refresh participants' understanding of parametric design, facilitating a subsequent discussion about its challenges.\finalDel{; and third, to allow observation and analysis of workflows, practices, and difficulties in creating parametric designs.}
After completing the first design, we asked the participants about their user experience, task difficulty, and specific challenges in the execution of parametric designs with \os.
We then introduced the new \os features using elements from their initial designs. This was followed by tasks like determining the parametric position of a cube's corner or positioning the bottom of a sphere on top of a cube to familiarize participants with the new features.
After about 10 minutes of practice and answering questions, participants embarked on the second design task, encouraged to utilize the new features where feasible.
The users then discussed their experience and the potential of such solutions for \pb \cadapps.

\subsection{Data collection}

We recorded the \os window activity while participants worked on both design tasks. Recording using the original version of \os
\finalDel{The recordings of the participants who created the first model with the original version \os were analyzed to find authoring strategies in the design.
Moreover, these recordings} were compared to the recordings of the modified version of \os to evaluate the potential and challenges of our solution.
In addition, upon completing each model, participants were asked to rank the task's difficulty.
At the end of the second design exercise, they provided comparative evaluations of both versions of \os.
Participants answered Likert scale questions focused on the functionality and usability of the new features and engaged in discussions about their perceptions of these solutions.

\subsection{Data analysis}

\finalReplace{We logged participants' actions while creating a parametric design in the original \os version to identify common errors when defining geometric properties. Additionally, w}{W}e evaluated participants' feedback on task difficulty, feature functionality, and usability. We summarize their responses. Furthermore, we recorded participants' \texttt{translate} statements and their verification attempts, comparing these results with those from designs in the original \os version.

\finalDel{
\textit{Parametric} \textit{design} \textit{in} \textit{\pb} \textit{\cadsh}
We identified common behaviors when participants designed with the original version of \os.}

\finalDel{
\textit{Parametric} \textit{design}.
The participants' approaches to creating parametric designs in \os revealed significant anticipatory mental processes. Initially, they focused on setting up parameters, with some forecasting all necessary parameters upfront and others focusing on specific sections, revisiting as needed. Additional parameters were often added as the design evolved.
All participants aimed to define geometric properties parametrically, avoiding raw numbers. They frequently inquired about the relationships between objects to generalize geometric definitions. For example, common questions included the position of spheres in model A relative to the cup height, or the position of windows in the box of model B.
}

\finalDel{
\textit{Design style}
Participants split the design into sub-parts and designed them separately, although the way to split the design varied.
For example, in the case of model B, some participants divided the design into 2: the box and the legs.
Nine participants opted to design the box as a cube with subtracted parts.
However, P2 and P10 created the box as a union of different walls.
Three participants showed the pattern of creating subparts in the origin and removing them from the scene by commenting on the code statement to continue with the next subpart.
When all subparts were completed, the participants started by placing the bottom part and creating \texttt{translate} statements to place each part on top of the other.
The rest of the participants created the designs cumulatively without removing elements.}

\finalDel{
\textit{Defining positions}
In creating parametric designs using \os, participants followed a consistent methodology and faced specific challenges. They began by mentally locating the object's coordinate system center, considering preceding spatial transformations and the object's center. They then determined the required translations axis by axis based on existing variables.
For example, in model A, participants initially constructed a cylinder for the base. This cylinder was then enclosed within a \texttt{difference} block to add cube-shaped cutouts along its edges. A non-centered cube (\texttt{center=false}) was created and translated along the positive x-axis. The cube's center was moved to the origin by adding the cylinder's radius, aligning it with the cylinder's edge, and then shifted back by half the cube's x-dimension to embed it halfway into the cylinder. The same method was applied to the y-axis, and for the z-axis, the cylinder's height was subtracted.
}

\finalDel{
This involved identifying the manipulated object's center, adjusting its position relative to other components by considering their dimensions, and iteratively applying translations, additions, or subtractions.
Common errors encountered in this process include:}

\finalDelEnumerate{
\item \sout{Neglecting the specific center that the \texttt{translate} operation targets. Spatial transformations can vary based on the object's geometric center. For instance, \texttt{cube(center = true)} centers the cube at the origin, while \texttt{cube()} or \texttt{cube(center = false)} places a corner at the origin. This requires adjusting the translation offset and rotation axis for centered versus non-centered objects.}
\item \sout{Misinterpreting the multipliers needed for positioning. For instance, participants occasionally miscalculate the offset using a quarter instead of half of the object's dimension, leading to trial and error to get the desired visual outcome.}
\item \sout{Misinterpreting the coordinate system's orientation and mistakenly applying the wrong sign to variables. Participants accurately identified the necessary variables and their multiplier factors but occasionally hesitated on the sign, adding when they should subtract, indicating a disconnection between spatial conceptualization and code expression.}
\item \sout{Confusing variables, particularly in complex expressions involving multiple variables, lead to the inclusion of incorrect variables in computations.}
\item \sout{Difficulties in mathematically deriving complex expressions when dealing with subtracted elements not visually represented in the model. To counteract this, some participants temporarily removed elements from \texttt{difference} blocks for verification or used modifiers for visual guidance, reintegrating elements when satisfied.}
}

\finalDel{
Strategies to address these errors typically involved trial and error with variable factors and signs and meticulous code examination to ensure accurate expression definition.
}
\finalDel{
\textit{Ensuring overlapping}
Another common strategy involved creating a variable with a minimal value to ensure necessary overlap.
Participants frequently utilized preview mode in \os due to its speed advantage.
However, this mode demonstrates limitations in \csg expression.
Given the abstract nature of \csg definitions, transitioning to geometric representation can exhibit unintended behaviors in preview mode.
Specifically, when elements theoretically align perfectly, their visual representation might not clearly depict this coincidence.
For instance, in scenarios where two cubes should intersect on a face, the application may fail to execute the intended subtraction if they are precisely coincident.
Participants introduced a "delta" variable to bypass this issue, slightly enlarging the elements to ensure overlapping.
This adjustment ensures that the geometric calculations accurately reflect the desired behavior, addressing the preview mode's imperfections.
}

\subsubsection{Perception on implemented features}

\begin{figure} [!htbp]
\centering
\footnotesize
\Description{The image shows the results of users ranking the difficulty of Models A and B in the original and modified version of OpenSCAD using the scale: 1 (Very Difficult), 2 (Difficult), 3 (Neutral), 4 (Easy), and 5 (Very Easy). It has 4 stacked bars for the combination with/without the features model A/model b}
\includegraphics[trim= 1cm 0.5cm 6cm 1cm, clip, width=\columnwidth]{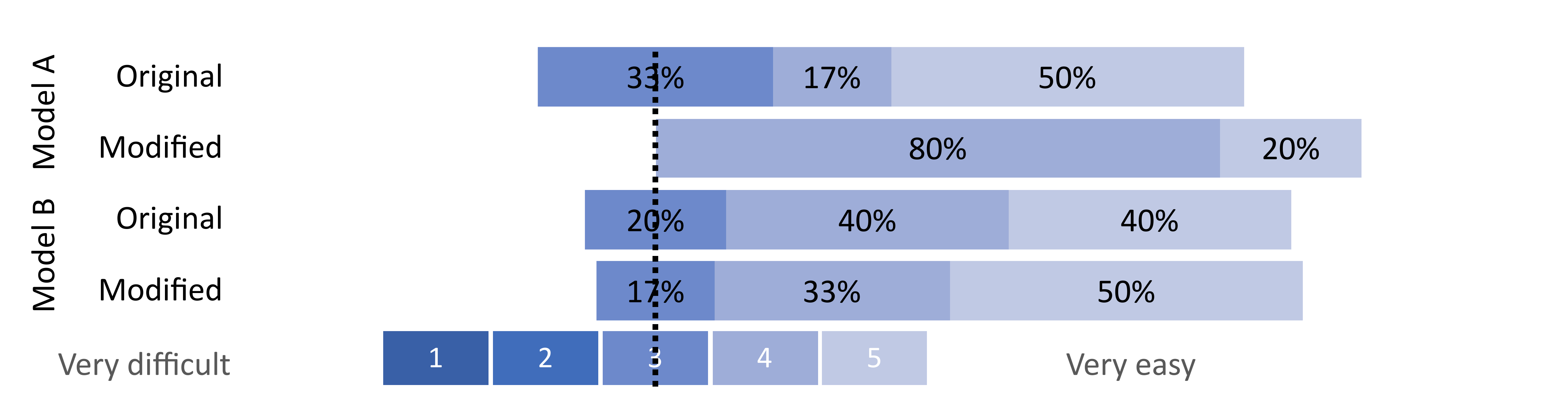}
\caption{Perceived difficulty of models in the original and modified version of \os using the scale: 1 (Very Difficult), 2 (Difficult), 3 (Neutral), 4 (Easy), and 5 (Very Easy).}
\label{fig:rankDifficultiesModel}
\end{figure}

Participants shared their experiences with the difficulty of creating the models and the functionality and usability of the implemented features.
\finalDel{Most participants rated both models relatively easy to create}
All participants rated the difficulty of both models between \textit{Neutral} (option 3), \textit{Easy} (option 4), and \textit{Very Easy} (option 5), as shown in Figure \ref{fig:rankDifficultiesModel}.
\finalDel{P1, P2, and P4 indicated that they perceived Model B as slightly more challenging than Model A.
However, a
Specifically, for Model A using the original version, two participants considered it neutral, one easy, and three very easy.
In the design of Model A with the modified version of \os, 4 participants considered easy, and one very easy.
For Model B using the original version of \os, one found it neutral, two easy, and two very easy.
In the design of Model B with the modified version of \os, one found it neutral, two easy, and three very easy.}

\begin{figure} [!htbp]
\centering
\footnotesize
\Description{The image shows the results of users answering answered if the modified version of OpenSCAD and the individual features made the design task easier or more difficult with the scale 1 – Much more difficult, 2 – Somewhat more difficult, 3 – About the same, 4 - Somewhat easier, 5 – Much easier. This answer is discriminated for the modified version of OpenSCAD and both features individually.}
\includegraphics[trim= 4.5cm 0.5cm 5cm 0cm, clip, width=\columnwidth]{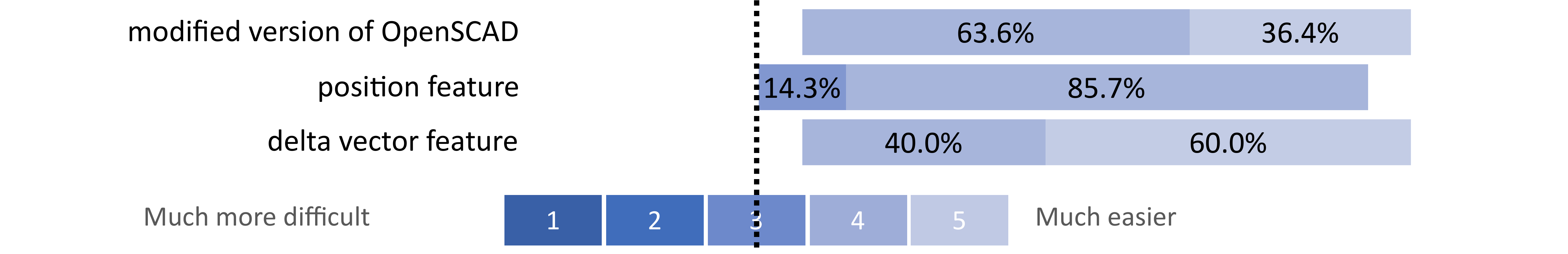}
\caption{Participants answered if the modified version of \os and each feature made the design easier or more difficult. 1-Much more difficult, 2–Somewhat more difficult, 3–About the same, 4-Somewhat easier, 5–Much easier}
\label{fig:questionModifiedDifficulty}
\end{figure}

After completing the design exercise using the implemented features, we asked participants if the modified version of \os made the design task easier or more difficult (Figure \ref{fig:questionModifiedDifficulty}).
All participants answered above \textit{About the same} (option 3), with seven participants with \textit{Somewhat Easier} (option 4) and four participants with \textit{Much easier} (option 5).
Then, we asked a similar question but individually targeted both features.
Not all participants used both features in the design exercise according to personal preference, so answers in Figure \ref{fig:questionModifiedDifficulty} report the percentage of the total of participants who used the feature and answered the question: seven participants for the \finalReplace{absolute location}{position} feature and ten participants for the \finalReplace{relative location}{delta vector} feature.
For the \finalReplace{absolute location}{position} feature, one participant (14.3\%) answered about the same, and six participants (85.7\%) answered somewhat easier.
Regarding the \finalReplace{relative location}{delta vector} feature, four participants (40\%) answered somewhat easier, and six participants (60\%) answered much easier.
\finalDel{Models A and B were perceived as similarly difficult, although some participants mentioned that model B was slightly more difficult.}

\begin{figure} [!htbp]
\centering
\footnotesize
\Description{The image shows the results of users answering answered how difficult was to use the implemented features individually with the scale 1 – Very difficult, 2 – Difficult, 3 – Neutral, 4 – Easy, 5 – Very easy}
\includegraphics[trim= 4.5cm 0.5cm 4.5cm 0cm, clip, width=\columnwidth]{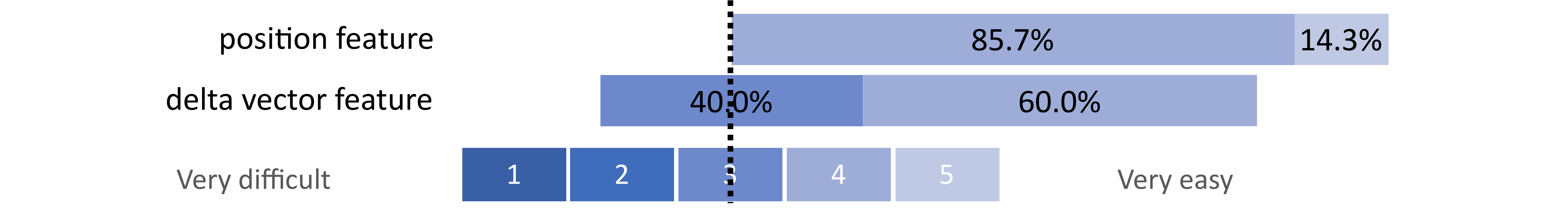}
\caption{Participants answered how difficult was to use the implemented features  with the scale 1 – Very difficult, 2 – Difficult, 3 – Neutral, 4 – Easy, 5 – Very easy}
\label{fig:featuresUsability}
\end{figure}

We also asked how difficult it was to use each feature regarding usability, as depicted in Figure \ref{fig:featuresUsability}.
Similar to the previous question, the answers report the total number of participants who used and answered the question about the feature.
In all cases, all participants answered between \textit{Neutral} (option 3), \textit{Easy} (option 4), and \textit{Very easy} (option 5).
For the \finalReplace{absolute location}{position} feature, six participants (58.7\%) answered easy, and one participant (14.3\%) answered very easy.
Regarding the \finalReplace{relative location}{delta vector} feature, four participants (44.4\%) answered neutral, and five participants (55.6\%) answered easy.
Three participants indicated that selecting control points with a right-click is inconvenient. P4
commented \uquote{My initial inclination is to left-click those handles; it's a little bit extra to remember to right-click the handle}. Furthermore, the control points did not scale when zoomed in, which was also reported as problematic.
For the delta vector, half of the participants answered Neutral (option 3) and the other half Easy (option 4).
Participants found the process with two objects difficult to remember and listed some problems.
For example, participants missed visual cues that guided them through the different steps.
P5 commented \uquote{There was no clear prompt indicating what was copied to the clipboard; it's unclear if it's correct. Upon pasting, the directionality, whether red goes into blue or vice versa, is confusing.}
\finalDel{A potential improvement from the rendering side could be to draw an animated arrow to indicate the directionality of the relative location, providing better guidance on its use.}
Further, P3 mentioned that using color to indicate the process can be difficult for some people \uquote{I'm not exactly colorblind, but it's hard for me to see colors. So it's nice for people like us if you have an indication that is not entirely dependent on colors.}
However, they found it easy overall, as commented P1 \uquote{It is not obvious but easy}.

Later, we asked participants if they thought that these features would help them in the design process in their normal modeling process.
All participants answered \textbf{Yes}.
Participants found several advantages.
P1 commented that it could help to avoid errors when designing parametrically: \uquote{A few days ago, I positioned objects by adding variables I believed would bring them to the correct position. It appeared to be correct in the preview \dots
However, when one value changed, the alignment was disrupted. One part was not where it was supposed to be. It was only correctly positioned when the variables coincidentally lined up.}.
P2 found that this is more interactive than other alternatives that try to include “anchors” selectable from the code.
\uquote{Tools like Cascade are being used by designers who are trying to add anchors to objects, making them selectable in the code. What you're doing is making a user interface more interactive, combining the interactivity of Fusion 360 with the capabilities of OpenSCAD, and putting together the advantages of both worlds. So I think this is a better solution to the problem.}
For instance, P4 and P9 found that such features could facilitate the transition of people with little experience into \pb \cadapps.
P9 commented \uquote{I think it would be incredibly valuable in helping users transition from normal CAD to scripted CAD, especially for those who don't have a rigorous background in computer science or math} respectively.
Finally, P2, P5, P6, P7, P8, and P10 mentioned that this would facilitate the deriving of mathematical expressions, making the design faster.
P2 said \uquote{When designing objects that need to be combined to form one design, I often do calculations to position things. This would mean fewer calculations for me to do}.

\subsubsection{Comparing original and modified version of \os}

Our analysis focused on participant approaches to defining \texttt{translate} statements in both the original and the modified versions of \os.
When participants defined a translation, they continued to render the result to verify the correctness of the code.
Each rendering \textit{attempt} was logged and categorized based on the outcome: \textit{Success} for correct placements, \textit{Wrong location} for incorrect placements, \textit{Uncertain/false positive} for when participants were unsure or incorrectly deemed the placement, and \textit{Syntax errors} for errors in the programming syntax.

\begin{figure}[!htbp]
\centering
\Description{The image depicts a chart illustrating the attempts using translate statements in OpenSCAD. The chart consists of four stacked bars divided into two groups: Model A and Model B. Each group includes two bars representing the attempts with the original version and the modified version of OpenSCAD for each model. Each stacked bar comprises four series: Success, Wrong location, Uncertain or false positive, and Syntax errors.
For Model A with the original version, the distribution is 62\% Success, 17\% Wrong location, 9\% Uncertain or false positive, and 12\% Syntax errors. For Model A with the modified version, the distribution is 79\% Success, 10\% Wrong location, 2\% Uncertain or false positive, and 8\% Syntax errors. For Model B with the original version, the distribution is 50\% Success, 34\% Wrong location, 8\% Uncertain or false positive, and 8\% Syntax errors. For Model B with the modified version, the distribution is 62\% Success, 30\% Wrong location, 1\% Uncertain or false positive, and 6\% Syntax errors. Each bar is described in this order for clarity.}
\scriptsize
\includegraphics[trim= 4.5cm 1cm 0cm 0cm, clip, width=\columnwidth]{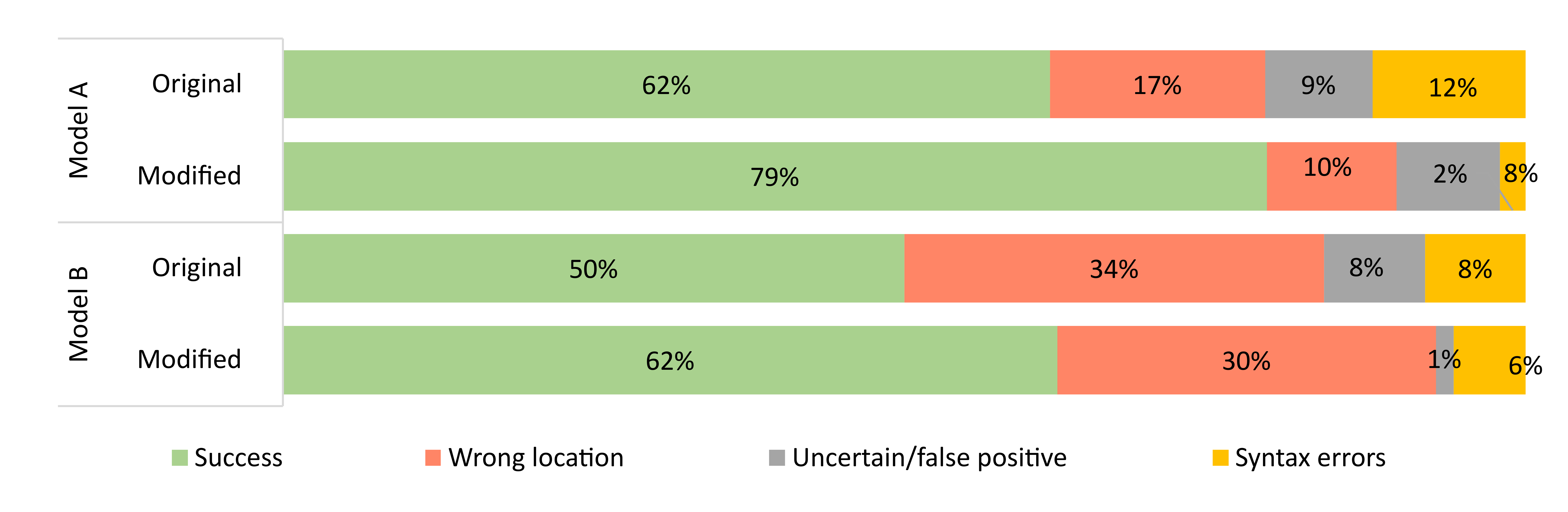}
\caption{Attempts to verify a \texttt{translate} statement upon rendering was logged and classified to compare versions.}\label{fig:resultsGeneral}
\end{figure}

Participants generated a total of 94 \texttt{translate} statements using the original \os version (52 in Model A and 42 in Model B) and 85 with the modified version (42 in Model A and 43 in Model B).
The original version had 155 rendering attempts (averaging 1.64 attempts per translation), while the modified version had 117 attempts (averaging 1.37 attempts per translation), possibly implying that with our modified version, participants required fewer attempts to reach a successfully \texttt{translate} definition.
The distribution of these attempts across different categories reveals significant insights.
As the number of spatial transformations differs between programming styles, we focused on investigating how difficult it is to correctly define the \texttt{translate} statements defined by the users in terms of the number of attempts per statement.
As illustrated in Figure \ref{fig:resultsGeneral}, the modified version demonstrated a higher success rate and fewer instances of incorrect placements, uncertain/false positives, and syntax errors compared to the original version.
\finalDel{Similar behavior is depicted in Figure \ref{subfig:resultsGeneral2} when the analysis is done per model.
Notably, Model B exhibited a lower success rate and an increase in wrong location attempts in both \os versions, aligning with participant feedback that Model B was slightly more challenging than Model A.}

\finalDelImage{gfx/OrigVsModV2_1.png}{Comparison between the original and modified versions of OpenSCAD for Models A and B}{subfig:resultsGeneral2}

\begin{figure}[!thbp]
\centering
\Description{The image depicts a chart illustrating the outcomes using translate statements with our modified version of OpenSCAD. The chart consists of eight bars divided into two groups: Model A and Model B. Each group includes two subgroups representing the successful and wrong placement attempts. Each subgroup has two bars for the number of outcomes using the features and not using the features.}
\footnotesize
\includegraphics[trim= 5cm 1cm 2cm 1cm, clip, width=\columnwidth]{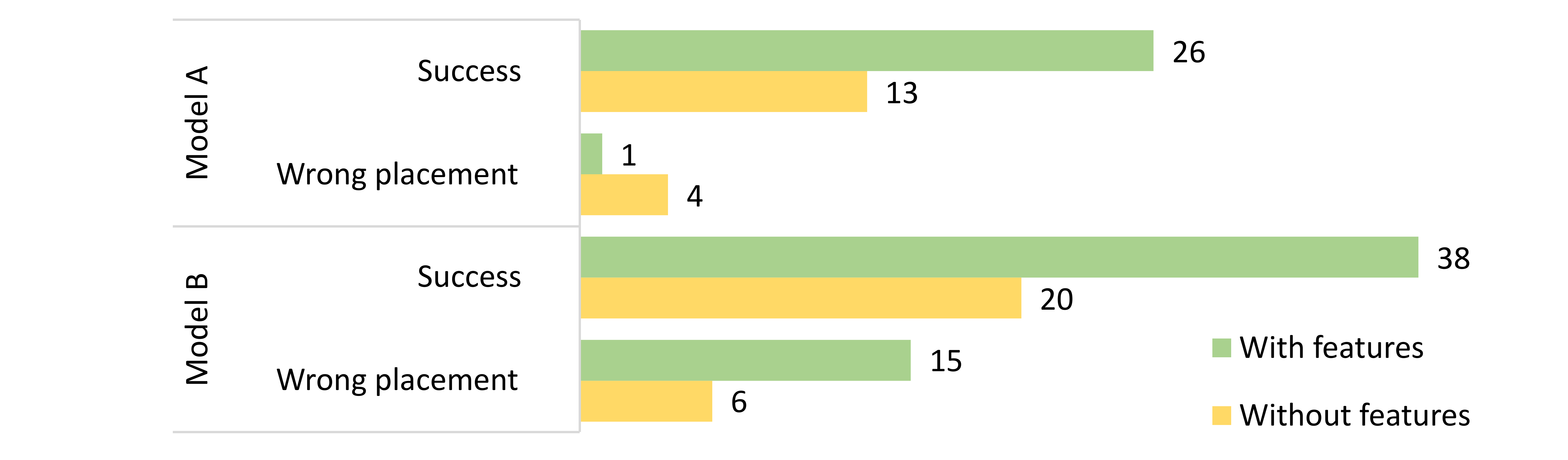}
\caption{Outcomes of attempts to verify \texttt{translate} using the modified version of \os}\label{fig:resultsFeatures}
\end{figure}

During the second design using the modified version of \os, not all translations were defined using the developed features.
Participants often opted to calculate expressions manually.
Figure \ref{fig:resultsFeatures} depicts the different attempts using the modified version of \os using the implemented features or describing the \code{translate} manually.
The participants actively tried to use the features.
Interestingly, Model B presented more errors using the features.
We perceived that Model B geometry presented more cases where the control points were hidden by geometries and cases where participants could not find a control point in the location they needed.
\finalDel{Moreover, Figure \ref{subfig:resFeatures2} shows the trials of using the implemented features.
In most of cases, users could use the features although in some cases they made mistakes or required assistance with specific questions.}

\finalDelImage{gfx/featuresRes2.png}{Outcomes when users attempted to use the features}{subfig:resFeatures2}

\section{Discussion}

We aim to facilitate the parametric designs in program\-ming-based \cadapps, particularly on the difficulty of defining objects' parametric geometric properties based on others' properties \cite{gonzalez_understanding_2024}.
We evaluated the potential and challenges of our proposed solution.

Participants unanimously agreed that our solution would ease the design process, helping avoid errors, simplifying mathematical definitions, enhancing interactivity, and lowering the skill barrier for newcomers in \pb \cad.
We identified two user scenarios: newcomers and experts.
Prior studies indicate newcomers often avoid design tasks due to high-skill requirements \cite{hudson_understanding_2016}, worsened in \pb \cad by the semantic gap between natural and programming languages \cite{rouhani_learning_2022,ko_six_2004,ma_investigating_2011}.
Effective design requires visual inspection to determine the next element's parametric position.
Our solution reduces this “gulf” between evaluation and execution \cite{wolf_taxonomic_1987} by allowing users to extract positions directly from the view, bypassing the need to interpret code.

Experts also face challenges with mathematical definitions \cite{gonzalez_understanding_2024}.
They follow mechanical workflows, yet hesitate over parameter accuracy in complex designs, resorting to trial and error.
Our solution speeds up this process by providing accurate positional descriptions, but challenges experienced users accustomed to coding workflows.
Users often find solutions restricted to the features of the programming language.
For instance, JSCAD offers functions to extract geometry boundary box information (\ie, minimum and maximum values in all dimensions) for later use in code \cite{jscad_user_group_design_2024}, but this still confines users to the code editor.
In contrast, our approach focuses on direct view interaction, allowing easy identification of positions without deriving them in the code, proven efficient in other applications~\cite{mathur_interactive_2020}.

The comparison between the original version of \os and our approach revealed that using the developed features, participants would require fewer attempts to reach the aimed geometric properties, resulting in a proportional lower rate of errors and higher rates of successful attempts.
Despite these benefits, experienced users showed resistance to changing established workflows.
P6 noted, \uquote{It seems easy to use, but changing the mindset to use the view for design is challenging.}  Similar opinions appeared in \os forums \cite{popkin_request_2014,mulier_functions_2013}, where discussions about integrating such solutions are limited to programming language features.
P11 \uquote{This is a philosophical difference among OpenSCAD users regarding textual programming and visual editing. Some users prefer doing everything in code.}
While resistance may occur, our study suggests the benefits can facilitate adaptation.

Our implementation has usability challenges, as noted by users.
Some control points were inaccessible due to overlaps with other volumes, and using right-click for selection was not intuitive.
These feedback points are crucial considerations for future refinements of our solution. Another issue involves removed geometries in \texttt{difference} statements.
Participants wanted to use the features to place subtracted elements, but these were not reachable from the view.
Some used background modifiers to make these geometries visible and selectable.
Typically, participants placed the cursor in the statement they were modifying.
Applications could make visible the element creating the code statement where the cursor is placed, using visual cues \cite{gonzalez_introducing_2023}.
This provides an explicit visual representation of the part being worked on, helpful for subtracted geometries.
A final challenge is understanding nested transformations.
Participants sometimes wanted to use the \finalReplace{absolute location}{position} feature to replace a \texttt{translate} statement, but these were often inside other \texttt{translate} statements.
Since the \finalReplace{absolute location}{position} feature gives the position relative to the \csg root, the retrieved definition is not useful inside another transformation. The application could consider the cursor's position to incorporate previous transformations when placing new definitions.

\vspace*{-3pt}

\section{Limitations}

Our study intended to compare the performance of the original and the modified version of \os.
However, the 15-minute practice session seemed to be insufficient for users to get used to the logic of the new features.
We concluded that a longer use time would be necessary to evaluate this factor and focused on the user experience, which we considered more important. Moreover, our solution only considers a limited set of cases.
Specifically, it does not include cases with spatial transformations other than \texttt{translate}.
Some of our recommendations are related to newcomers, although none of the participants were newcomers.
Further exploration with beginner users must be carried out to confirm our suggestions.

\vspace*{-3pt}

\section{Conclusion}

We hypothesized a general structure for creating geometric properties in parametric designs within \pb \cadapps.
We conducted a formative study to test our hypothesis, analyzing the code of thirty \os models sourced from Thingiverse, which validated our initial assumptions.
Subsequently, we proposed a design goal centered on a bidirectional programming approach to streamline the creation of parametric models in \pb \cadapps.
We modified the source code of \os to achieve this goal, implementing features that align with our design objectives.
To validate our solution, we conducted an experimental study involving eleven \os users who created parametric designs using both the original and our modified versions of \os.
They evaluated their experience and discussed the challenges and potential of such solutions.

Our findings indicate that allowing users to retrieve information directly from the view using direct manipulation interactions to reuse in the code holds significant promise.
This approach could notably reduce design errors, enhance the interactivity and appeal of the design process, and facilitate the entry for newcomers by reducing the mathematical skills requirements typically associated with \pb \cadapps.

\vspace*{-3pt}

\begin{acks} 

This work was supported and funded by the National Sciences and Engineering Research Council of Canada (NSERC) through a Discovery grant (2017-06300) and an Alliance (557253-2020). It was also supported and funded by the Région Hauts-de-France.

\end{acks}

\clearpage

\bibliographystyle{ACM-Reference-Format}
\bibliography{references.bib}


\appendix
\clearpage
\renewcommand{\thesubfigure}{\roman{subfigure}}

\onecolumn

\section{Appendix}\label{sec::appendixModels}

\begin{figure*}[!htbp]
\Description{The image presents a table with 6 rows and 5 columns, containing a total of 30 cells. Each cell contains a visual representation of one of the 30 models used in the formative study.}
\footnotesize
\centering
\subfloat[Multi-Color Pencil Cup Remix, Thing:\href{https://www.thingiverse.com/thing:6291495}{6291495}]{\includegraphics[width=.185\linewidth]{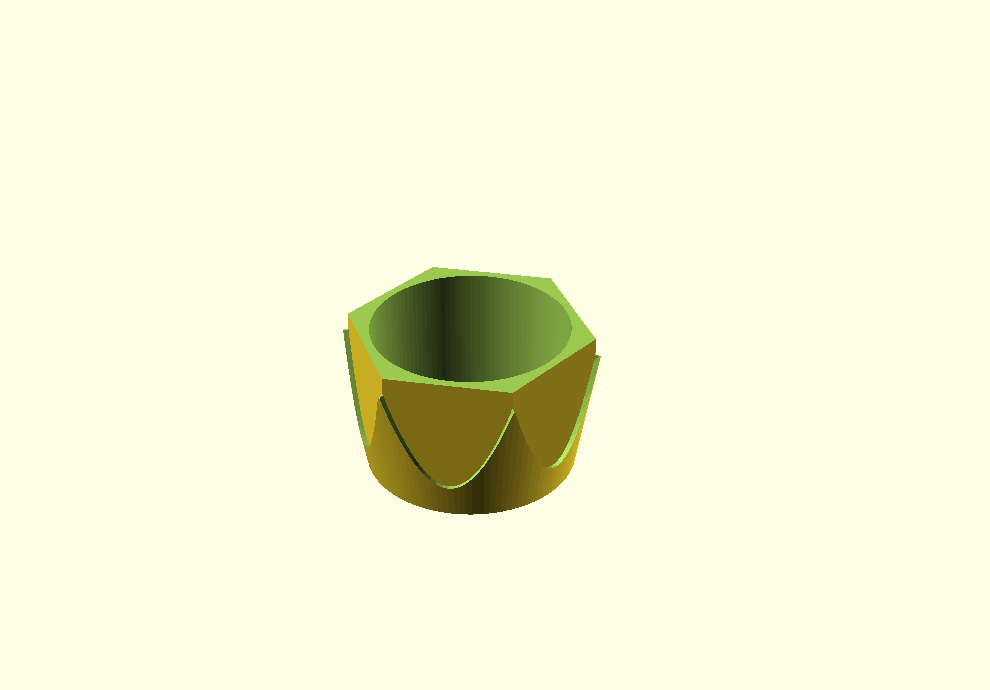} \captionsetup{font=tiny}} \quad
\subfloat[Master Lock M1 Key, Thing:\href{https://www.thingiverse.com/thing:6289846}{6289846}]{\includegraphics[width=.185\linewidth]{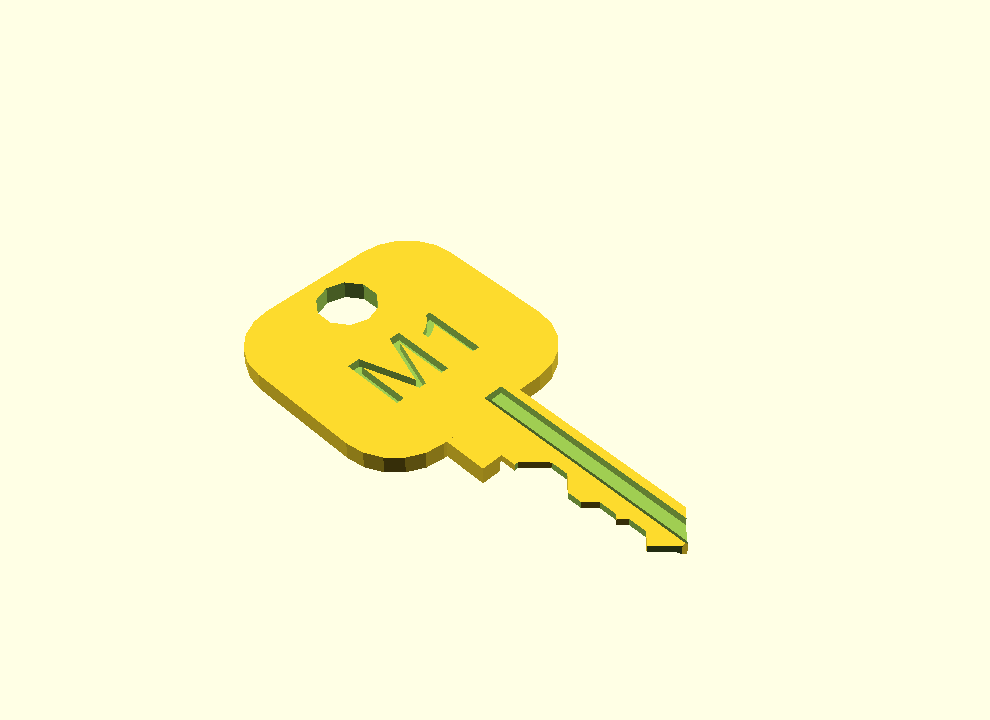} \captionsetup{font=tiny}} \quad
\subfloat[Gridfinity Kcup Holder, Thing:\href{https://www.thingiverse.com/thing:6284181}{6284181}]{\includegraphics[width=.185\linewidth]{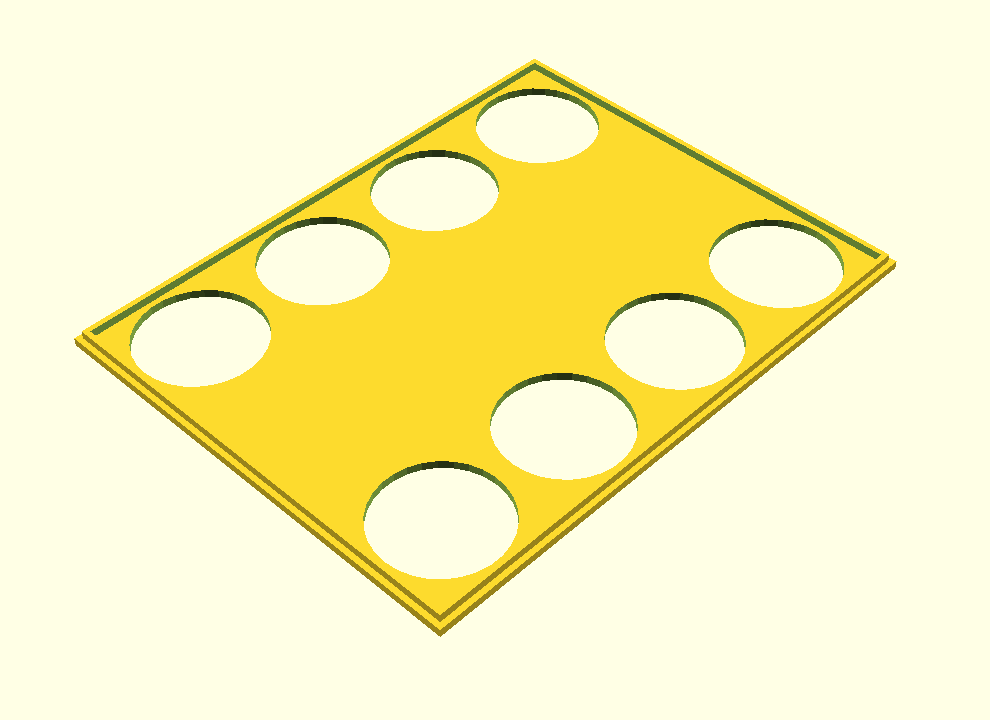} \captionsetup{font=tiny}} \quad
\subfloat[Kumihimo Disk, Thing:\href{https://www.thingiverse.com/thing:6290477}{6290477}]{\includegraphics[width=.185\linewidth]{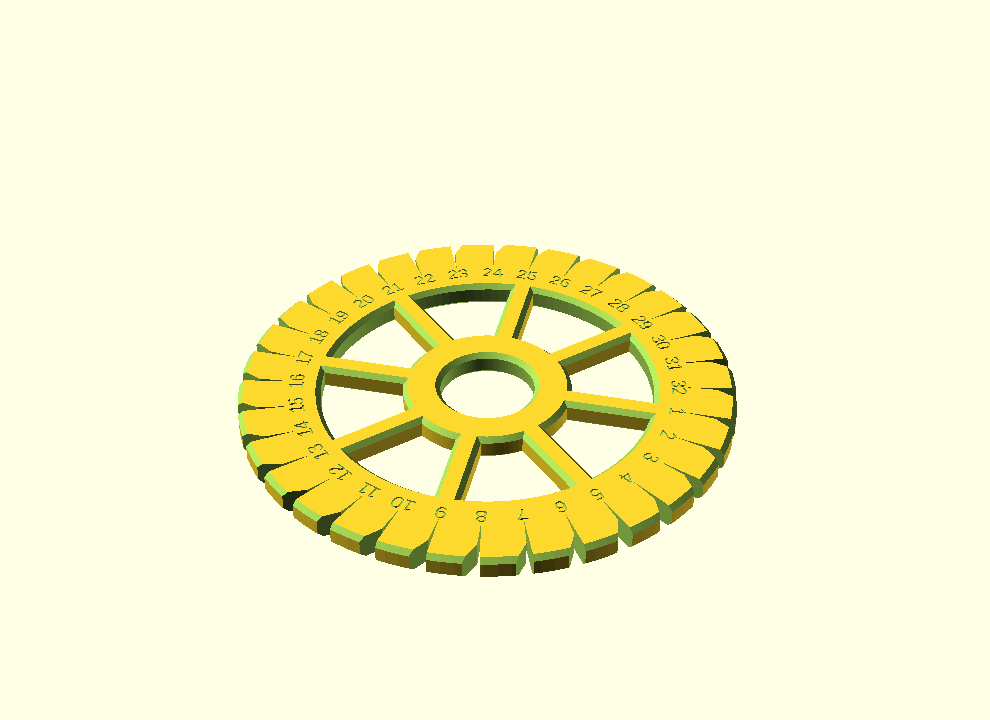} \captionsetup{font=tiny}} \quad
\subfloat[Spare Peg For Ikea Hammer, Thing:\href{https://www.thingiverse.com/thing:6287601}{6287601}]{\includegraphics[width=.185\linewidth]{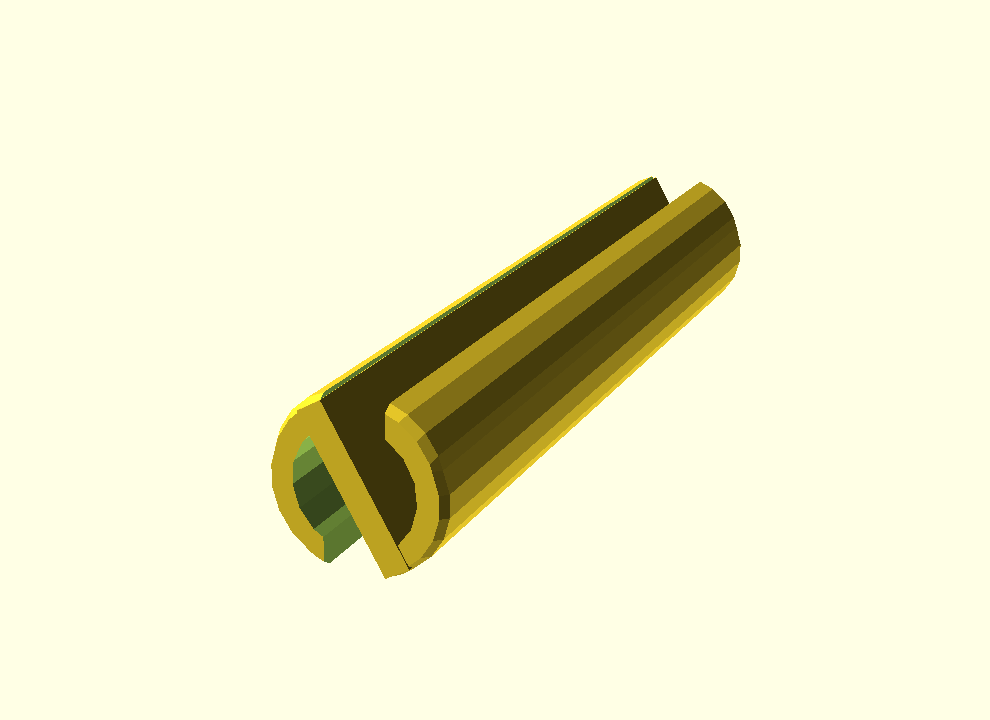} \captionsetup{font=tiny}} \quad
\subfloat[Customized Kettle Whistle, Thing:\href{https://www.thingiverse.com/thing:6291759}{6291759}]{\includegraphics[width=.185\linewidth]{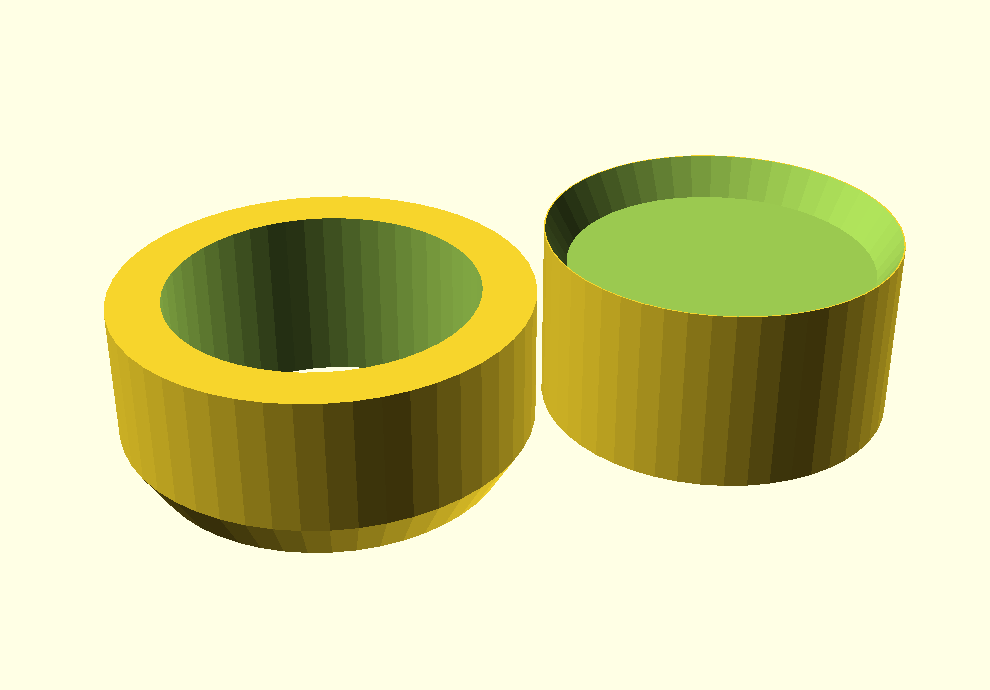} \captionsetup{font=tiny}} \quad
\subfloat[Hook On A Plate, Thing:\href{https://www.thingiverse.com/thing:6287141}{6287141}]{\includegraphics[width=.185\linewidth]{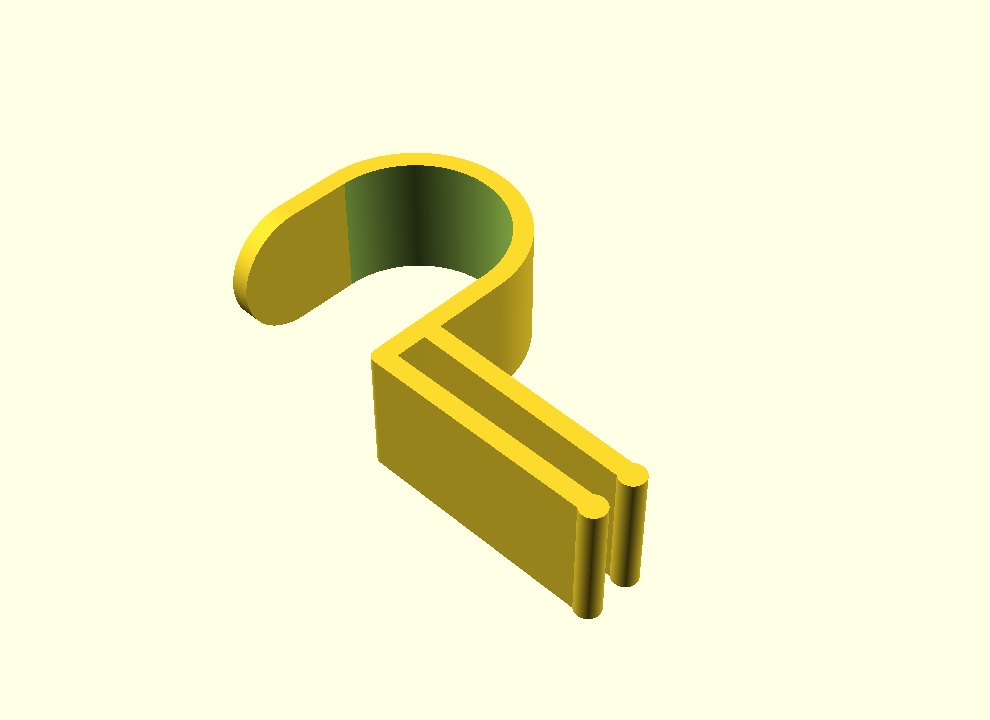} \captionsetup{font=tiny}} \quad
\subfloat[Turret Cap Generator, Thing:\href{https://www.thingiverse.com/thing:6292455}{6292455}]{\includegraphics[width=.185\linewidth]{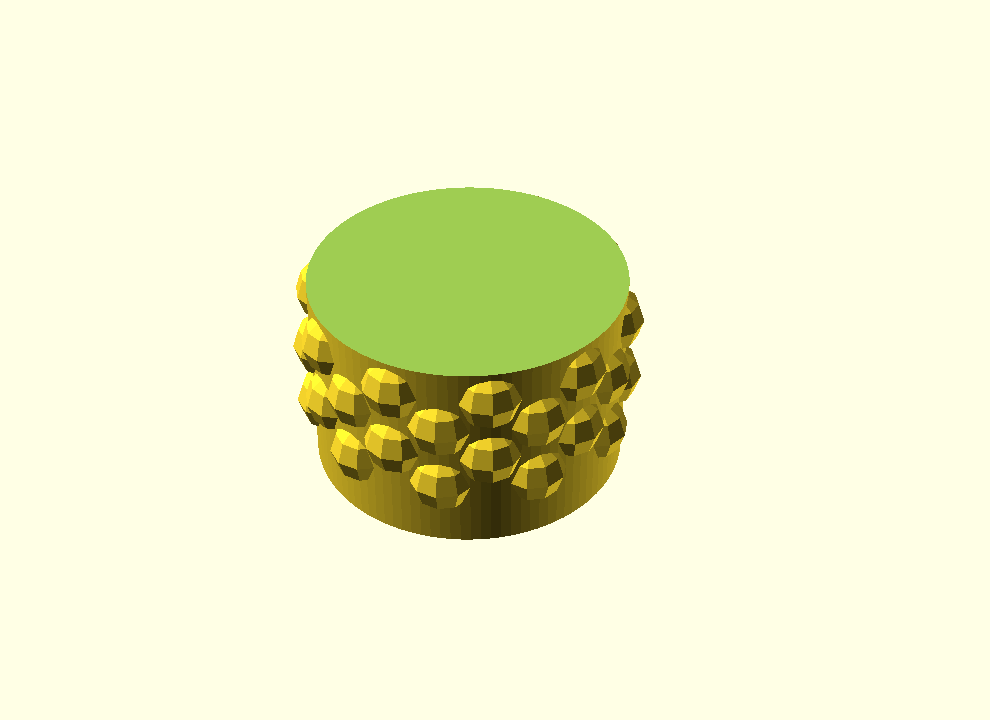} \captionsetup{font=tiny}} \quad
\subfloat[Cable Hook, Thing:\href{https://www.thingiverse.com/thing:6288817}{6288817}]{\includegraphics[width=.185\linewidth]{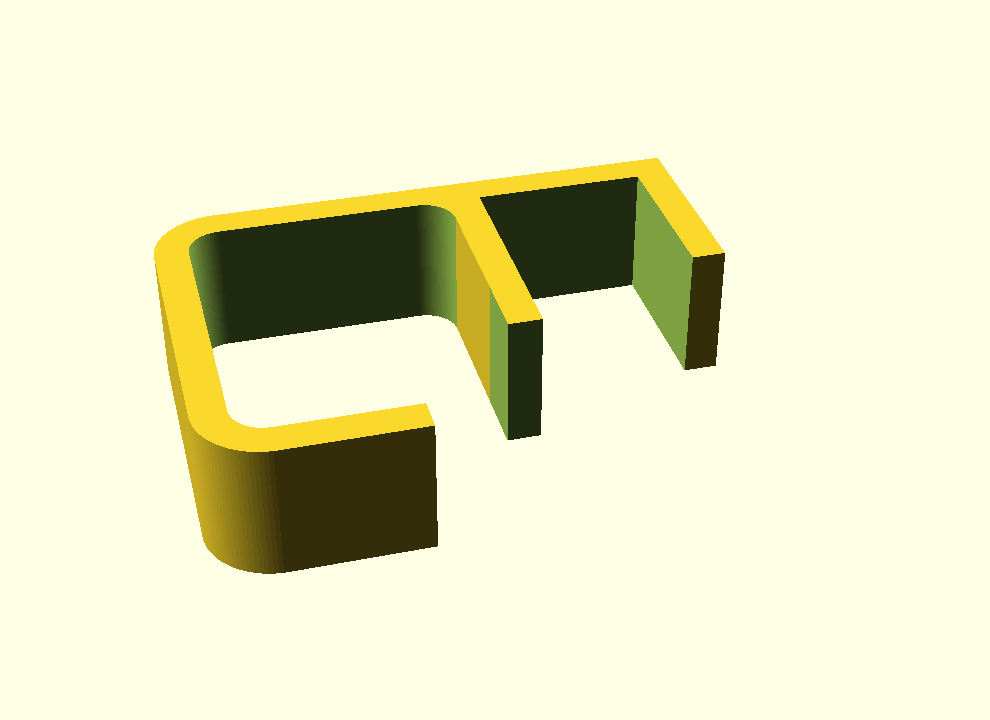} \captionsetup{font=tiny}} \quad
\subfloat[Door Hook 1 Angle, Thing:\href{https://www.thingiverse.com/thing:6287795}{6287795}]{\includegraphics[width=.185\linewidth]{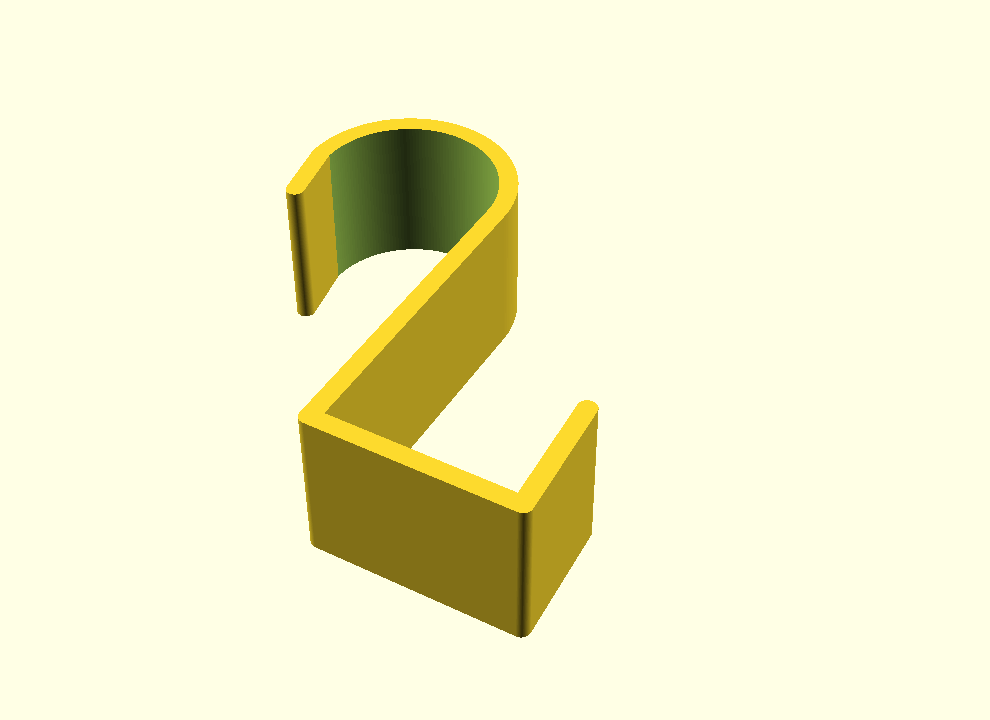} \captionsetup{font=tiny}} \quad
\subfloat[Parametric Porch Hook, Thing:\href{https://www.thingiverse.com/thing:6286541}{6286541}]{\includegraphics[width=.185\linewidth]{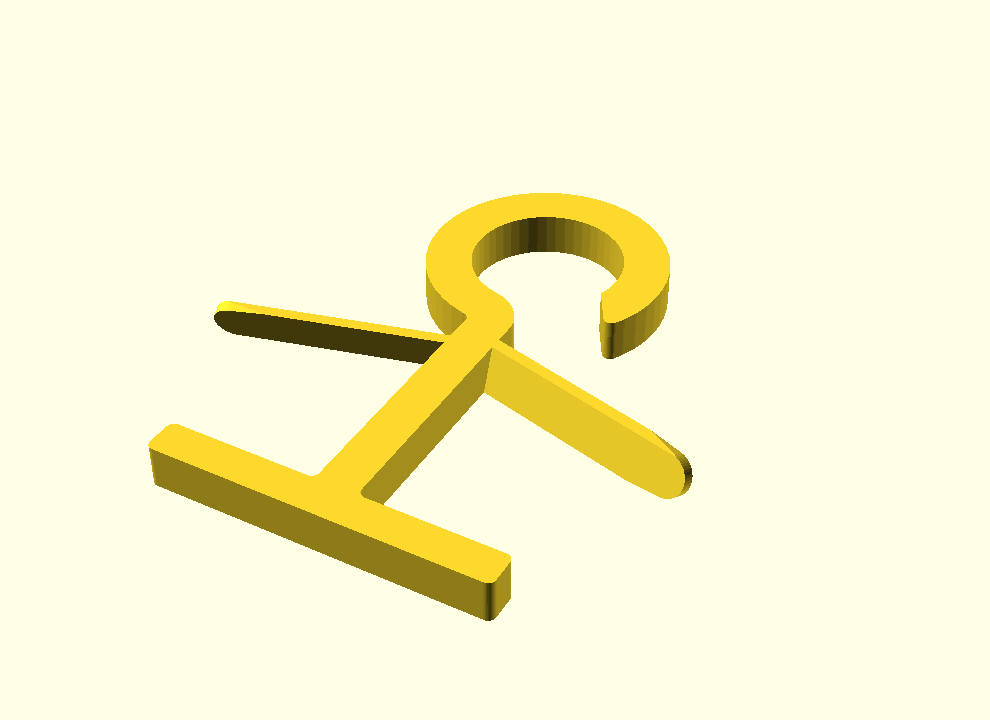} \captionsetup{font=tiny}} \quad
\subfloat[Filament Spool Holder Frame, Thing:\href{https://www.thingiverse.com/thing:6255969}{6255969}]{\includegraphics[width=.185\linewidth]{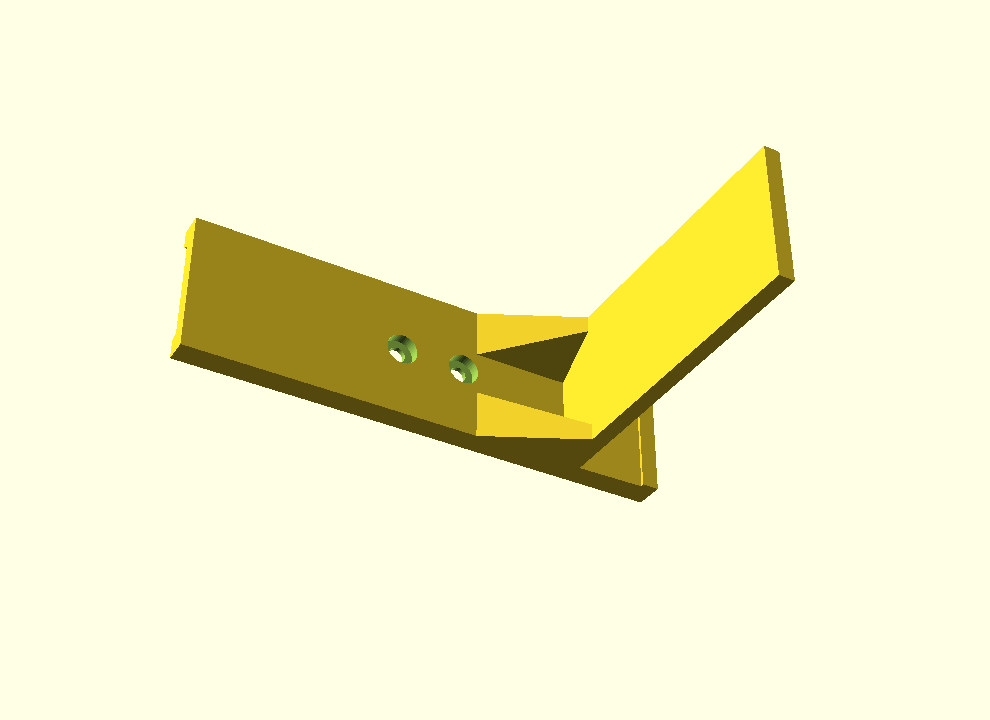} \captionsetup{font=tiny}} \quad
\subfloat[Battery End Caps, Thing:\href{https://www.thingiverse.com/thing:6248029}{6248029}]{\includegraphics[width=.185\linewidth]{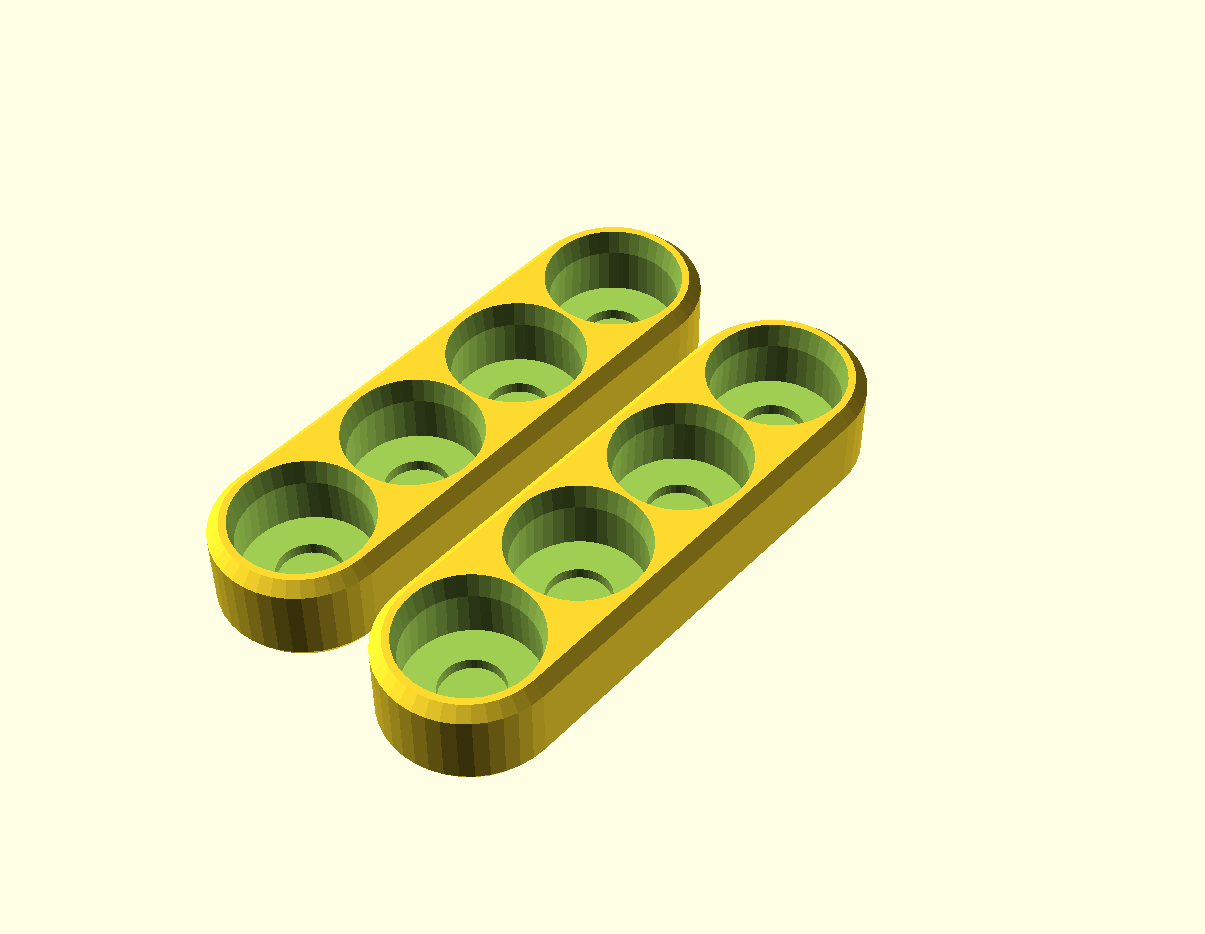} \captionsetup{font=tiny}} \quad
\subfloat[Small Customizable Box, Thing:\href{https://www.thingiverse.com/thing:6266913}{6266913}]{\includegraphics[width=.185\linewidth]{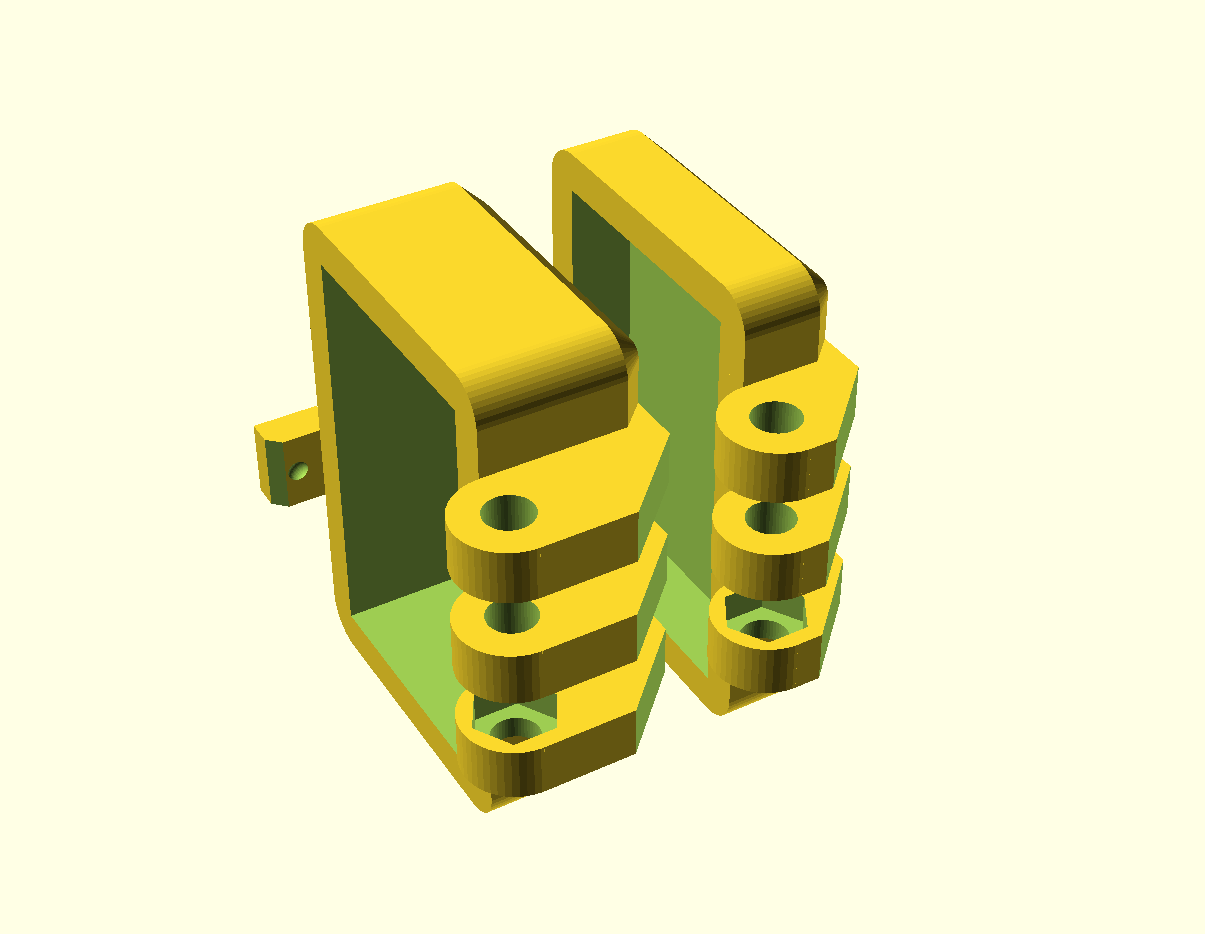} \captionsetup{font=tiny}} \quad
\subfloat[Infinity Cube Customizer, Thing:\href{https://www.thingiverse.com/thing:6249758}{6249758}]{\includegraphics[width=.185\linewidth]{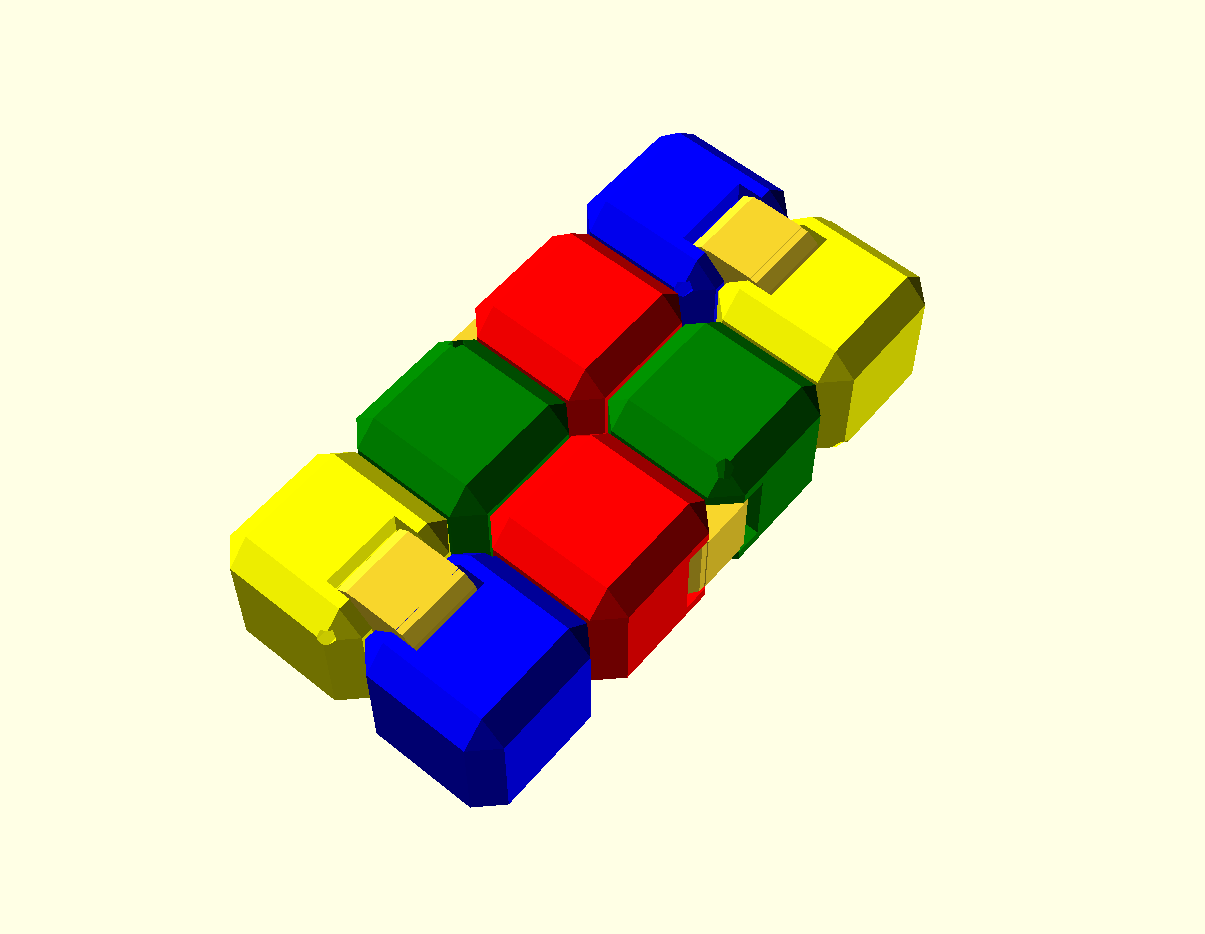} \captionsetup{font=tiny}} \quad
\subfloat[Customizable Key Tag, Thing:\href{https://www.thingiverse.com/thing:6249968}{6249968}]{\includegraphics[width=.185\linewidth]{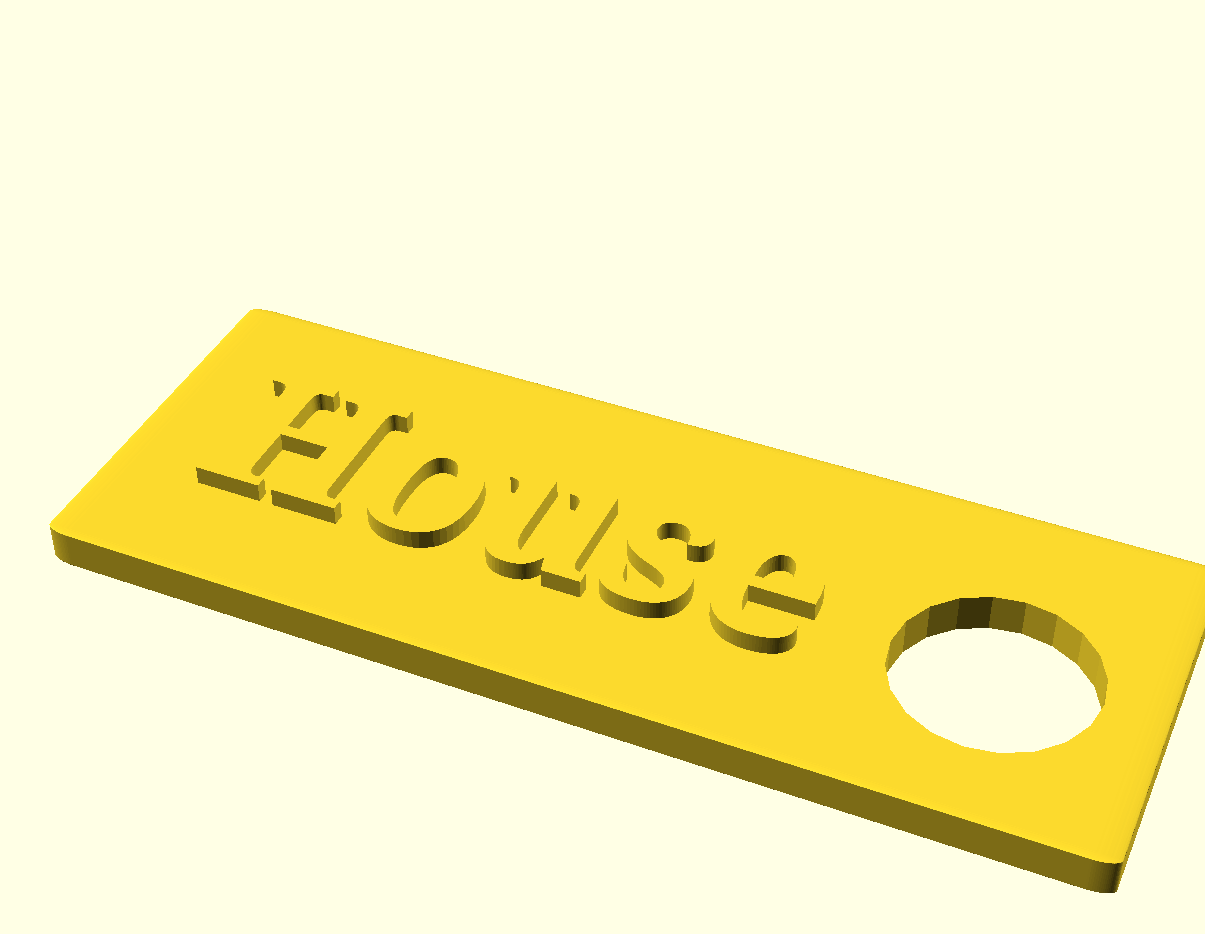} \captionsetup{font=tiny}} \quad
\subfloat[Halloween Spider, Thing:\href{https://www.thingiverse.com/thing:6253273}{6253273}]{\includegraphics[width=.185\linewidth]{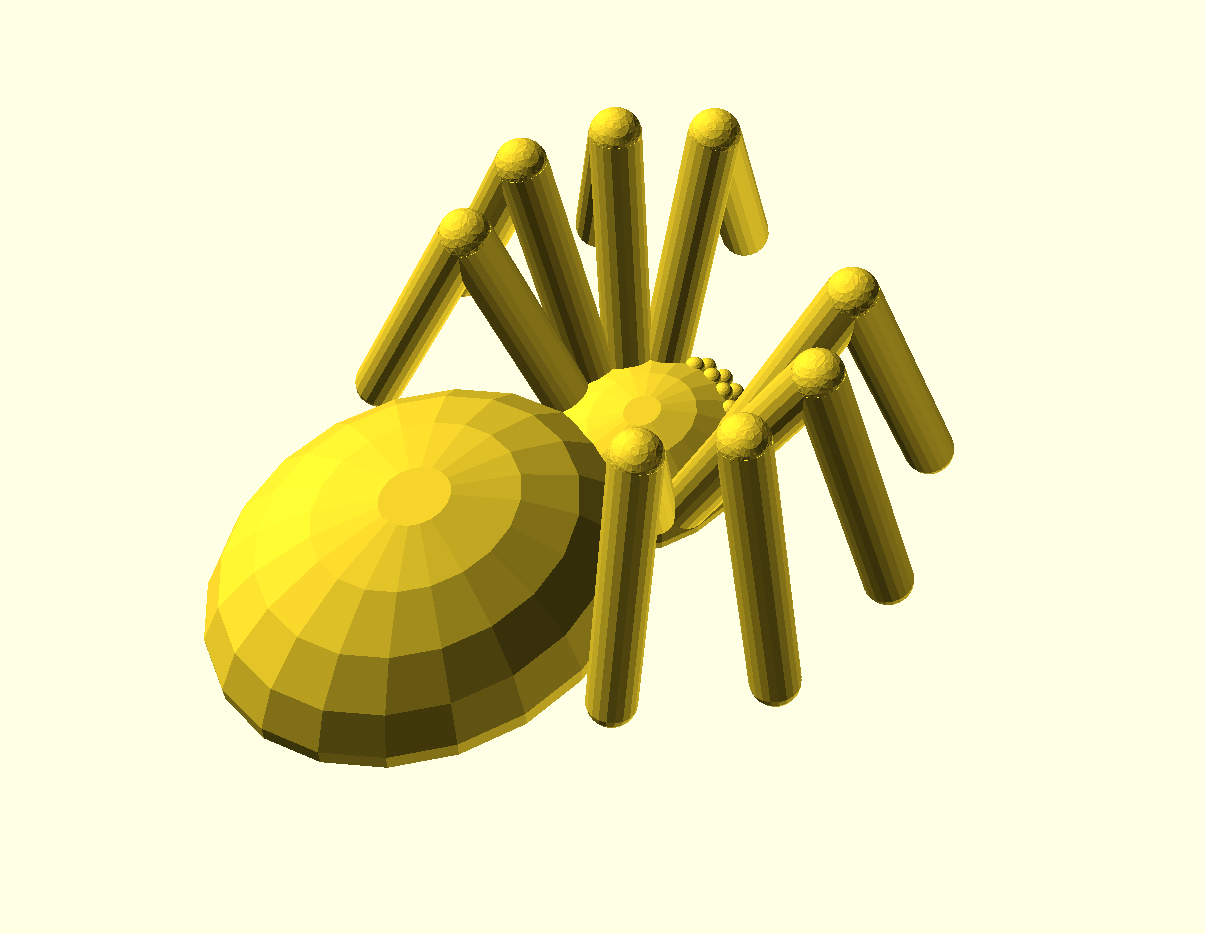} \captionsetup{font=tiny}} \quad
\subfloat[Filter Basket For Fluval Evo, Thing:\href{https://www.thingiverse.com/thing:6267303}{6267303}]{\includegraphics[width=.185\linewidth]{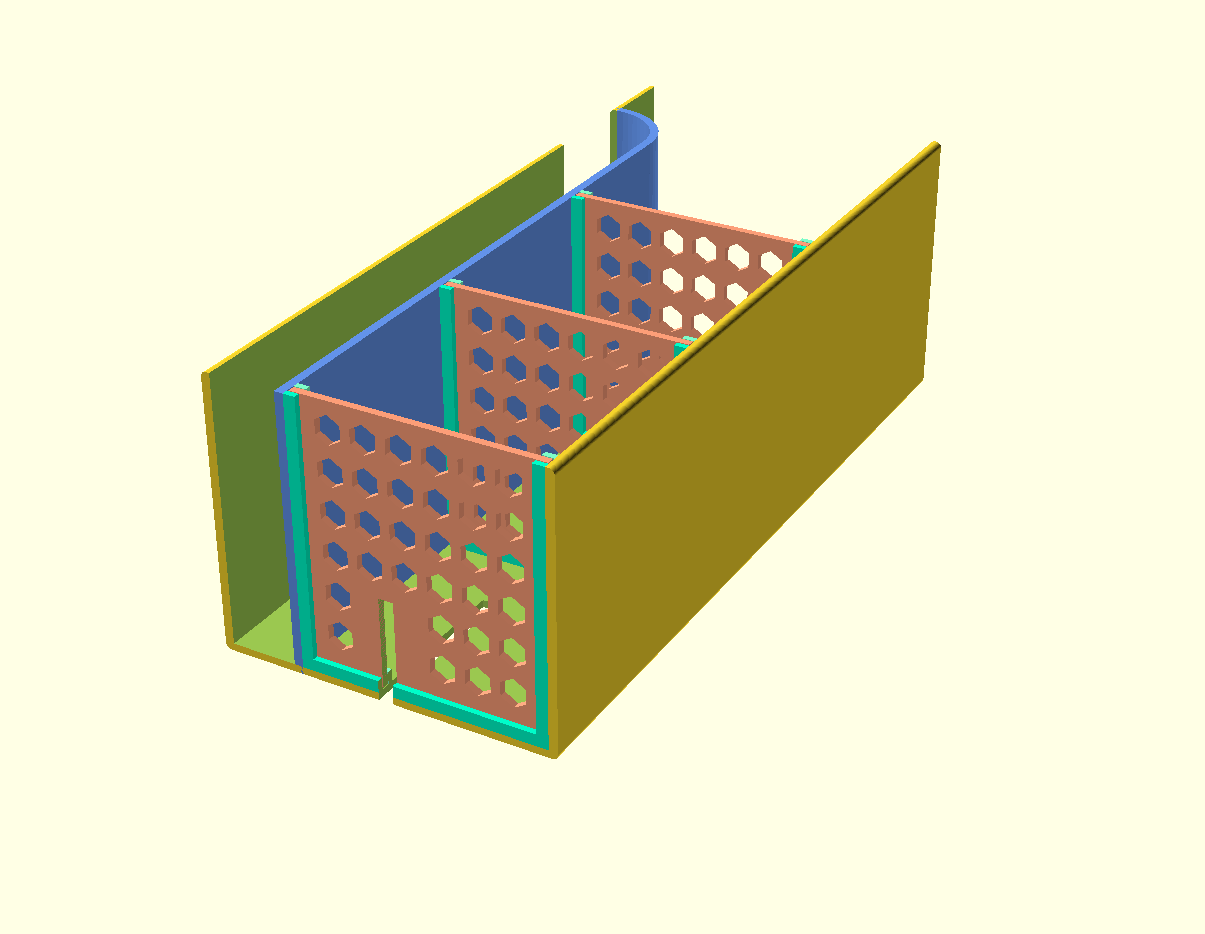} \captionsetup{font=tiny}} \quad
\subfloat[Halloween Candle, Thing:\href{https://www.thingiverse.com/thing:6267335}{6267335}]{\includegraphics[width=.185\linewidth]{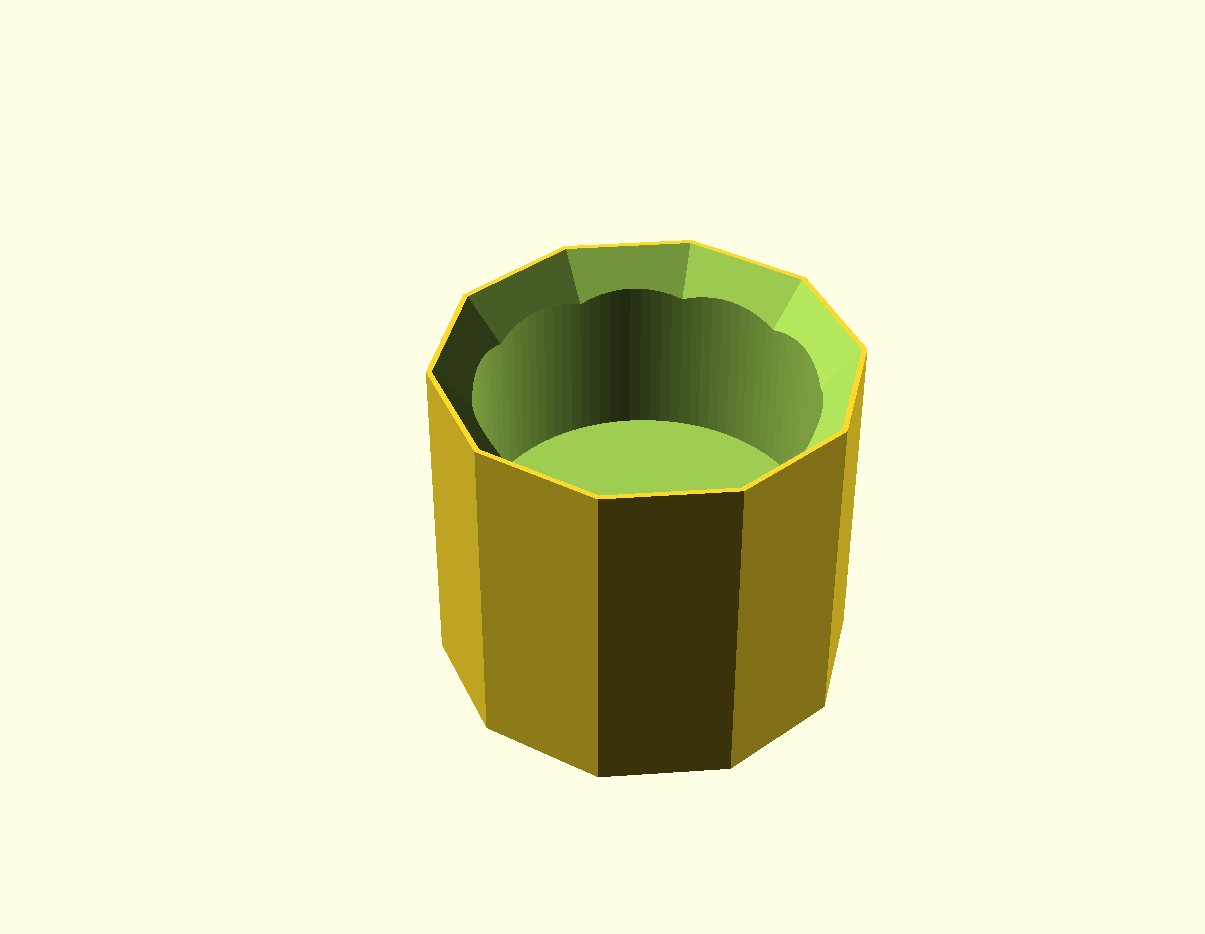} \captionsetup{font=tiny}} \quad
\subfloat[Modular Drawer Divider, Thing:\href{https://www.thingiverse.com/thing:6250410}{6250410}]{\includegraphics[width=.185\linewidth]{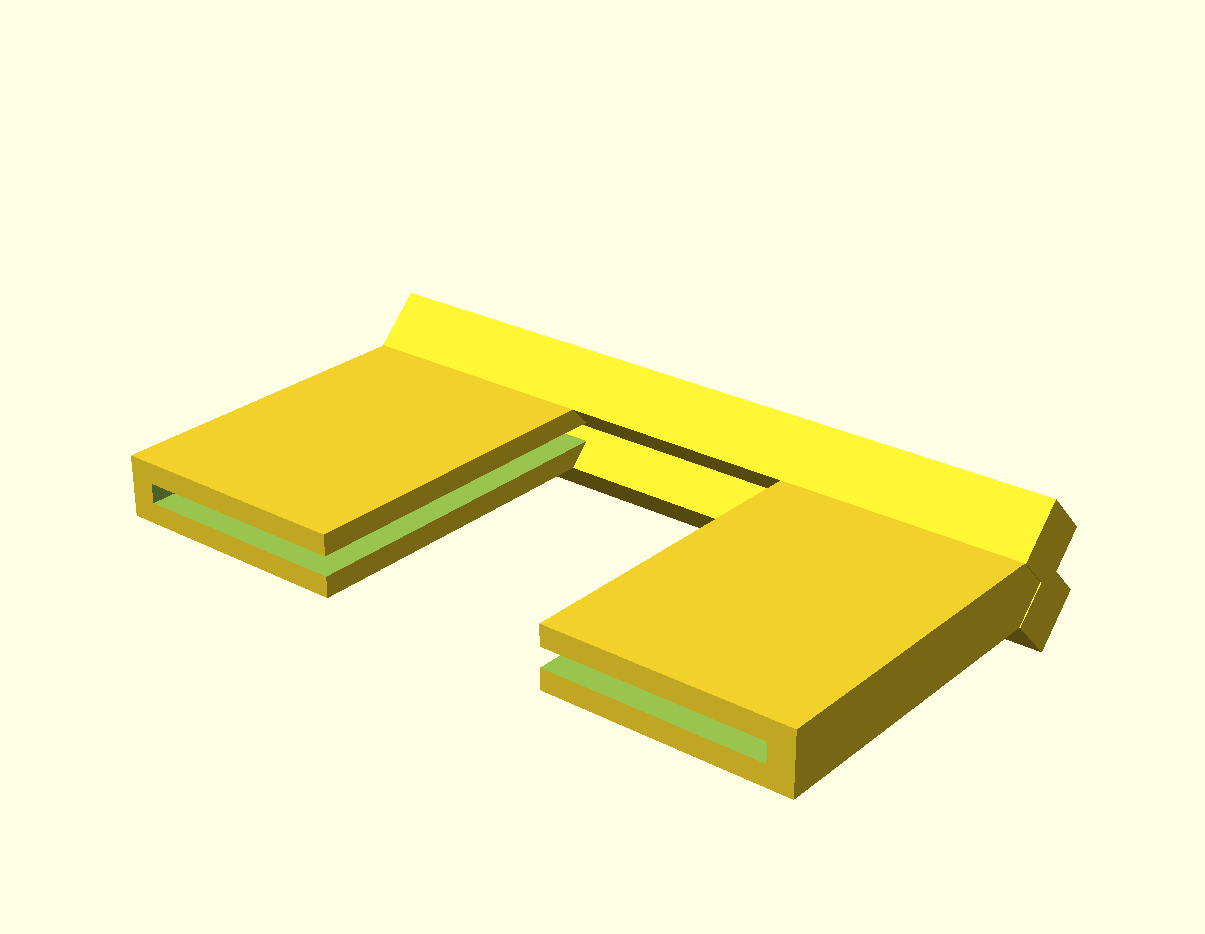} \captionsetup{font=tiny}} \quad
\subfloat[Twist-Lock Hose Flange, Thing:\href{https://www.thingiverse.com/thing:5988719}{5988719}]{\includegraphics[width=.185\linewidth]{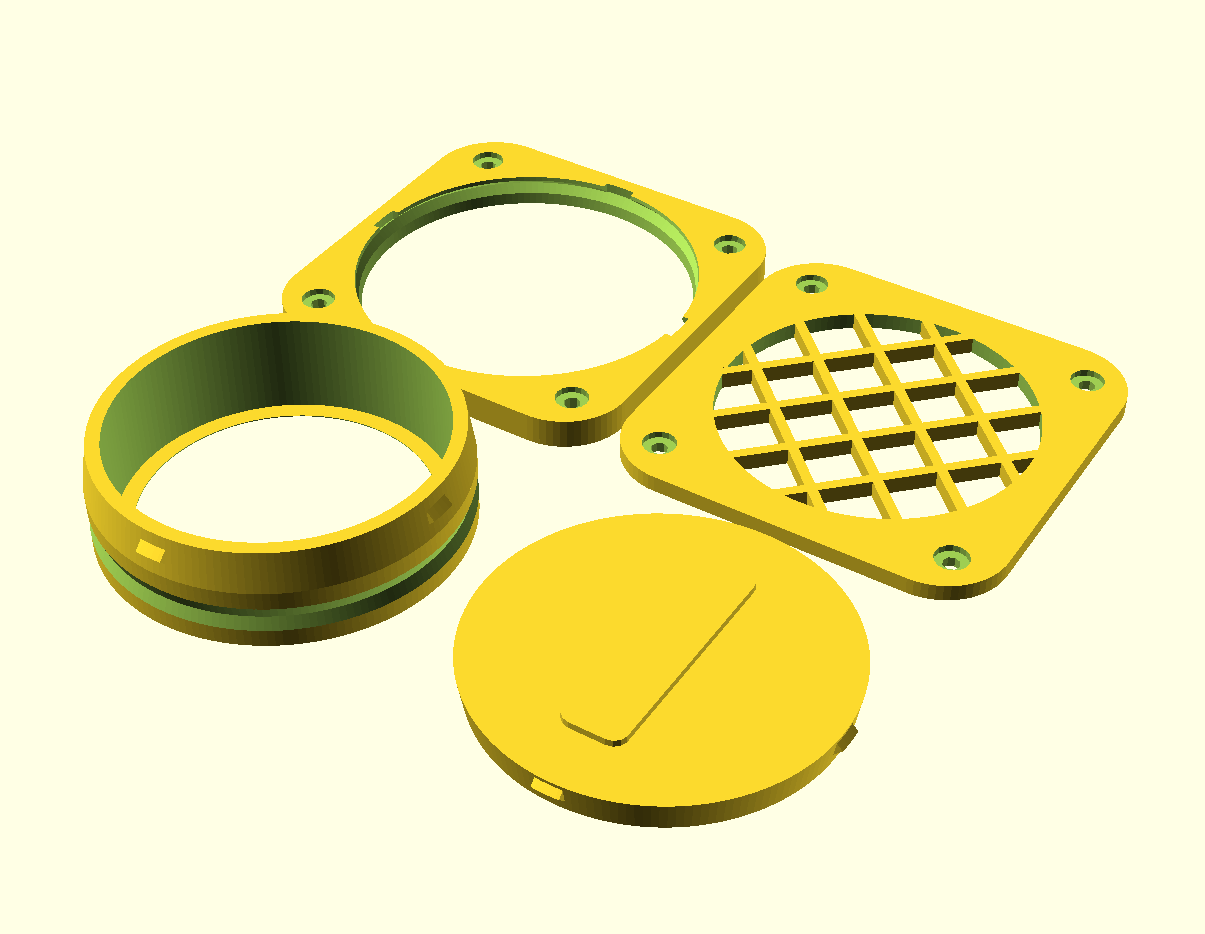} \captionsetup{font=tiny}} \quad
\subfloat[Knurled Screw Lid Container, Thing:\href{https://www.thingiverse.com/thing:6095952}{6095952}]{\includegraphics[width=.185\linewidth]{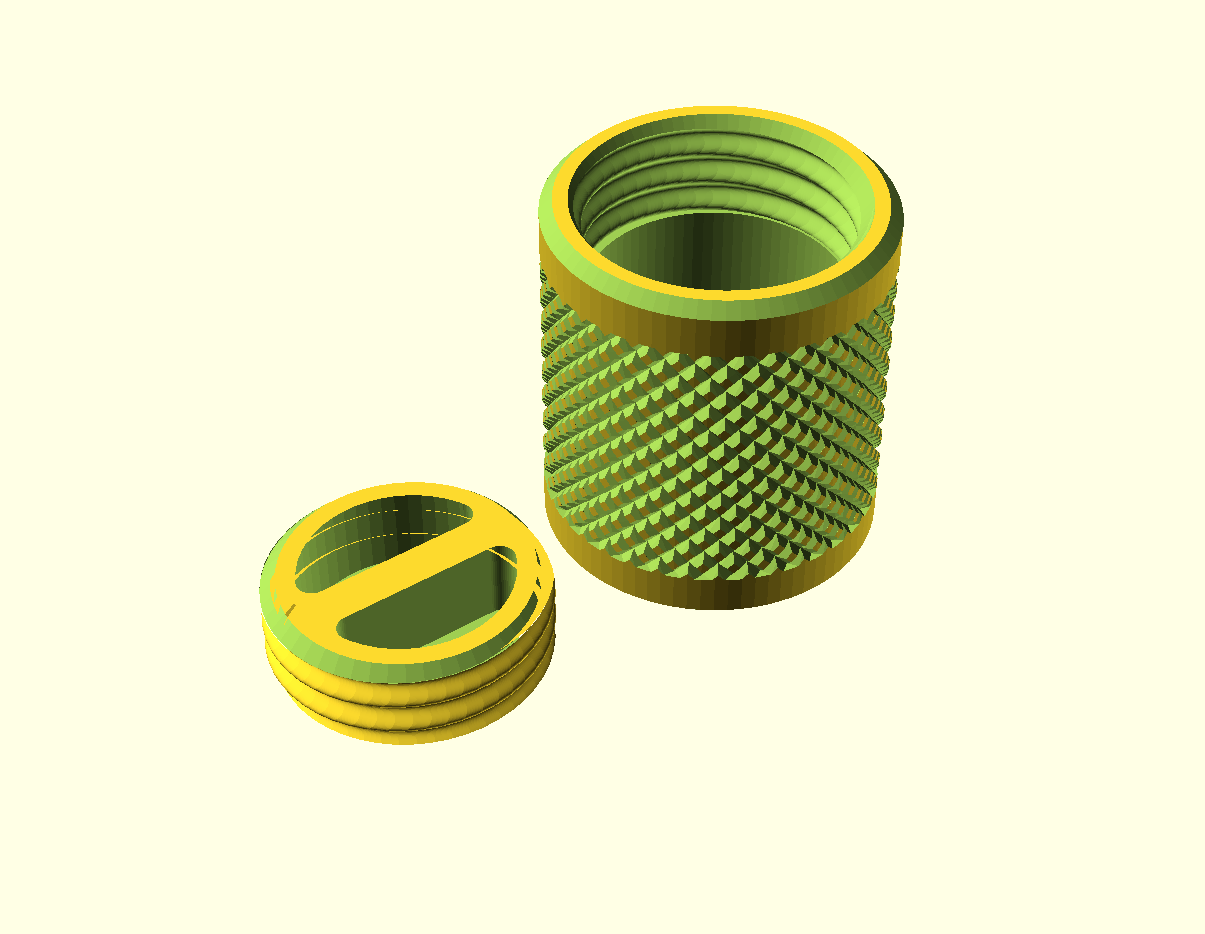} \captionsetup{font=tiny}} \quad
\subfloat[3D Snowflake Ornament, Thing:\href{https://www.thingiverse.com/thing:5673707}{5673707}]{\includegraphics[width=.185\linewidth]{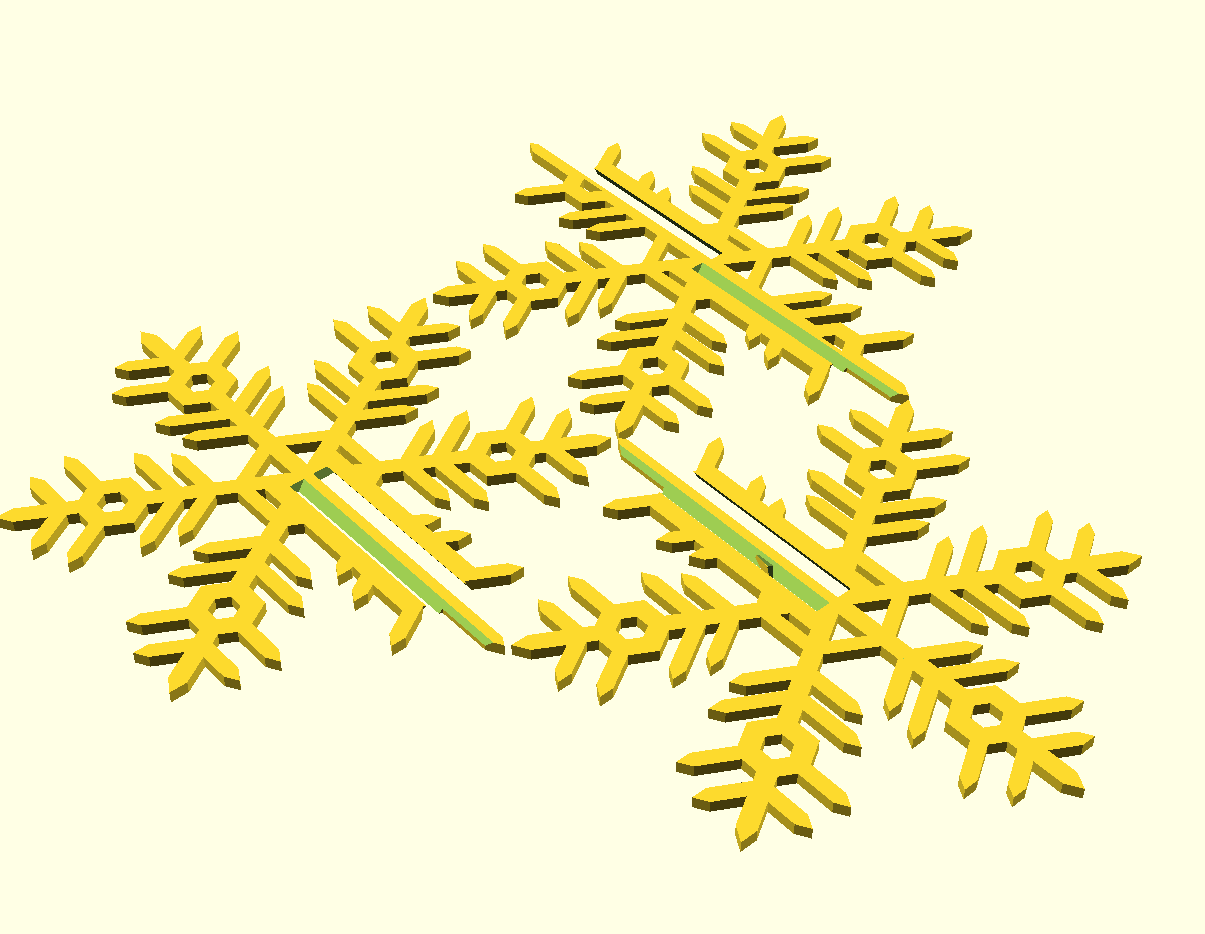} \captionsetup{font=tiny}} \quad
\subfloat[Shotgun Shell Case, Thing:\href{https://www.thingiverse.com/thing:6153068}{6153068}]{\includegraphics[width=.185\linewidth]{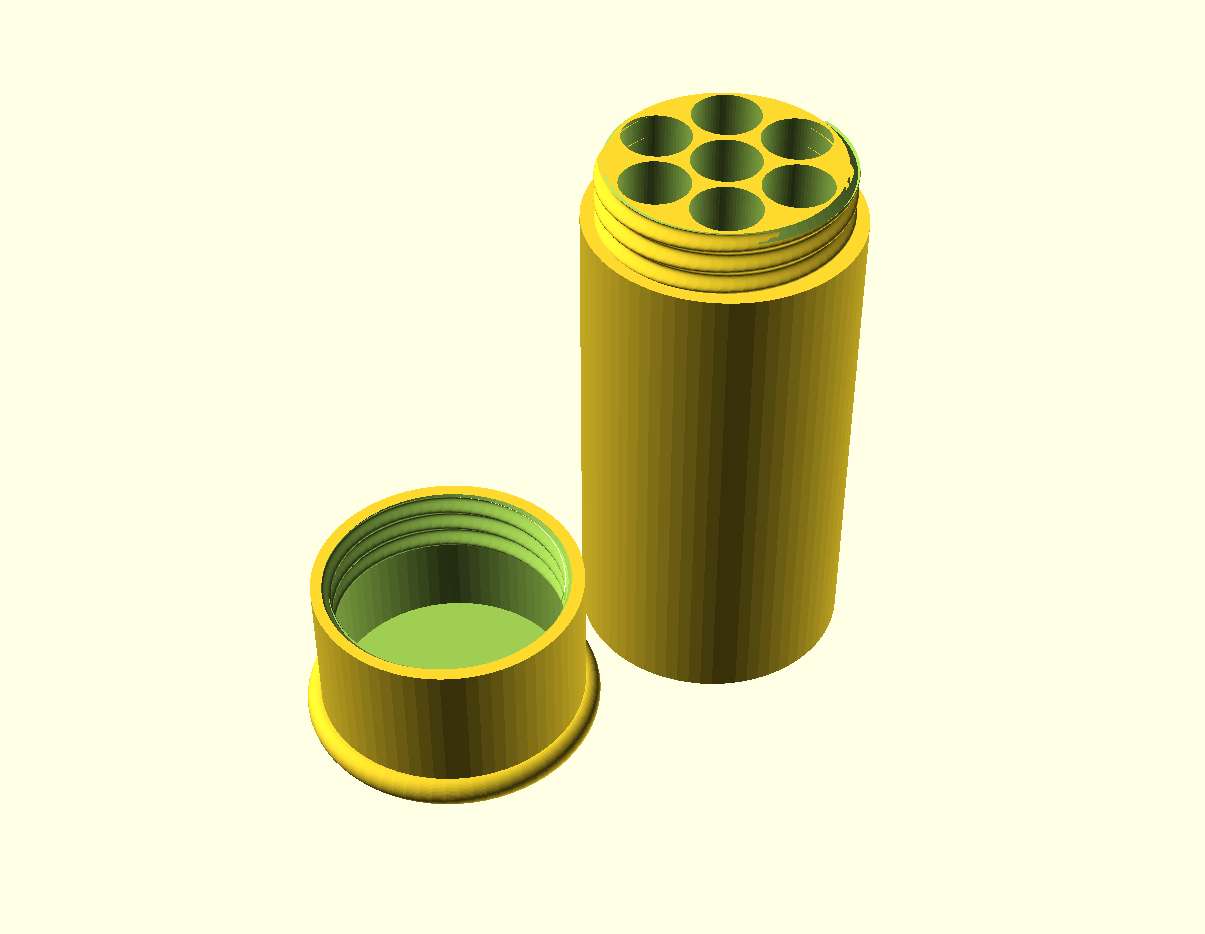} \captionsetup{font=tiny}} \quad
\subfloat[Customizable Jar/Bottle, Thing:\href{https://www.thingiverse.com/thing:6211287}{6211287}]{\includegraphics[width=.185\linewidth]{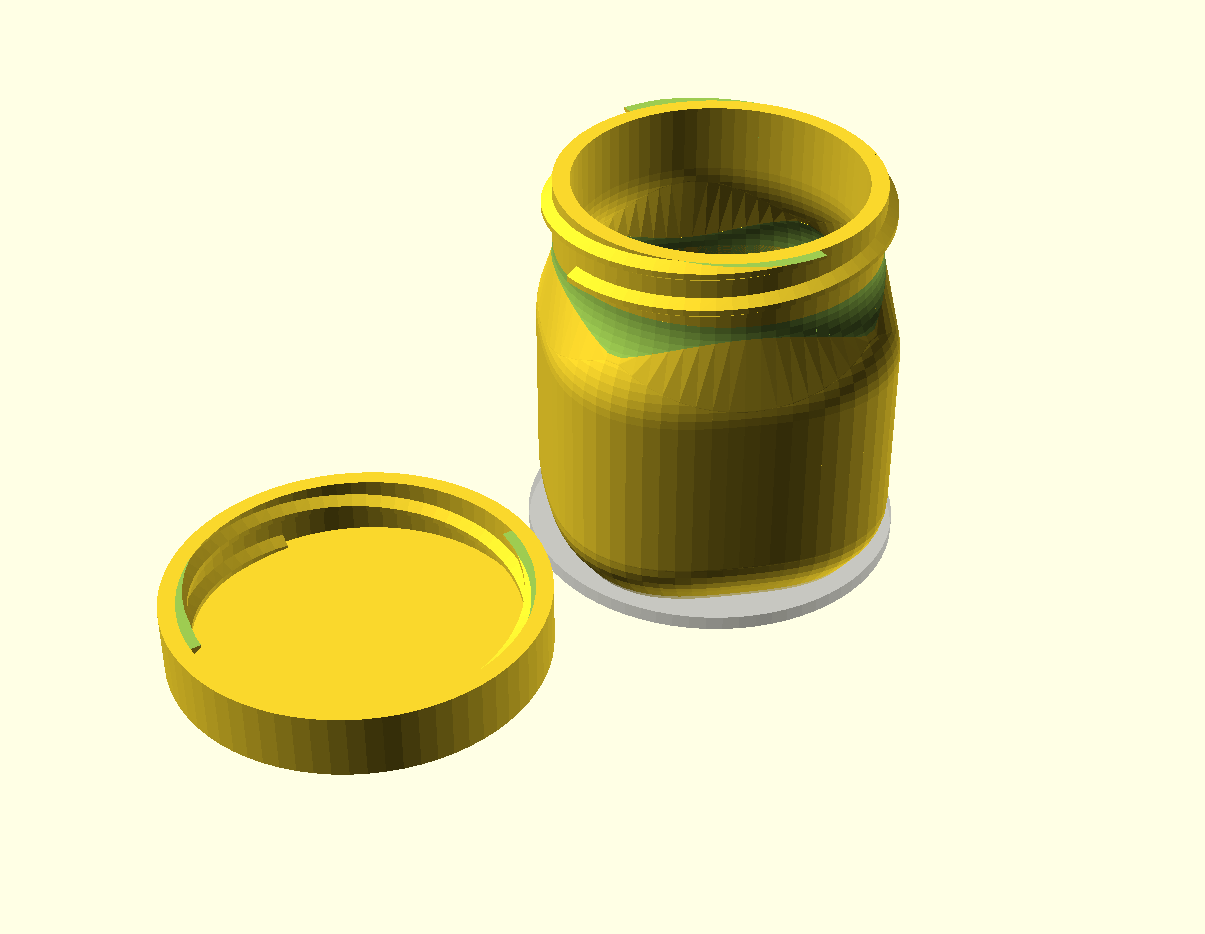} \captionsetup{font=tiny}} \quad
\subfloat[Parametrizable Rugged Box, Thing:\href{https://www.thingiverse.com/thing:5983067}{5983067}]{\includegraphics[width=.185\linewidth]{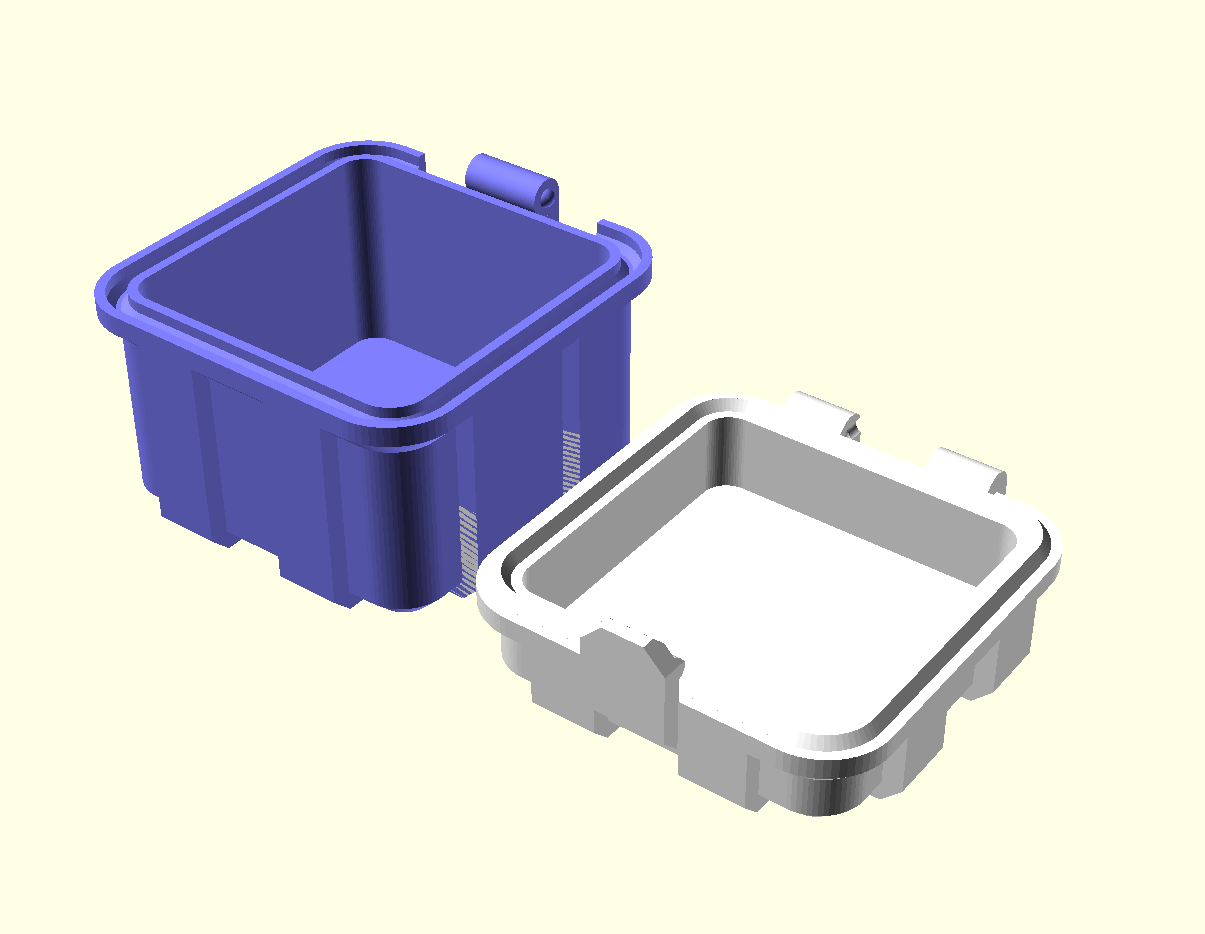} \captionsetup{font=tiny}} \quad
\subfloat[Customizable Cable Tie, Thing:\href{https://www.thingiverse.com/thing:5789087}{5789087}]{\includegraphics[width=.185\linewidth]{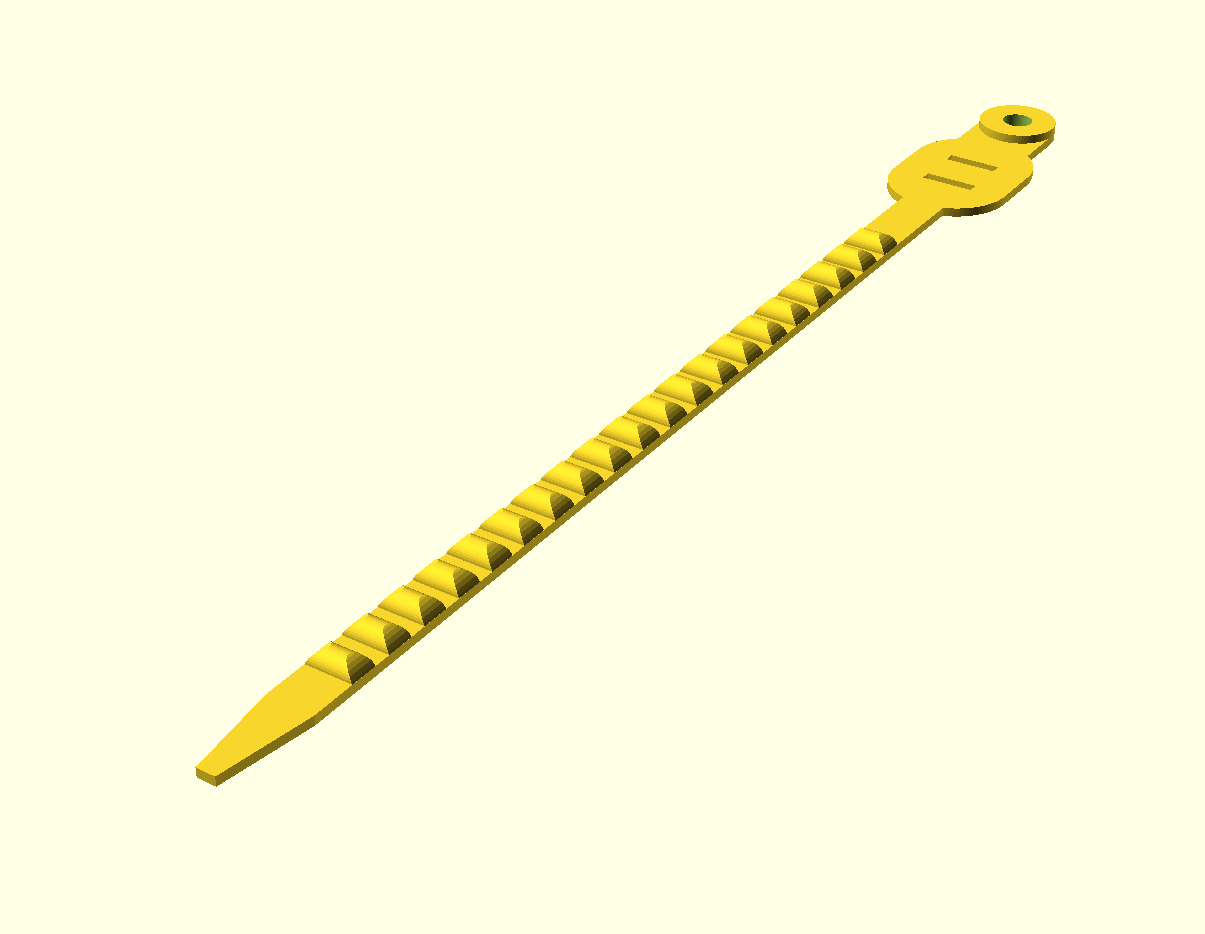} \captionsetup{font=tiny}} \quad
\subfloat[Phone Mount, Thing:\href{https://www.thingiverse.com/thing:5816088}{5816088}]{\includegraphics[width=.185\linewidth]{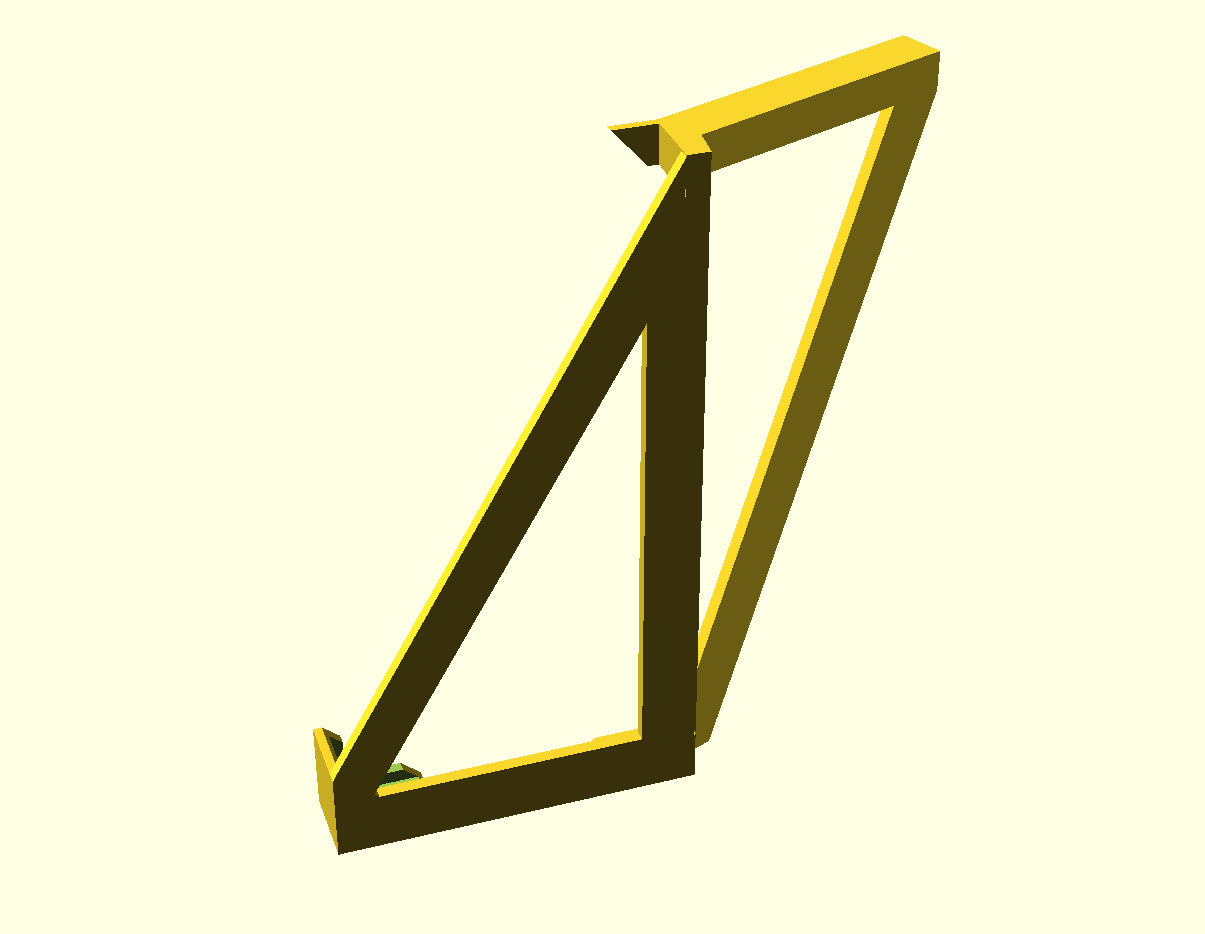} \captionsetup{font=tiny}} \quad
\subfloat[Folding Utility Knife, Thing:\href{https://www.thingiverse.com/thing:6117454}{6117454}]{\includegraphics[width=.185\linewidth]{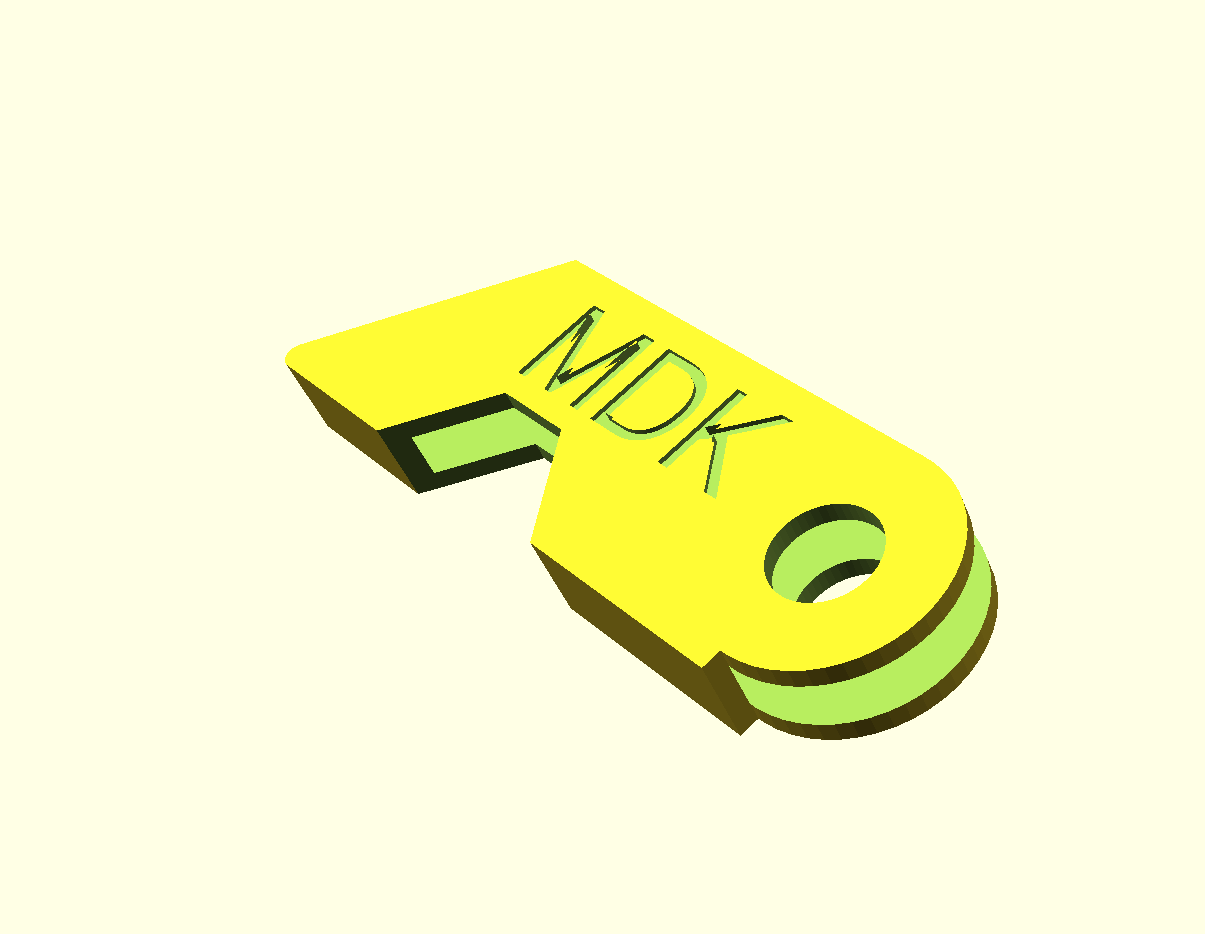} \captionsetup{font=tiny}} \quad
\subfloat[Halloween Jack-O'-Lantern, Thing:\href{https://www.thingiverse.com/thing:6221369}{6221369}]{\includegraphics[width=.185\linewidth]{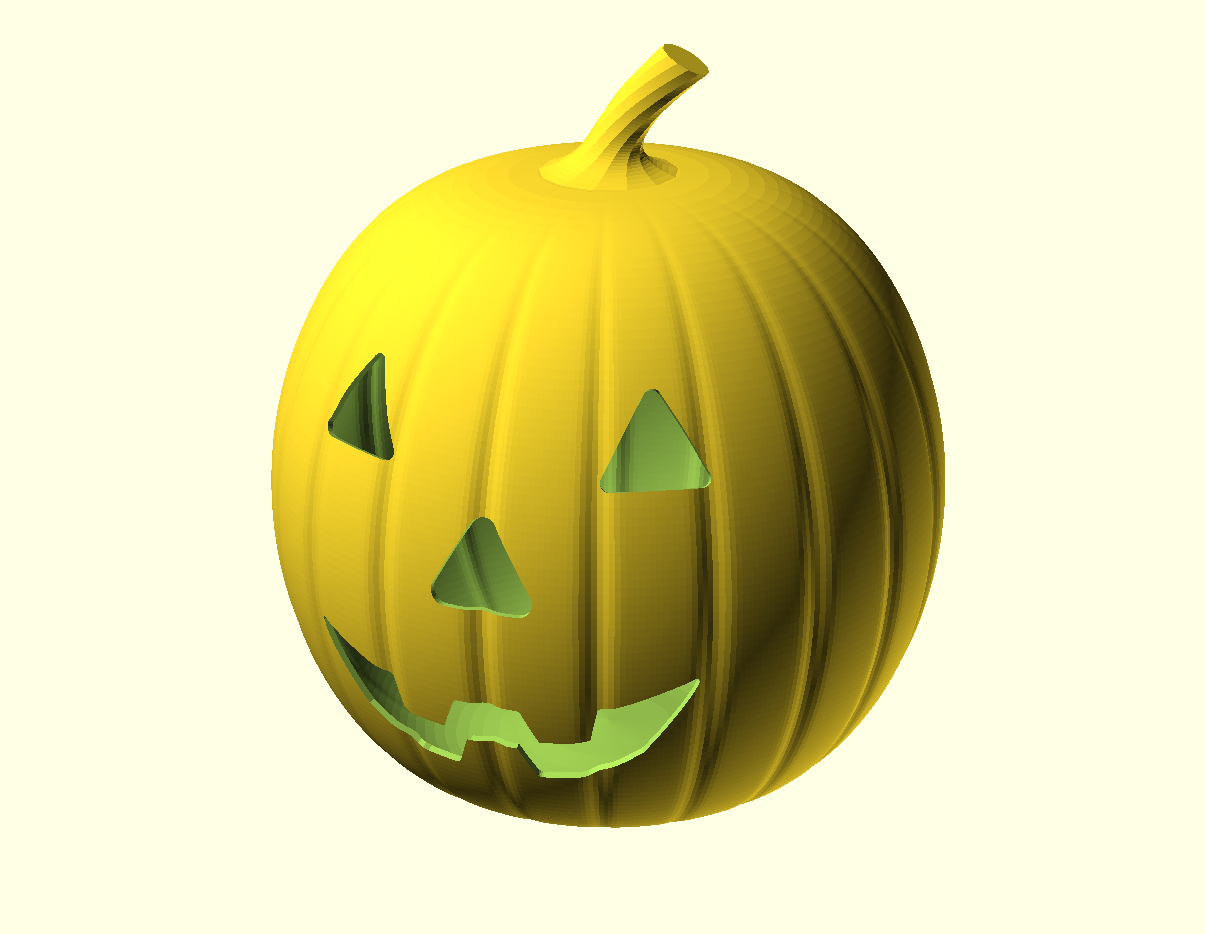} \captionsetup{font=tiny}} \quad

\caption{\os models taken from Thingiverse for the formative study.}
\label{fig:figmodels}
\end{figure*}

\aptLtoX[graphic=no,type=html]{\begin{table}
\centering
\Description{Description of Handles for Primitive Shapes: This table describes the distribution of the handles in each primitive. The first column contains the name of the primitive and a description of the distribution of the handles. The second column includes an example image of the primitive and the handles.}
\caption{Description of Handles for Primitive Shapes.}
\label{tab:control_points}
\begin{tabular}{|p{23pc}|c|}

\hline

\textbf{Cube (27 points):} Cube nodes create a grid of 3x3x3 points, including 1 point in the center, 1 at each corner (8 points), 1 in the center of each face (6 points), and 1 in the middle of each edge (12 points).

&
\vspace{2pt}
\includegraphics[width=0.65\linewidth]{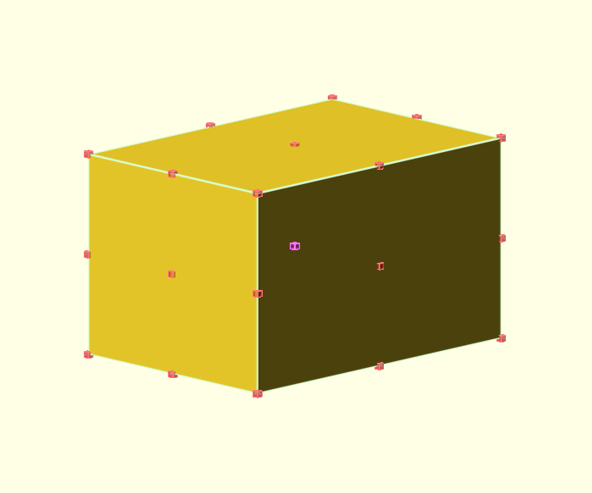}

\\
\hline

\textbf{Sphere (27 points):} Spheres create a boundary cube using diameter as size for height, width and depth. The node places handles on the boundary cube following the same distribution as the cube nodes.

&
\vspace{2pt}
\includegraphics[width=0.65\linewidth]{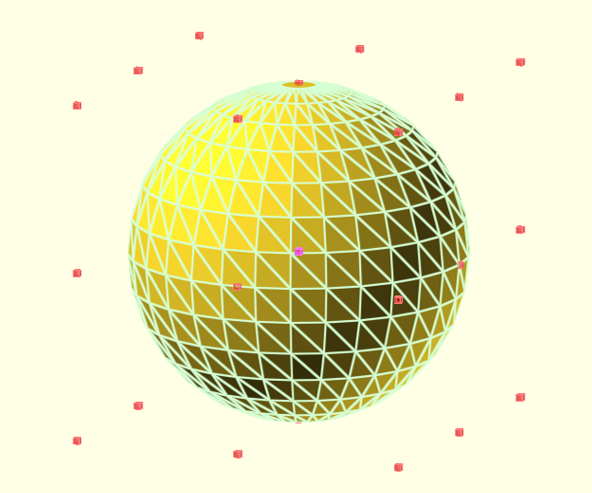}

\\
\hline

\textbf{Cylinder (27 points):} Cylinder nodes form a boundary cuboid, with its bottom and top square faces sized by the cylinder parameters \texttt{d1} and \texttt{d2}, respectively. The height of the cuboid is determined by \texttt{h}. Handles are positioned on the cuboid's boundary, following the cube nodes' distribution pattern.

&
\vspace{2pt}
\includegraphics[width=0.68\linewidth]{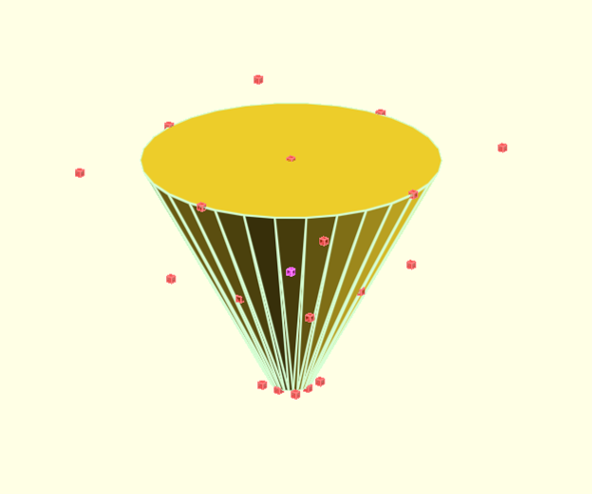}

\\
\hline

\textbf{Square (9 points):} Square nodes create a grid of 3x3 points, including 1 point in the center, 1 at each corner (4 points), and 1 at the middle of each edge (4 points).

&
\vspace{2pt}
\includegraphics[width=0.65\linewidth]{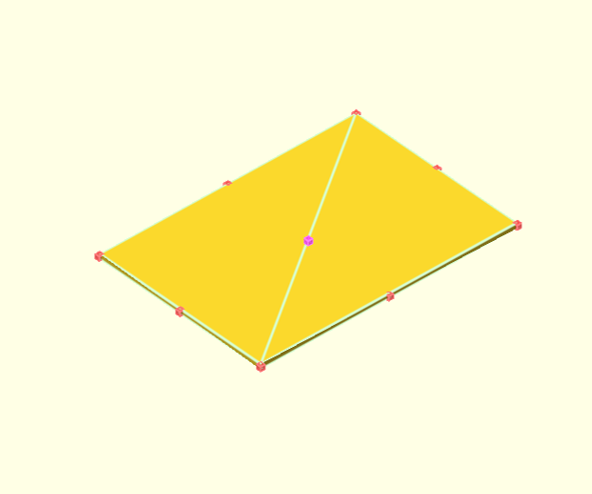}

\\
\hline
\textbf{Circle (9 points):} Circle nodes create a boundary square using the diameter as the size of the height and width. 1 point in the center and 1 at each extreme along each axis (4 points).

&
\vspace{2pt}
\includegraphics[width=0.65\linewidth]{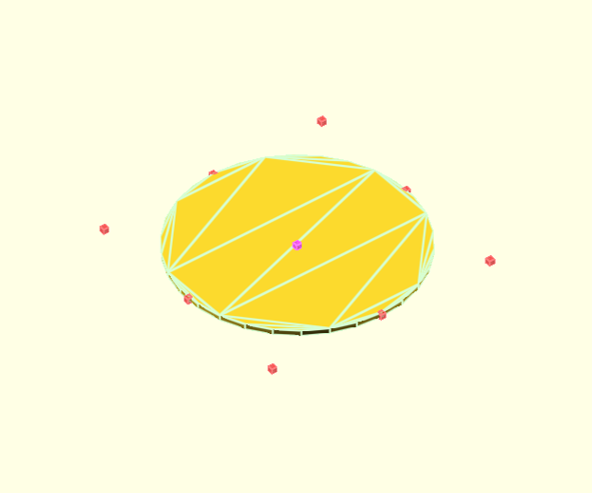}

\\
\hline

\end{tabular}
\end{table}}{\begin{table*}[!htbp]
\centering
\caption{Description of Handles for Primitive Shapes.}
\label{tab:control_points}
\resizebox{\textwidth}{!}{
\begin{tabular}{|c|c|}

\hline
\begin{minipage}{0.55\textwidth}
\textbf{Cube (27 points):} Cube nodes create a grid of 3x3x3 points, including 1 point in the center, 1 at each corner (8 points), 1 in the center of each face (6 points), and 1 in the middle of each edge (12 points).
\end{minipage}

&
\begin{minipage}{0.40\textwidth}
\centering
\vspace{2pt}
\includegraphics[width=0.65\textwidth]{gfx/controlPointsCube.png}
\vspace{2pt}
\end{minipage}

\\
\hline
\begin{minipage}{0.55\textwidth}
\textbf{Sphere (27 points):} Spheres create a boundary cube using the diameter as the size for height, width and depth. The node places handles on the boundary cube following the same distribution as the cube nodes.
\end{minipage}

&
\begin{minipage}{0.4\textwidth}
\centering
\vspace{2pt}
\includegraphics[width=0.65\textwidth]{gfx/controlPointsSphere.png}
\vspace{2pt}
\end{minipage}

\\
\hline
\begin{minipage}{0.55\textwidth}
\textbf{Cylinder (27 points):} Cylinder nodes form a boundary cuboid, with its bottom and top square faces sized by the cylinder parameters \texttt{d1} and \texttt{d2}, respectively. The height of the cuboid is determined by \texttt{h}. Handles are positioned on the cuboid's boundary, following the cube nodes' distribution pattern.
\end{minipage}

&
\begin{minipage}{0.4\textwidth}
\centering
\vspace{2pt}
\includegraphics[width=0.65\textwidth]{gfx/controlPointsCylynder.png}
\vspace{2pt}
\end{minipage}

\\
\hline
\begin{minipage}{0.55\textwidth}
\textbf{Square (9 points):} Square nodes create a grid of 3x3 points, including 1 point in the center, 1 at each corner (4 points), and 1 at the middle of each edge (4 points).
\end{minipage}

&
\begin{minipage}{0.4\textwidth}
\centering
\vspace{2pt}
\includegraphics[width=0.65\textwidth]{gfx/controlPointsSquare.png}\vspace{2pt}
\end{minipage}

\\
\hline
\begin{minipage}{0.55\textwidth}
\textbf{Circle (9 points):} Circle nodes create a boundary square using the diameter as the size of the height and width. 1 point in the center and 1 at each extreme along each axis (4 points).
\end{minipage}

&
\begin{minipage}{0.4\textwidth}
\centering
\vspace{2pt}
\includegraphics[width=0.65\textwidth]{gfx/controlPointsCircle.png}
\vspace{2pt}
\end{minipage}

\\
\hline

\end{tabular}
}
\end{table*}}
\clearpage
\includepdf[pages=-]{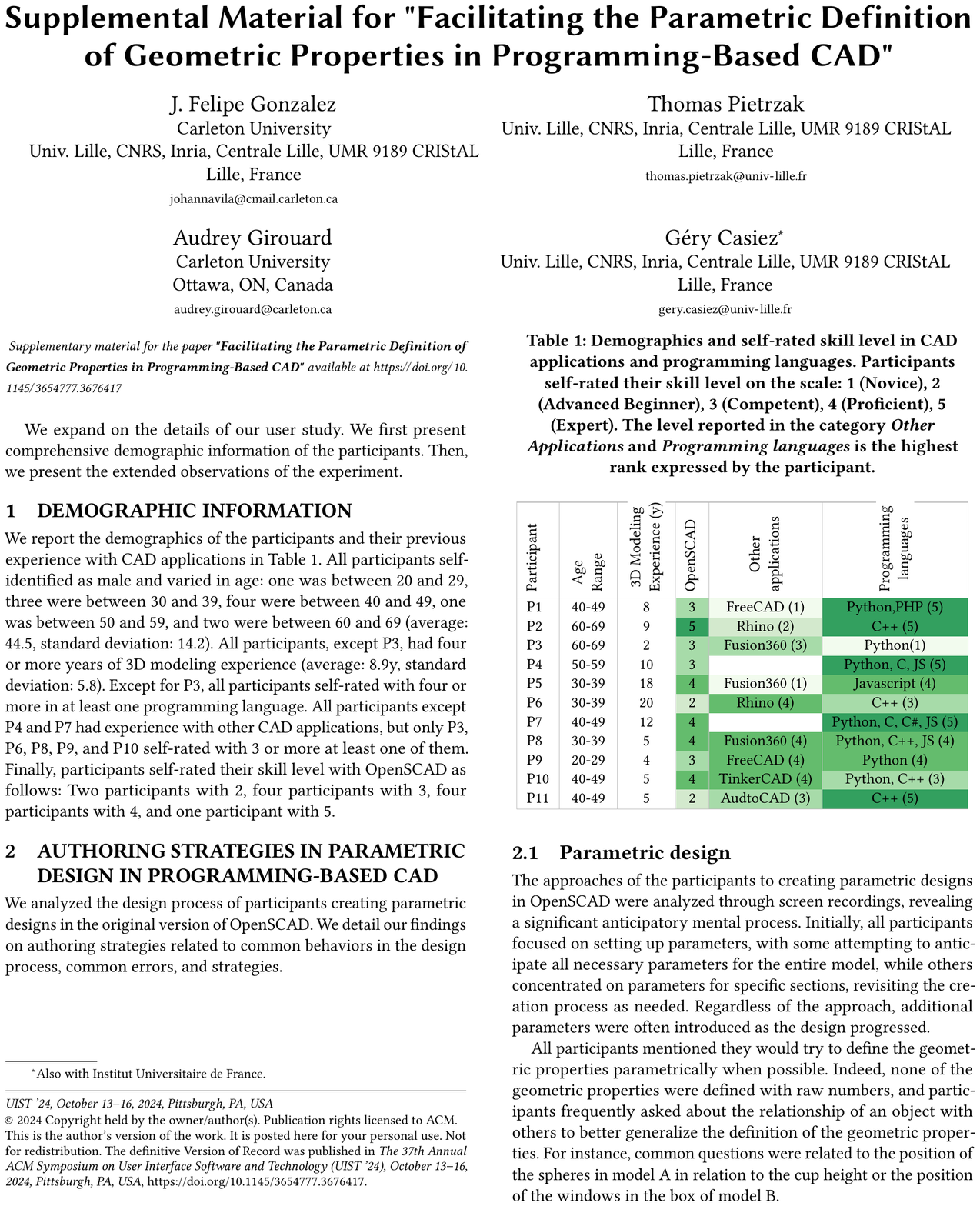}

\end{document}